\documentclass[10pt,twocolumn]{article}
\usepackage[top=1.8cm, bottom=1.8cm, left=1.8cm, right=1.8cm]{geometry}

\usepackage{amsthm}
\usepackage[keeplastbox]{flushend}
\usepackage{times}
\usepackage{booktabs}
\usepackage{url}

\usepackage[ruled]{algorithm2e}
\usepackage{fancyhdr}
\usepackage{enumitem}
\usepackage{eurosym}
\usepackage{graphicx}
\usepackage[small,bf]{caption}
\usepackage{subcaption}
\usepackage{color}
\usepackage{physics}
\newtheorem{thm}{Definition}[section]

\newtheorem{KeywordsList}[thm]{Definition}
\newtheorem{Levenshtein}[thm]{Definition}
\newtheorem{Kolmogorov}[thm]{Definition}
\newtheorem{ProbabilisticClassifiers}[thm]{Definition}
\newtheorem{TreeClassifiers}[thm]{Definition}
\newtheorem{FleissKappa}[thm]{Definition}

\usepackage{times}

\usepackage[compact]{titlesec}
\titlespacing*{\section}{0pt}{*3}{3pt}
\titlespacing{\subsection}{0pt}{*2}{2pt}
\titlespacing{\subsubsection}{0pt}{*2}{2pt}

\definecolor{linkcol}{rgb}{0,0,0.5}
\definecolor{citecol}{rgb}{0,0.5,0.3}
\definecolor{urlcol}{rgb}{0.3,0,0}

\newcommand{\descr}[1]{\smallskip\noindent\textbf{#1}}
\newcommand{\descremph}[1]{\smallskip\noindent\emph{#1}}
\newcommand{\dsrk}[1]{{\sf{\//r9k\//}}\xspace}
\newcommand{\dspol}[1]{{\sf{\//pol\//}}\xspace}

\newif\ifcomment
\commenttrue 

\ifcomment

\else  

\fi

\definecolor{heraldBlue}{rgb}{0.0,0.0,0.8}
\definecolor{heraldRed}{rgb}{0.8,0.0,0.0}
\definecolor{heraldGray}{rgb}{0.4,0.4,0.4}
\definecolor{heraldBlack}{rgb}{0.0,0.0,0.0}
\definecolor{heraldGreen}{rgb}{0.0,0.4,0.0}

\newenvironment {squishlist}
{\begin{list}{$\bullet$}
  { \setlength{\itemsep}{1pt}
     \setlength{\parsep}{1pt}
     \setlength{\topsep}{1pt}
     \setlength{\partopsep}{1pt}
     \setlength{\leftmargin}{1.5em}
     \setlength{\labelwidth}{1em}
     \setlength{\labelsep}{0.5em} } }
{\end{list}}

\renewenvironment{thebibliography}[1]{
  \begin{oldthebibliography}{#1}
    \setlength{\itemsep}{0.0em}
    \setlength{\parskip}{0.0em}
}
{
  \end{oldthebibliography}
}

\renewcommand{\footnoterule}{%
  \kern -3pt
  \hrule width 1in
  \kern 2pt
}

\makeatletter
\def\url@leostyle{%
  \@ifundefined{selectfont}{\def\UrlFont{}}%
  {\def\UrlFont{}}%
}
\makeatother
\usepackage[hyphenbreaks]{breakurl}

\definecolor{darkred}{RGB}{153,0,0}
\definecolor{darkblue}{RGB}{0,0,99}
\usepackage[colorlinks=true, linkcolor = darkred,   citecolor = darkred, urlcolor = darkblue]{hyperref}

\begin{document}
\title{\bf Detecting Cyberbullying and Cyberaggression in Social Media\thanks{This is an extended work of the authors' prior publications presented in~\cite{Chatzakou2017MeanBD}, \cite{chatzakou2017hypertext}, and \cite{chatzakou2017measuring}. Work done while the 1st author was with the Aristotle University of Thessaloniki.}}  
\date{}

\author{\Large Despoina Chatzakou$^{1}$, Ilias Leontiadis$^2$, Jeremy Blackburn$^3$, Emiliano De Cristofaro$^4$,\\
\Large Gianluca Stringhini$^5$, Athena Vakali$^6$, and Nicolas Kourtellis$^7$\\[1ex]
$^1$Center for Research and Technology Hellas, $^2$Samsung AI, $^3$SUNY Binghamton,\\ 
$^4$UCL, $^5$Boston University, $^6$Aristotle University of Thessaloniki, $^7$Telefonica Research}

\maketitle

\begin{abstract}

Cyberbullying and cyberaggression are increasingly worrisome phenomena affecting people across all demographics. 
More than half of young social media users worldwide have been exposed to such prolonged and/or coordinated digital harassment.
Victims can experience a wide range of emotions, with negative consequences such as embarrassment, depression, isolation from other community members, which embed the risk to lead to even more critical consequences, such as suicide attempts.

In this work, we take the first concrete steps to understand the characteristics of abusive behavior in Twitter, one of today's largest social media platforms.
We analyze $1.2$ million users and $2.1$ million tweets, comparing users participating in discussions around seemingly normal topics like the NBA, to those more likely to be hate-related, such as the Gamergate controversy, or the gender pay inequality at the BBC station.
We also explore specific manifestations of abusive behavior, i.e., cyberbullying and cyberaggression, in one of the hate-related communities (Gamergate).
We present a robust methodology to distinguish bullies and aggressors from normal Twitter users by considering text, user, and network-based attributes.
Using various state-of-the-art machine learning algorithms, we classify these accounts with over $90\%$ accuracy and AUC.
Finally, we discuss the current status of Twitter user accounts marked as abusive by our methodology, and study the performance of potential mechanisms that can be used by Twitter to suspend users in the future.
\end{abstract}

\section{Introduction}\label{sec:intro}

In today's digital society, cyberbullying and cyberaggression are serious and widespread issues affecting an increasingly high number of Internet users, mostly at their sensitive teen and young age.
In a way, while physical bullying is somewhat limited to particular places or times of the day (e.g., school hours), its digital counterpart can instead occur anytime, anywhere, with just a few taps on a keyboard.
Cyberbullying and cyberaggression can take many forms; there is no generally accepted definition, and cyberaggression is often considered a form of cyberbullying~\cite{grigg2010cyber,Smith2008CyberbullyingNature,Tokunaga2010CriticalReviewCyberbullyingVictimization}.
Overall, the former typically denotes repeated and hostile behavior performed by a group or an individual, while the latter intentional harm delivered via electronic means to a person or a group of people who perceive such acts as offensive, derogatory, harmful, or unwanted~\cite{grigg2010cyber}.
Similar to bullying in face-to-face social interactions, two main characteristics indicative of cyberbullying behavior are the repetition intensity over time and the power imbalance between the victims and the bullies.

Cyberbullying was not taken seriously in the early Web era: the typical advice was to ``just turn off the screen'' or ``disconnect'' your device~\cite{screenbullying1}.
However, as Web's proliferation and the extent of its consequences reached epidemic levels~\cite{cyberbullyingResearchCenter}, such behavior can no longer be ignored; in 2017, based on a survey conducted from the Pew Research Center~\cite{cyberbullying-facts-adults2017}, 41\% of Americans have been personally subjected to harassing behavior online, while 66\% have witnessed these behaviors directed at others.
Furthermore, about 50\% of young social media users reported being bullied online in various forms~\cite{factsaboutbullying}.
Overall, 27\% of students report being cyberbullied at some point in their lifetimes~\cite{cyberbullying-facts}.
Even more worrisome is that 15\% of high school students of in grades 9 to 12, and 9\% of students in grades 6 to 12, have experienced cyberbullying.

\subsection{Challenges} These concerns motivate the demand to design methods and tools for early detection and prevention of such abusive behavior, especially as it evolves in social media platforms.
Many complexities are involved in developing efficient and effective methods for detecting such online phenomena, due to the:
(i)  heterogeneity of users with respect to their culture norms and their demographics, 
(ii)  transient nature of the problem (such phenomena are unpredictable and they may spike or stop unexpectedly), 
(iii) anonymity capability offered in social media which may ease bullies to launch attacks without fearing consequences, and 
(iv) multiple aggression and bullying forms beyond just obviously abusive language (e.g., via constant sarcasm and trolling).

Such strong challenges hinder the effective design of automatic detection methods.
However in this paper, we address most of these difficulties by an effective approach focused on Twitter, which presents some characteristics posing even additional difficulties w.r.t.~cyberbullying and cyberaggression prevention.
These difficulties are due to:
(a) \textit{non formal content}, since tweets source is a short text often full of grammar and syntactic flaws, making it harder to rely on natural language processing methods to extract text-based attributes and characterize user interactions;
(b) the provision of fairly \textit{limited context} in each tweet, thus, taken on its own, an aggressive tweet may be disregarded as normal text, whereas, read along with other tweets, either from the same user or in the context of aggressive behavior from multiple users, the same tweet could be characterized as bullying, 
(c) \textit{spam accounts intensity} since despite extensive work on spam detection in social media~\cite{GiatsoglouCSFV15,stringhini2010detecting,Wang2010SpamDetectionTwitter}, Twitter is still full of spam accounts~\cite{Chen2015SpamTweets}, 
often using vulgar language and exhibiting behavior (repeated posts with similar content, mentions, or hashtags) that could also be considered as aggressive or bullying actions.

In particular, in this paper we identify and address the following open research questions:
\begin{itemize}
\item  \textbf{RQ1: } What characteristics differentiate abusive from normal users based on their activity on diverse Twitter communities?
\item  \textbf{RQ2: } Can we design a machine learning methodology to automatically and effectively detect such abusive behavior and users?
\item  \textbf{RQ3: } How has Twitter addressed the problem of abusive users in its platform? What are the characteristics of users who were suspended? Can we approximate this suspension mechanism?
\end{itemize}

This study builds upon our previous work on detecting abusive behavior on Twitter~\cite{chatzakou2017measuring,chatzakou2017hypertext,Chatzakou2017MeanBD}.
These works laid the foundations for gaining an initial understanding of what abusive activity looks like on Twitter, by comparing the posts and users of a hate-related community vs. randomly selected users.
Furthermore, past works investigated the core differences among distinct user categories (i.e., bullies, aggressors, spam, and normal users), as well as the behavioral patterns of abusive vs. normal users in relation to Twitter's statuses (i.e., deleted and suspended).

\subsection{Contributions}
Overall, this paper makes the following contributions:

\textit{RQ1:} We examine the behavior of users participating in different types of groups, from well-established and popular communities discussing general issues and topics (like NBA), to more hate-related, but still well organized, communities that focus on specific issues (like the Gamergate controversy~\cite{Massanari09102015}).
We also consider communities that are newer and less organized, and focus on trending topics like gender inequality in salaries.
We investigate a set of attributes for the users in these communities, and how all such extracted attributes are associated with abusive behavioral patterns.
In fact, the study is performed in relation to activity and emotional properties of these communities and users, providing a better conceptualization of abusive behavior on Twitter.

\textit{RQ2:} We use the learnings of this investigation to inform a methodology able to distinguish abusive users (i.e., bullies and aggressors) from the normal.
To automatically detect abusive users, we build a ground truth dataset, and train and test machine learning methods, from more traditional classifiers to deep neural networks, based on a wide variety of behavioral attributes.

\textit{RQ3:} We study in more depth the Twitter suspension mechanism, in relation to the observed users' behavioral patterns, testing at the same time whether we can emulate Twitter's suspension mechanism using typical machine learning methods.

\subsection{Roadmap} 
We start, in Section~\ref{sec:data}, with an overview of the datasets used in the paper, as well as the steps taken for the data collection and preprocessing.
Specifically, for the data collection we rely on the Twitter social media platform and its streaming API.
The preprocessing step consists of both cleaning of texts (e.g., removal of stop words and punctuation marks), as well as spam content removal.
Then, in Section~\ref{sec:data-analysis}, we analyze the behavioral patterns exhibited by users involved in different communities (i.e., Gamergate, NBA, and BBCpay), and what differentiates them from the baseline (random) users.
The analysis is performed on a set of activity attributes (e.g., number of posted tweets, number of followers and friends, account age) and emotional attributes (e.g., sentiment and emotions).
The objective of such an analysis is to understand better the characteristic ways in which users from diverse communities act.

Section~\ref{sec:ground-truth} presents the process followed to create a ground-truth dataset suitable for distinguishing among bullies, aggressors, spammers, and normal users.
Since analyzing single tweets does not provide enough context to discern if a user is behaving in an aggressive or bullying way, we group tweets from the same user, based on time clusters, into {\em sessions}.
This allows us to analyze contents of sessions rather than single tweets.
Thus, based on the sessionized tweets, we build ground truth (needed for machine learning classification, as explained next) using human annotators.
For the building of ground truth, we use a crowdsourced approach by recruiting workers who are provided with a set of tweets from a user, and are asked to classify them according to the previously mentioned four labels.

Section~\ref{sec:features} discusses a set of 38 features extracted from the annotated dataset against the four classes of users considered.
These features are extracted from both tweets and user profiles.
More specifically, we extract user-, text-, and network-based features such as the number of followers, tweets, hashtags, etc.
This analysis helps us select appropriate features to speed up the training of the machine learning model, and improve its quality~\cite{Kira1992PracticalApproachToFeatureSelections}.
Based on this analysis, we focus on the attributes that help distinguish better the considered user categories, and can be fed into the classification process.
In contrast to the previous sections (Sections~\ref{sec:data-analysis} and~\ref{sec:ground-truth}) where the analysis is conducted on tweet-level, in Sections \ref{sec:features} and \ref{sec:machine-learning} the analysis is performed on user-level, i.e., the ground truth dataset is used where we group tweets under the same user.

\begin{table*}[!t]
\centering
\resizebox{.75\textwidth}{!}{
\centering
\begin{tabular}{lllllll}
\hline
               & \textbf{Period}  & \textbf{Tweets} & \textbf{\#Users} & \textbf{Size (cleaned)} & \textbf{Users (cleaned)} & \textbf{Sections} \\ \hline
Baseline       & June-August 2016 & $1M$          & $610k$           & 70\%                    & 73\%                     & 3, 4, 5, 6, 7          \\
Gamergate      & June-August 2016 & $600k$        & $312k$           & 69\%                    & 58\%                     & 3, 4, 5, 6, 7          \\
NBA            & July 2017        & $400k$        & $202k$           & 57\%                    & 66\%                     & 3, 7.1, 7.2            \\
BBC gender pay & July 2017        & $100k$        & $64k$            & 69\%                    & 75\%                     & 3, 7.1, 7.2            \\ \hline
\end{tabular}}
\vspace{-0.2cm}
\caption{An overview of our datasets, with number of tweets, users involved, collection period, and where in the paper each dataset was used.}
\label{tbl:datasets}
\vspace{-0.2cm}
\end{table*}

Section~\ref{sec:machine-learning} presents the classification process, that is, the machine learning techniques used to model and predict online bullying and aggressive user behavior.
The classification process is performed using the aforementioned attributes and the created ground truth dataset.
Four classification setups are tested to assess the feasibility of detecting abusive behavior on Twitter: (i) detecting bullies and aggressors out of spam and normal users,  (ii) detecting bullies, aggressors, and normal users - having eliminated spammers from our dataset, (iii) distinguishing offensive users overall, by assuming that bullying and aggression is a united, abusive behavior, and (iv) distinguishing offensive users from normal after eliminating spammers from our dataset.

In Section~\ref{sec:twitter-reaction}, we present an analysis of the users' account status on Twitter, and compare this status with the labels provided by the annotators.
Furthermore, and building on our knowledge of features extracted from tweets and user profiles, we build a new machine learning method based on similar features presented earlier; this method attempts to emulate the Twitter suspension mechanism.
We discuss our classification results, and recent efforts performed by Twitter to eliminate abusive behavior from the platform.

In Section~\ref{sec:related-work}, we review the related work, and provide a discussion on how the present manuscript contributes to the state-of-art.
Finally, we conclude in Section~\ref{sec:conclusions} with a set of learnings and extended discussion.

\section{Datasets}\label{sec:data}
For this study, we gathered texts from a highly popular social media platform, i.e., Twitter, which is comprised of more than $300M$ connected users on a monthly basis~\cite{twitterStats} and supports the daily broadcasting of short burst posts (i.e., tweets) to the online community.
More specifically, we rely on Twitter's Streaming API which provides free access to $1\%$ of all tweets. 
The API returns each tweet in a JSON format, with the content of the tweet, some metadata, such as creation time and whether it is a reply or a retweet, as well as information about the poster, such as username, followers, friends, and number of total posted tweets.

Overall, two types of data were gathered: (i) two datasets that are more prone to contain abuse and hate-related cases, with $700k$ tweets in total (i.e., Gamergate: $600k$, BBCpay: $100k$), and (ii) two datasets which are expected to be of more ``normal'' behavior, with a total of $1.4M$ tweets (i.e., NBA: $400k$, baseline: $1M$).
The gathering of two abusive related datasets is made in order to ensure that the used method for detecting online abusive cases is generalizable and does not only apply at a specific topic under investigation.
The selection of a less hate-related, but still very popular, topic (i.e., NBA) was made to test whether the hate-related attributes are significantly related or not with any popular online topic on Twitter, and not just to the abusive-related topics.
Table~\ref{tbl:datasets} provides an overview of the datasets used in this paper (the `cleaned' columns indicate the percentage of tweets and users remained after the data preprocessing - see Section~\ref{subsec:datapreprocessing}, while the `reference sections' column indicates which datasets were used in each section).

Sections~\ref{subsec:datacollection} and~\ref{subsec:datapreprocessing} summarize all fundamental concepts and processes required for creating the aforementioned datasets, with an emphasis on the data collection and preprocessing, respectively.
Finally, Section~\ref{subsec:dataoverview} overviews the collected data by analyzing the users' posting behavior, as well as the discussed topics.

\subsection{Data Collection}\label{subsec:datacollection}
To create an abuse-related dataset, i.e., a dataset which contains abusive behaviors with high probability, previous works rely on a number of (seed) words which are highly related with the manifestation of abusive/aggressive events.
In this sense, next we present a step-by-step process which can be followed to ensure that the collected data will contain an adequate number of abusive cases.

\descremph{Seed keyword(s).}
The first step is to manually select one or more seed keywords, which are likely related to the occurrence of abusive incidents.
Good examples are the \#GamerGate, \#BlackLivesMatter, and \#PizzaGate.
In addition to such seed words, a set of hate- or curse-related words can also be used, e.g., words manually extracted from the Hatebase database (HB)~\cite{hatebase}, to start collecting possible abusive texts from social media sources.
Therefore, at time $t_1$, the list of words to be used for filtering posted texts includes only the seed word(s), i.e., $L(t_1) = <seed(s)>$.

In our case, we focus on Twitter for the data collection process.
So, initially we obtain the sample public tweets and parse them to select all tweets containing the seed word depending on the dataset to be built.
More specifically, for the dataset built around the Gamergate controversy, the \#GamerGate was used as a seed word, for the \emph{BBC gender pay controversy} related dataset the \#BBCpay, while for the \emph{NBA} dataset the \#NBA.
Such hashtags serve as seeds for an automatic snowball sampling of other hashtags likely associated with abusive behavior.

\descremph{Dynamic list of keywords.}
In addition to the seed keyword(s), further filtering keywords can be used to select abusive-related content.
The list of the additional keywords can be updated dynamically in consecutive time intervals based on the posted texts during these intervals.
Definition~\ref{KeywordsList} shows the state of the keyword list, $L(t_i)$, at a specific time interval, $t_i$.
Depending on the topic under examination, i.e., if it is a popular topic or not, the creation of the dynamic keywords list can be split to different consecutive time intervals.

\theoremstyle{Keywords list}
\begin{KeywordsList}{\textbf{Keywords list.}}
\label{KeywordsList}
In time instances $t_i$ within the set $T = \{t_1, t_2, ..., t_n\}$, and assuming a list of seed words $seed(s)$, the keywords list $L(t_i)$ equals to:
\begin{center}$L(t_i) = <seed(s), kw_1, kw_2, kw_N>$,\\\end{center}
\noindent
where $kw_j$ is the $j$th top keyword in time period $\Delta T = t_i - t_{i-1}$.
\end{KeywordsList}

To maintain the dynamic list of keywords for the time period $t_{i-1} \rightarrow t_{i}$, we should investigate the texts posted in this time period.
The $N$ keywords that were found during that time should be extracted, to compute then their frequency and rank them into a temporary list $LT(t_i)$.
So, then the dynamic list is adjusted $L(t_i)$ with entries from the temporary list $LT(t_i)$ to create a new dynamic list that contains the up-to-date top $N$ keywords along with the seed words.
This new list is used in the next time period $t_{i} \rightarrow t_{i+1}$ for the filtering of posted text.
Such process can be followed until an adequate number of instances has been collected.
The dynamic list can be updated automatically (i.e., without any manual inspection process to be involved) following the above described process.
The full list of tags will be made available upon request.

Here, we include tweets with hashtags appearing in the same tweets as \#GamerGate, \#BBCpay, and \#NBA depending on the dataset under consideration.
For the Gamergate controversy we reach $308$ hashtags during the data collection period, where a manual examination of these hashtags reveals that they do contain a number of hate words, e.g., \#InternationalOffendAFeministDay, \#IStandWithHateSpeech, and \#KillAllNiggers.
For the BBC gender pay controversy the dynamic list of keywords is comprised of $113$ hashtags, where indicative included hashtags are the following: \#equalityforall, \#genderpaygap, \#greedy.
Finally, for the NBA dataset we reach $526$ hashtags, with the following to be some indicative examples: \#basketball, \#jordan, and \#lakers.

\subsection{Data Preprocessing}\label{subsec:datapreprocessing}
In order to proceed with the content analysis of the collected datasets, a set of preprocessing tasks takes place, i.e., cleaning and spam removal to conclude with less noisy datasets.

\descr{Cleaning.} 
The first step is to clean the data of noise, i.e., removing numbers, stop words, and punctuations, as well as converting all characters to lower case.

\descr{Removing spammers.}
Even though extensive work has been done on spam detection in social media, e.g.,~\cite{stringhini2010detecting,Wang2010SpamDetectionTwitter}, Twitter is still plagued by spam accounts~\cite{Chen2015SpamTweets}.
Two main indications of spam behavior are~\cite{Wang2010SpamDetectionTwitter}: (i)~the large number of hashtags within a user's posts, as it permits the broader broadcast of such posts, and (ii)~the population of large amounts of (almost) similar posts.
Authors in~\cite{Amleshwaram2013spamtwitterduplicates}, in addition to the similarity of tweets as an indication of spam behavior, they also studied the URLs and domain names similarity.
Apart from the content duplication, based on~\cite{McCord2011SpamDetectionTraditionalClassifiers}, spammers often use popular hashtags on their tweets to persuade legitimate users to read their tweets.
The phenomenon of posting duplicate content and its connection to spam behavior it is also apparent to online activity outside the Twitter community.
For instance, in~\cite{Fetterly2004duplicatewebpages, Fetterly2005phraselevelduplication} the authors studied content duplication and found that clusters with duplicate content are often spam-related.

Considering the aforementioned behaviors as indication of spam activity, in our datasets the distributions of hashtags and duplications of posts are examined to detect the cutoff-limit above which a user will be characterized as spammer and consequently will be removed from the datasets.

\descr{Hashtags.} Studying the hashtags distribution, we observe that users use on average $0$ to $17$ hashtags.
Building on this, we examine various cutoffs to select a proper one above which we can characterize a user as spammer.
In the end, after a manual inspection, we observed that in most of the cases where the number of hashtags was $5$ or more, the text was mostly related to inappropriate content.
So, the limit of $5$ hashtags is used, and consequently we remove those users that have more than $5$ hashtags on average in their tweets.

\begin{table*}[!t]
\centering
\resizebox{.65\textwidth}{!}{
\centering
\begin{tabular}{@{}ll@{}}
\toprule
 & \textbf{Topic}                                                                                  \\ \midrule
1   & porn, milf, porno, cougar, fuck, nude, sexy, bikini, watch, photo \\
2   & boobs, liveoncam, sexcams, camgirls, milf, tits, naked, cumming, tatyanast, hardcore \\
3   & followme, ayala, followmejp, followback, crisethq, babykathniel, nylinam, chinadorables, obcessed, follow \\
4   & teamfollowback, follow, null, december, nightybutera, otwolmanilainlove, vote, pushaw, immadam, leonizamagic \\\bottomrule
\end{tabular}
}
\vspace{-0.2cm}
\caption{Popular topics in spam content.}
\label{tbl:spam_topics}
\vspace{-0.2cm}
\end{table*}

\begin{figure}[!t]
	\centering
	\includegraphics[width=0.3\textwidth]{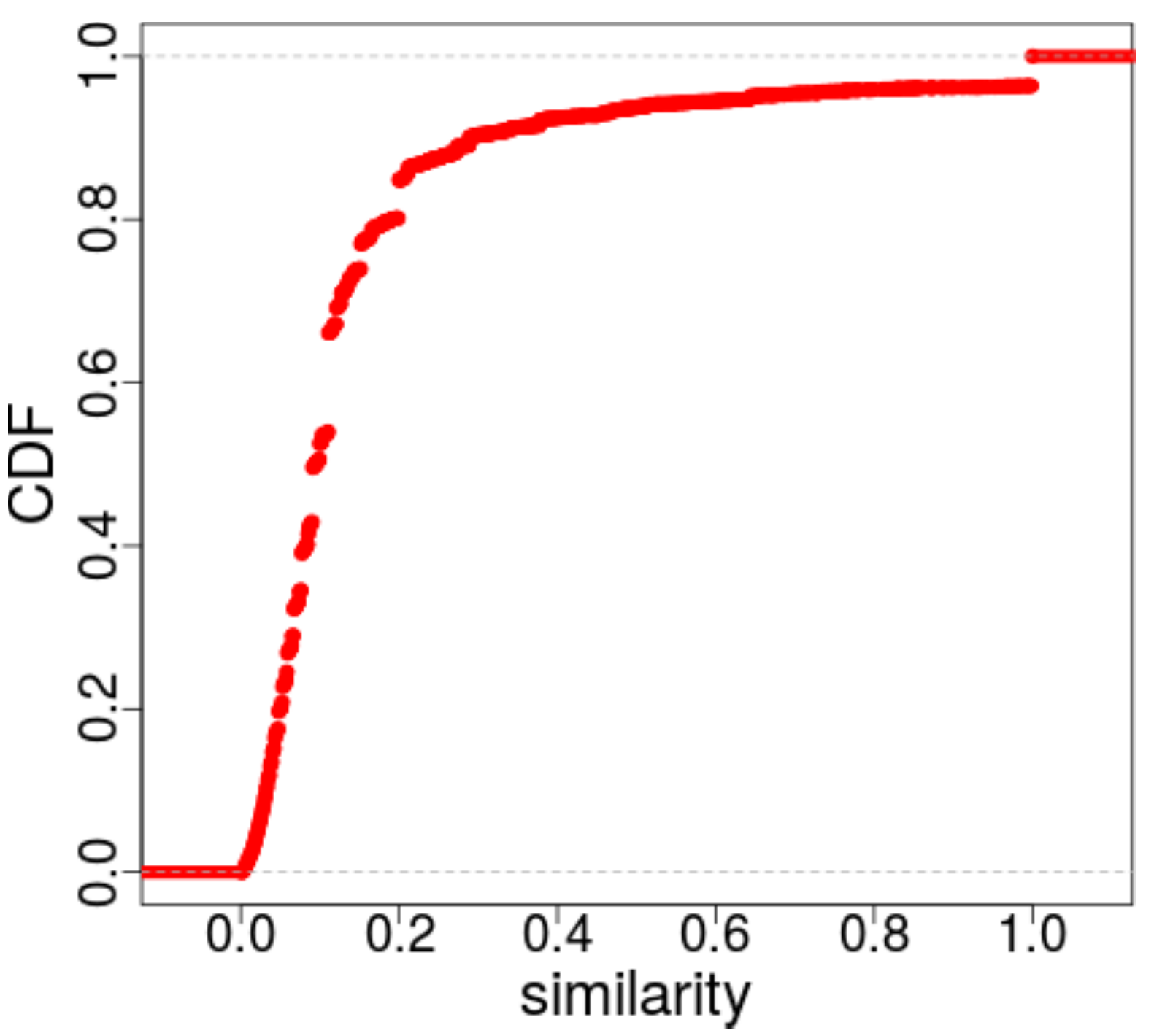}
	\vspace{-0.2cm}
	\caption{Similarity distribution of duplicate posts across the datasets.}
	\label{fig:duplications}
\end{figure}

\descr{Duplications.} In many cases, a user's texts are (almost) the same, with only the listed mentioned users modified.
So, in addition to the previously mentioned cleaning processes, we also remove all mentions.
Then, to estimate the similarity of a user's posts we proceed with the Levenshtein distance~\cite{Navarro2001ApproximateStringMatching} which counts the minimum number of single-character edits needed to convert one string into another, averaging it out over all pairs of their tweets (Definition~\ref{Levenshtein}).
Initially, for each user we calculate the intra-tweets similarity.
Thus, for a user with $x$ tweets, we arrive at a set of $n$ similarity scores, where $n = x (x - 1) / 2$, and an average intra-tweet similarity per user.
Then, all users with average intra-tweets similarity above $0.8$ (about $5\%$, see Figure~\ref{fig:duplications}) are excluded from the datasets.

\theoremstyle{Levenshtein distance}
\begin{Levenshtein}{\textbf{Levenshtein distance (lev).}}
\label{Levenshtein}
\small
The Levenshtein distance between two strings a, b (of length $\abs{a}$ and $\abs{b}$, respectively) is defined as:
\begin{center}$lev_{a,b}(i,j) = \begin{cases}max(i,j) & if \ min(i,j) = 0
\\min\begin{cases}lev_{a,b}(i-1,j) +1\\lev_{a,b}(i,j-1)+1\\lev_{a,b}(i-1,j-1)+1_{a_i }\neq b_j\end{cases} & otherwise\end{cases}$\\\end{center}
\noindent
\end{Levenshtein}

Table~\ref{tbl:spam_topics} provides some indicative examples of the most popular topics of the content that has been characterized as spam.
We followed the Latent Dirichlet Allocation (LDA) topic detection process in all the datasets, combined.
LDA~\cite{blei2003latent} is a generative statistical model where its objective is to find distinct topics in document collections.
It is a generative process that models each document as a mixture of latent topics, where a topic is described by a distribution over words.
The topic extraction was made based on the JSAT~\cite{JSAT2017}, i.e., a Java statistical analysis tool.
The tool provides an implementation of LDA which is based on the Stochastic Variational Inference.
To run the LDA model we set the following parameters: (i) batch size: 256, (ii) $\kappa$: 0.6, and (iii) $tau_o$ : 1.
The $\kappa$ value indicates the `forgetfulness' factor in the learning rate, where larger values increase the rate at which old information is `forgotten'.
The $tau_o$ is a learning rate constant that controls the influence of early iterations on the solution.
In this case, larger values reduce the influence of earlier iterations, while smaller ones increase the weight of earlier iterations.
Finally, the number of the training epochs was set to $10$.

We observe a high posting activity of inappropriate content, or an effort to attract more followers, which are common examples of spam behavior.
Across online social networks there are various types of spam users, such as~\cite{LeeHoneypots2010}: (i) duplicate spammers who post a series of nearly identical links, (ii) pornographic spammers where their data contain adult content, and (iii) friend infiltrators who follow many users and try to accumulate many followers to start then the spam activity.
So, based on the existing types of the spam users and the extracted examples we can conclude that the followed spam removal approach is suitable for removing at least the highly spam content.

\subsection{Datasets Overview}\label{subsec:dataoverview}

\begin{figure}[!t]
	\centering
	\includegraphics[width=0.35\textwidth]{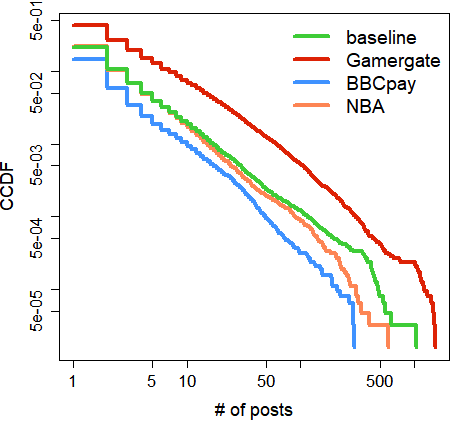}
	\vspace{-0.2cm}
		\caption{CCDF plots (log-log scale) of the Baseline, Gamergate, BBCpay, and NBA datasets.}
	\label{fig:ccdf-datasets}
	\vspace{-0.2cm}
\end{figure}

In the next paragraphs, we overview the collected datasets (i.e., Gamergate, BBCPay, NBA, and baseline) with respect to the frequency of posting per user, as well as the type of topics discussed.
For our initial analysis, we plot the Complementary Cumulative Distribution Function (CCDF) of the number of posts made by the users in the considered datasets (Figure~\ref{fig:ccdf-datasets}).
Since there is an important difference in the numbers of the participated users in the four considered datasets, before proceeding with the CCDF plot we randomly select equal-sized sample of users, i.e., $64k$, in alignment with the smallest dataset.
CCDF studies how often the random variable $X$ is above a particular level $x$, and is defined as follows:
\[
\widehat{F(x)} = P(X > x) = 1 - P(X\leq x)
\]
In fact, the Cumulative Distribution Function $F(x)$ of a random variable $X$ is defined as the following probability, for $x$ being any specified number:
\[
F(x) = P(X\leq x)
\]
We observe that the posting behavior of users follows a typical power-law distribution, in which many users have published a few posts, and a few users have posted a lot.
In fact, this behavior is consistent across all three communities (i.e., communities created by the people involved in the Gamergate, BBCPay, and NBA discussion groups), and is in line with~\cite{nilizadeh2017poised} which shows how communities are formulated around a specific topic, while also studies how similar are the messages that are exchanged from the involved participants.
Also, authors in~\cite{bruns2013towards} observed a long-tail distribution pattern for both user activity and user visibility, and more specifically they saw that a handful of leading users are disproportionately active or visible by comparison with the vast majority of their peers.
Here, we should state that we use a more abstract definition of the term `online community' which can be considered as `a community that exists online, mainly on the Internet, where its members with similar interests, experience, or shared values take part in social interactions such as share information, knowledge, and experience with other members'~\cite{zou2015trust}.

To further understand the content that has been posted in the collected datasets, we proceed with the aforementioned LDA model.
We apply this technique in each of the datasets and compute the top topics discussed in each one.
To provide a clear indication of the most popular topics discussed in the datasets under consideration, we decided to define a small number of topics N to be extracted based on the LDA method.
We experimented with various values of $N$=$[2,15]$ and concluded that at most five topics resulted to a clear set of distinct topics.

\subsubsection{Gamergate Controversy}
Gamergate controversy is one of the most well documented large-scale cases of bullying/aggressive behavior we are aware of~\cite{Massanari09102015}.
It started with a blog post by an ex-boyfriend of independent game developer Zoe Quinn, alleging sexual improprieties.
4chan boards like \dsrk~\cite{r9kthread} and \dspol~\cite{polthread}, turned it into a narrative about ``ethical'' concerns in video game journalism and began organizing harassment campaigns~\cite{hine2016longitudinal}.
It quickly grew into a larger campaign centered around sexism, feminism, and social justice, taking place on social media like Twitter~\cite{guberman2017}.
Although held up as a pseudo-political movement by its adherents, there is substantial evidence that Gamergate is more accurately described as an organized campaign of hate and harassment~\cite{theguardianfeliciaday}.
What started as ``mere'' denigration of women in the gaming industry, eventually evolved into directed threats of violence, rape, and murder~\cite{nytimessarkeesiandeaththreats}.
With individuals on both sides of the controversy using it, and extreme cases of bullying and aggressive behavior associated with it (e.g., direct threats of rape and murder), Gamergate controversy, and more specifically the \#GamerGate, can serve as a relatively unambiguous hashtag associated with texts that are likely to involve abusive/aggressive behavior from a fairly mature and hateful online community.
In~\cite{Mortensen2016} the author shows that \#GamerGate can be likened to hooliganism, i.e., a leisure-centered aggression were fans are organized in groups to attack another group's members.
Also,~\cite{guberman2017} aims to detect toxicity on Twitter, considering \#GamerGate to collect a sufficient number of harassment-related posts.

Table~\ref{tbl:gamergate_topics} presents some popular topics extracted based on the LDA topic detection process.
One of the most discussed issues is the one related to the `Black Lives Matter' international activist movement, which has its origin in the African-American community.
Such a movement campaigns against violence and systemic racism towards black people and it has spurred the interest of the Gamergate community as well.
Moreover, based on the detected popular topics, we observe that there is an increased interest in political issues,
such as {\em Brexit} and the USA presidential elections in 2016, as well as hostility against Hillary Clinton.

\begin{table}[!t]
\centering
\resizebox{.5\textwidth}{!}{
\centering
\begin{tabular}{@{}ll@{}}
\toprule
 & \textbf{Topic}                                                                             \\ \midrule
1   & blacklivesmatter, white, black, nude, blonde, legs, west, believe, showing, means \\
2   & lovewins, boobs, gamergate, love, booty, follow, remain, young, lady, leave       \\
3   & imwithher, hillary, booty, hillaryclinton, horny, whore, women, bernie, slut, feelthebern \\
4   & euref, lgbt, people, cameron, model, referendum, david, really, guys, believe     \\
5   & brexit, farage, feminism, nigel, beautiful, read, anti, model, voters, culo       \\ \bottomrule
\end{tabular}}
\vspace{-0.2cm}
\caption{Popular topics discussed in the Gamergate dataset.}
\label{tbl:gamergate_topics}
\vspace{-0.2cm}
\end{table}

Zooming in on the Gamergate users' activity, we observe that there are some heavy contributors to this community's activity (i.e., more than 20 posts).
Table~\ref{tbl:gamergate_topics_top_contributors} summarizes some popular topics discussed by such users.
Here, we observe again that the `Black Lives Matter' movement and `Hillary Clinton' are among the most discussed topics.
Of course, as in any (online) community, there are also peripheral users who, though they contribute content relevant to the community, perhaps it does not become highly influential.
Later on in our investigation on cyberbullying (Sections~\ref{sec:ground-truth}-\ref{sec:twitter-reaction}), we particularly focus on this mature community, as it exhibits more stable characteristics with respect to users involved and can enable the automatic detection of its abusive users.
 
\begin{table}[!t]
\centering
\resizebox{.5\textwidth}{!}{
\centering
\begin{tabular}{@{}ll@{}}
\toprule
 & \textbf{Topic}                                                                                  \\ \midrule
1   & blacklivesmatter, internationalwomensday, black, pawg, bullying, games, mulher, seriesbrazil, better, youtube \\
2   & imwithher, hillaryclinton, culo, busty, tetas, make, rtpig, life, honey, sega      \\
3   & women, internationalwomensday, love, happy, fuck, housewives, demdebate, asian, lesbian, sexe \\ \bottomrule
\end{tabular}}
\vspace{-0.2cm}
\caption{Popular topics from top contributors in the Gamergate dataset.}
\label{tbl:gamergate_topics_top_contributors}
\vspace{-0.2cm}
\end{table}

\subsubsection{BBC gender pay controversy (BBCpay)}
The BBC gender pay controversy first appeared in summer of 2017 when BBC published the annual salary report which revealed that the highest salaries were delivered to men~\cite{bbcpaycontroversy}.
After the publishing of such report, a storm of tweets produced with some posters for instance to express an objection on behalf of those people whose personal information was exposed, or those who directly attacked BBC for not well behaving towards gender equality.
Table~\ref{tbl:bbc_topics} depicts five popular topics as they were discussed in the BBCpay dataset.
We observe that among the most popular topics is the one related with the salary inequality among women and men in the BBC.
For instance, there is an increased reference to Gary Lineker's salary since he is one of the BBC's highest paid on-air talent~\cite{garylineker}.
Of course, as it is expected, there are also other popular discussed topics, such as {\em Brexit}, which has dominated the discussions all over the news for a long time period.

Considering the nature of the dataset, we speculate that even though it will contain an important number of aggressive instances, their intensity will probably be lower than that of the Gamergate dataset.
Overall, even though the BBC gender pay controversy is considered as a trending topic with hate and abusive elements within it, it is very new and much less organized.
Thus, although it may carry some similarities with the Gamergate controversy, the latter is an older, more mature and organized community.

\begin{table}[!t]
\centering
\resizebox{.5\textwidth}{!}{
\centering
\begin{tabular}{@{}ll@{}}
\toprule
 & \textbf{Topic}                                                                                                  \\ \midrule
1   & bbcsalaries, salary, garylineker, vine, jeremy, issue, bbcpaygap, never, uncomfortable, chatting \\
2   & bbcpay, fcbars, year, bbcsalaries, revealed, paying, wages, penny, eddiemair, angry \\
3   & trump, news, mike, fakenews, president, business, much, more, again, maga                         \\
4   & brexit, british, europe, hard, actually, tourists, voters, rest, debate, leaving \\
5   & news, iran, world, video, wednesdaywisdom, travel, more, most, bitcoin, july                      \\ \bottomrule
\end{tabular}}
\vspace{-0.2cm}
\caption{Popular topics discussed in the BBCpay dataset.}
\label{tbl:bbc_topics}
\vspace{-0.2cm}
\end{table}

\subsubsection{NBA \& Baseline}
The random sample of $1M$ tweets serves as a baseline, since it is less prone to contain abusive content, and therefore it provides a way to compare and distinguish among abusive and typical user behavioral patterns.
From Table~\ref{tbl:baseline_topics}, as it is expected, we observe that a variety of topics is discussed in the baseline dataset, such as about popular singers, posts related to job opportunities, or tech related information.
Similar to the random sample, the NBA dataset can be considered to include less hateful and/or aggressive comments, compared to the Gamergate and BBC gender pay controversy datasets.
Nevertheless, a common characteristic of the NBA dataset with the more hate-related ones is the sense of community among its members, as they are interested in a specific topic which allows them to differentiate from randomly selected users.
Table~\ref{tbl:nba_topics} provides an indication of the topics that are discussed in NBA dataset.
There is a lot of discussion around various popular sports, such as football and baseball, and brands, such as Adidas, as well as hot events, like the NBA 2018 summer league.
Therefore, the purpose of collecting such a dataset (NBA) is to enable us to study the similarities and differences among both the less hate-related datasets, and those in which strong groups are formed among their members but are not strongly hate-related.

\begin{table}[!t]
\centering
\resizebox{.5\textwidth}{!}{
\centering
\begin{tabular}{@{}ll@{}}
\toprule
 & \textbf{Topic}                                                                                       \\ \midrule
1   & ariana, grande, videomtv, mtvstars, vídeo, justin, voting, bieber, yaass, votar \\
2   & androidgames, gameinsight, android, food, harvested, tribez, bank, consulting, united, join \\
3   & jobs, apply, hiring, sales, needed, engineer, getalljobs, nurse, analyst, assistant \\
4   & thanksgivingwithblackfamilies, cousin, grandma, plate, house, cousins, girl, make, come, grown \\
5   & news, love, money, bitcoin, best, tech, great, fashion, death, photography \\ \bottomrule
\end{tabular}}
\vspace{-0.2cm}
\caption{Popular topics discussed in the baseline dataset.}
\label{tbl:baseline_topics}
\vspace{-0.2cm}
\end{table}

\begin{table}[!t]
\centering
\resizebox{.5\textwidth}{!}{
\centering
\begin{tabular}{@{}ll@{}}
\toprule
 & \textbf{Topic}                                                                                       \\ \midrule
1   & game, jersey, sales, want, play, tickets, home, social, nbasummer, thanks              \\
2   & food, free, ebay, football, great, mondaymotivation, sport, live, money, theopen       \\
3   & card, hiring, baseball, today, careerarc, cards, jersey, newyork, recommend, instagram \\
4   & news, chance, instantwingame, world, points, like, thanks, love, sywsweeps, sport      \\
5   & adidas, league, coupon, champion, nbadraft, shirt, vintage, best, nbasummer, love     \\ \bottomrule
\end{tabular}}
\vspace{-0.2cm}
\caption{Popular topics discussed in the NBA dataset.}
\label{tbl:nba_topics}
\vspace{-0.2cm}
\end{table}

\begin{figure*}[!t]
	\centering
	\begin{subfigure}[b]{0.32\textwidth}
		\includegraphics[width=\textwidth]{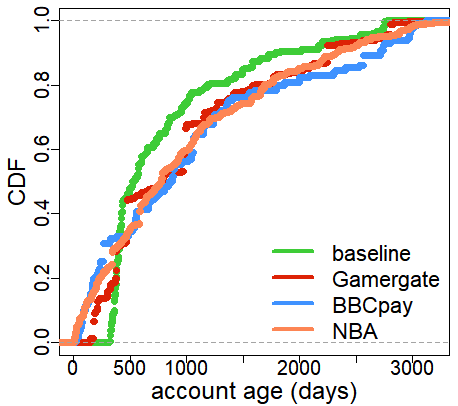}
		\captionsetup{font=footnotesize}
		\caption{Account age}
		\label{fig:data_age}
	\end{subfigure}
	\begin{subfigure}[b]{0.32\textwidth}
		\includegraphics[width=\textwidth]{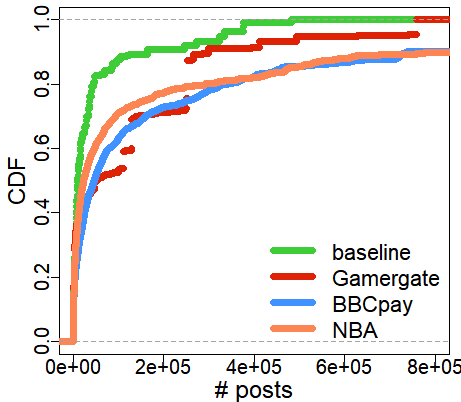}
		\captionsetup{font=footnotesize}
		\caption{Number of posts}
		\label{fig:data_posts}
	\end{subfigure}
	\begin{subfigure}[b]{0.32\textwidth}
		\includegraphics[width=\textwidth]{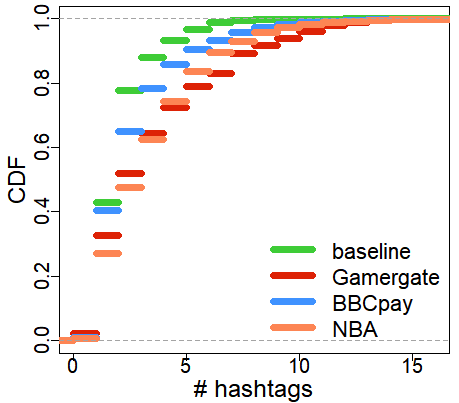}
		\captionsetup{font=footnotesize}
		\caption{Number of hashtags}
		\label{fig:data_hashtags}
	\end{subfigure}
	
	\begin{subfigure}[b]{0.32\textwidth}
		\includegraphics[width=\textwidth]{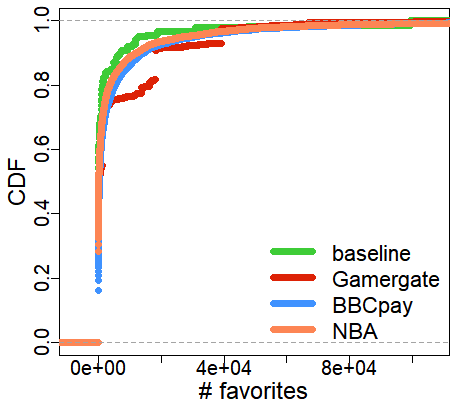}
		\captionsetup{font=footnotesize}
		\caption{Number of favorites}
		\label{fig:data_favorites}
	\end{subfigure}
	\begin{subfigure}[b]{0.32\textwidth}
		\includegraphics[width=\textwidth]{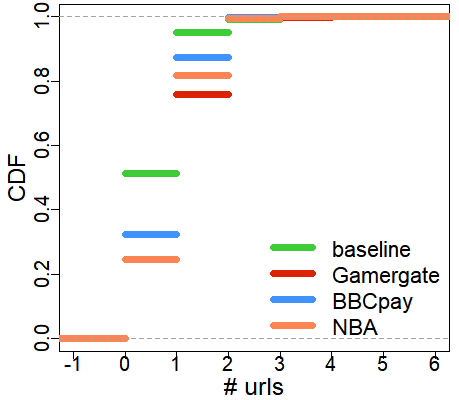}
		\captionsetup{font=footnotesize}
		\caption{Number of urls}
		\label{fig:data_urls}
	\end{subfigure}
	\begin{subfigure}[b]{0.32\textwidth}
		\includegraphics[width=\textwidth]{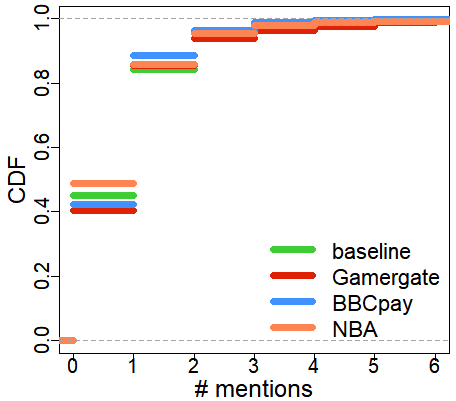}
		\captionsetup{font=footnotesize}
		\caption{Number of mentions}
		\label{fig:data_mentions}
	\end{subfigure}	
	\vspace{-0.2cm}	
	\caption{CDF distribution for various user profile features: (a) Account age, (b) Number of posts, (c) Hashtags, (d) Favorites, (e) Urls, and (f) Mentions.}
	\vspace{-0.2cm}
	\end{figure*}

\section{A systematic analysis of users behavior traits}\label{sec:data-analysis}
In this section we consider various dimensions including: user attributes, posting activity, and content semantics, to offer a systematic measurement-based characterization, comparing the baseline and the Gamergate datasets, as well as the observed differences with the BBC gender pay and NBA (the cases with the most significant differences are presented).
 
To examine the significance of differences among the distributions presented next, we use the two-sample Kolmogorov-Smirnov test (Definition~\ref{Kolmogorov}), a non-parametric statistical test, to compare the probability distributions of different samples.
This test is proposed since it enables assessing whether two samples come from the same distribution by building upon their empirical distribution function (ECDF)\footnote{A fraction of data-points that are less than or equal to some predetermined values within a set of random numbers.}.
We consider as statistically significant all cases with $p < 0.01$ and report only these cases.

\theoremstyle{Two-sample Kolmogorov-Smirnov test}
\begin{Kolmogorov}{\textbf{Two-sample Kolmogorov-Smirnov test.}}
\label{Kolmogorov}
Let's suppose that the first sample has size $n$ with an observed cumulative distribution function $E_1(x)$ and the second sample has size $m$ with an observed cumulative distribution function $E_2(x)$.
Then, the Kolmogorov-Smirnov statistic $D_{m,n}$ is defined as:
\begin{center}$D_{n,m} = max_x|E_1(x) - E_2(x)|$ \end{center}
\end{Kolmogorov}
\noindent
The two hypotheses under test are the following:\\
- $H_o$ (Null hypothesis): the two samples come from a common distribution;\\
- $H_a$ (Otherwise): the two samples do not come from a common distribution.\\
We reject the $H_o$ at significance level $\alpha$, if $D_{n,m} > \sqrt{-\frac{ln\alpha}{2}} \sqrt{\frac{n+m}{nm}}$.

\subsection{Activity Characteristics}

Here, we study the differences among the aforementioned datasets on the following activity related characteristics: (i) account age, (ii) number of posted tweets, (iii) number of hashtags, (iv) number of favorites, (v) number of urls, (vi) number of mentions, and (vii) number of followers and friends.
Table~\ref{tbl:statsActivity} summarizes the mean, median, and standard deviation values for the user categories under discussion in accordance to the activity-related characteristics.
In the CDFs presented next, we trim the plots when necessary to improve readability.

\begin{table*}[!t]
\centering
\centering
\resizebox{.775\textwidth}{!}{
\centering
\begin{tabular}{@{}lllll@{}}
\toprule
Metric				& \textbf{Gamergate}                & \textbf{BBCpay} & \textbf{NBA} & \textbf{Baseline}              \\
\midrule
Account age (days)		& 982.94 / 788 / 772.49 	        & 1,043 / 865 / 913.04              & 996.58 / 780 / 837.07  	  & 834.39 / 522 / 652.42    \\
Tweets			& 135,618 / 48,587 / 185,997   & 236,972 / 50,405 / 408,088  & 82,134 / 28,241 / 126,206    & 49,342 / 9,429 / 97,457 \\
Hashtags			& 3.47 / 2 / 2.97                        & 2.57 / 2 / 2.19      		     & 3.33 / 3 / 2.48   		  & 2.02 / 2 / 1.35                 \\
Favorites (tweets)		& 7,005 / 297 / 15,350              & 3,741 / 38 / 19,507                & 5,559 / 121 / 4,434   	  & 22,393 / 39 / 13,298     \\
Urls				& 0.998 / 1 / 0.711                    & 0.805 / 1 / 0.646      	     & 0.940 / 1 / 0.662		  &  0.544 / 0 / 0.622             \\
Mentions			& 0.760 /1.0 / 0.916                  & 0.726 /1.0 / 0.937      	     & 0.744 /1.0 / 0.917   		  & 0.774 /1.0 / 0.923            \\
Followers			& 4,931 / 490 / 123,411            & 8,222 / 587 / 100,299            & 9,380 / 576 / 175,631       	  & 1,515 / 120 / 44,069    \\
Friends			& 5,609 / 540 / 14,823              & 5,573 / 793 / 36,509              & 5,750 / 788 / 31,574             & 4,319 / 109 / 28,376    \\
\bottomrule
\end{tabular}}
\vspace{-0.2cm}
\caption{Statistical overview of the activity characteristics (mean/median/standard deviation).}
\label{tbl:statsActivity}
\vspace{-0.2cm}
\end{table*}

\descr{Account Age.}
Figure~\ref{fig:data_age} shows the distribution of account age for Gamergate, BBCpay, NBA participants, and baseline Twitter users.
For the most part, based on Table~\ref{tbl:statsActivity} we observe that Gamergate users tend to have older accounts than baseline Twitter users ($D = 0.20$).
Among such two sets of data, the oldest account belongs to a Gamergate user, while only $26.64\%$ of baseline users have account ages older than the mean value of the Gamergate users.
Concerning the other two sets of data, we observe that their living age on Twitter is more relevant to the Gamergate users with the older accounts belonging to BBCpay users ($D = 0.18$ between Gamergate and BBCpay users).
The users of the NBA community are quite expected to have a long duration on Twitter as well, since this topic revolves around very active and popular events for several years.
Finally, the BBCpay dataset apart from regular users, also may contain reporters and commentators on topical issues with long activity on Twitter since in many cases Twitter serves as a mean of broadcasting news~\cite{Bhattacharya2015NewsAgencyTwitter}.
Overall, we see that users with longer-running accounts on Twitter tend to be involved more on discussions around specific topics, i.e., Gamergate, BBC gender pay controversy, or the NBA.

\descr{Tweets and Hashtags.} 
In Figure~\ref{fig:data_posts} we plot the distribution of the number of tweets made by Gamergate, BBCpay, NBA, and baseline users.
We observe that the first three user categories are significantly more active than baseline Twitter users ($D = 0.35$, $D = 0.31$, $D = 0.22$, respectively).
Focusing more on the Gamergate and baseline users, in Table~\ref{tbl:statsActivity} we see that the mean and STD values are $135,618$ (baseline: $49,342$) and $185,997$ (baseline: $97,457$) posts, respectively, which justifies the image observed in the aforementioned figure.

Figure~\ref{fig:data_hashtags} reports the CDF of the number of hashtags found in users' tweets for both Gamergate and the baseline sample, finding that Gamergate users use significantly ($D = 0.26$) more hashtags than baseline Twitter users.
Additionally, we observe that the Gamergate users follow some of the trends of other communities, i.e., the NBA community, where it seems that an important number of hashtags is used in their posts, most probably for dissemination reasons.
 
 \begin{figure*}[!t]
	\centering
	\begin{subfigure}[b]{0.32\textwidth}
		\includegraphics[width=\textwidth]{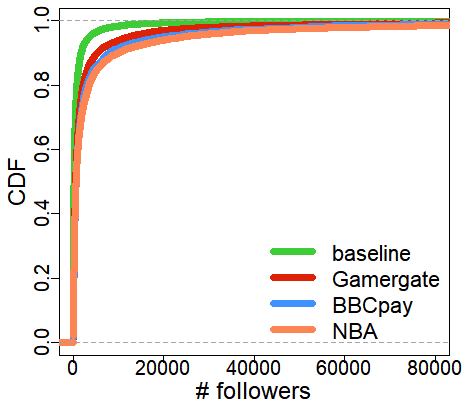}
		\captionsetup{font=footnotesize}
		\caption{Number of followers}
		\label{fig:data_followers}
	\end{subfigure}
	\begin{subfigure}[b]{0.32\textwidth}
		\includegraphics[width=\textwidth]{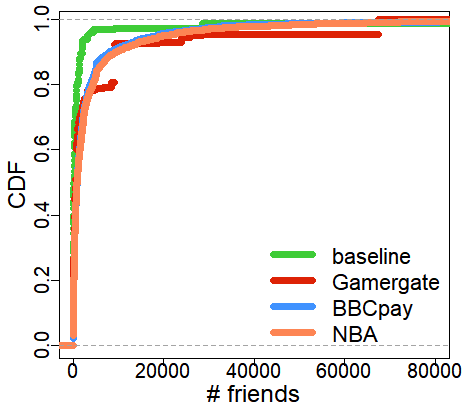}
		\captionsetup{font=footnotesize}
		\caption{Number of friends}
		\label{fig:data_friends}
	\end{subfigure}
	\vspace{-0.2cm}
	\caption{CDF distribution of (a) Number of followers and (b) friends.}
	\vspace{-0.2cm}
\end{figure*}

\descr{Favorites, Urls, and Mentions.} 
Figure~\ref{fig:data_favorites} shows the CDFs of favorites declared in the users' profiles.
In the median case, Gamergate users are similar to baseline users, but on the tail end ($30\%$ of users), Gamergate users have more favorites declared than baseline users (from Table~\ref{tbl:statsActivity} the mean value of the Gamergate users favorited tweets is almost the double that the baseline users).
Both the users of the NBA and BBCpay communities follow a more similar behavior to the baseline ones, in terms of their favoring activity.

Then, Figure~\ref{fig:data_urls} reports the CDF of the number of URLs found in tweets by Gamergate, BBCpay, NBA, and baseline users.
Baseline users post fewer URLs (the median indicates a difference of 1-2 URLs, $D = 0.27$), while Gamergate users post more in an attempt to disseminate information about their ``cause,'' somewhat using Twitter like a news service.
The use of urls on users posts shows the existence of a similar pattern with the number of used hashtags from the four different user categories with the users of the NBA community to be more similar to the Gamergate and the BBCpay users to the baseline ones.

Finally, Figure~\ref{fig:data_mentions} shows that the mentioning activity between the Gamergate and baseline users is quite similar, with the difference in the mean value to be only $0.014$.
As far as the BBCpay and NBA participants tend to use fewer mentions in their posts, while their distinction from the Gamergate users in both cases is statistically significant (with BBCPay: $D=0.032$, with NBA: $D=0.083$).

\begin{figure*}[!t]
	\centering
	\begin{subfigure}[b]{0.32\textwidth}
		\includegraphics[width=\textwidth]{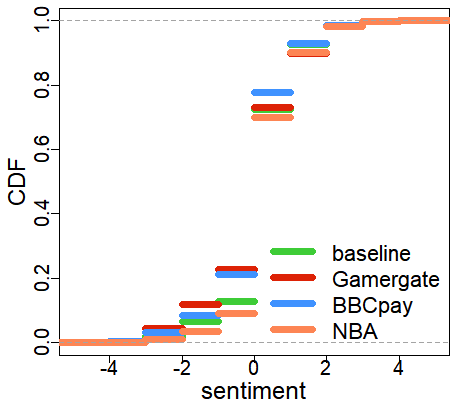}
		\captionsetup{font=footnotesize}
		\caption{Sentiment}
		\label{fig:data_sentiment}
	\end{subfigure}
	\begin{subfigure}[b]{0.32\textwidth}
		\includegraphics[width=\textwidth]{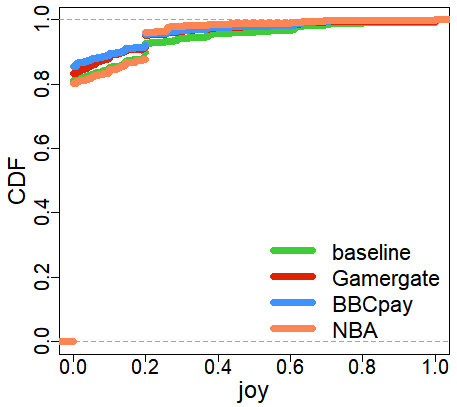}
		\captionsetup{font=footnotesize}
		\caption{Joy}
		\label{fig:data_joy}
	\end{subfigure}
	\begin{subfigure}[b]{0.32\textwidth}
		\includegraphics[width=\textwidth]{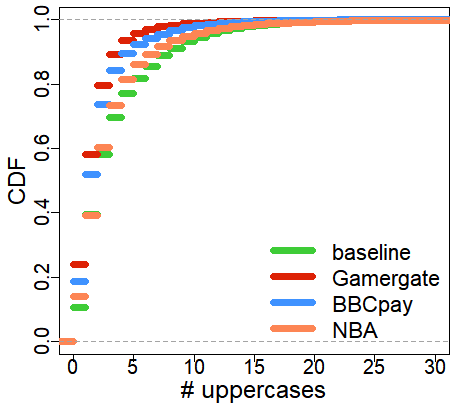}
		\captionsetup{font=footnotesize}
		\caption{Uppercase letters}
		\label{fig:data_uppercases}
	\end{subfigure}
	\vspace{-0.2cm}	
	\caption{CDF distribution of (a) Sentiment, (b) Joy, and (c) Uppercases.}
	\vspace{-0.2cm}
	\end{figure*}

\descr{Followers and Friends.} 
Gamergate users are involved in anti-social behavior.
However, this is somewhat at odds with the fact that their activity takes place primarily on social media. 
Aiming to give an idea of how ``social'' Gamergate users are, in Figures~\ref{fig:data_followers} and~\ref{fig:data_friends} we plot the distribution of followers and friends for the four user categories.
We observe that, perhaps surprisingly, Gamergate users tend to have more followers and friends than the baseline users ($0.39$ and $D=0.34$).
More specifically, from Table~\ref{tbl:statsActivity} we observe that Gamergate users have almost $70\%$ and $23\%$ more followers, respectively, than the baseline users (based on the mean values).
Although this might be somewhat counter-intuitive, the reality is that Gamergate was born on social media and the controversy appears to be a clear ``us vs. them'' situation.
This leads to easy identification of in-group membership, thus heightening the likelihood of relationship formation.

Statistical significant is also the difference of both the number of followers (BBCpay: $D=0.29$ and NBA: $0.28$) and friends ($D=0.37$) concerning the BBCpay and NBA participants with the baseline users, who also seem to be more social and well connected than baseline users.
Overall, the users of the NBA community tend to have the highest number of followers, which is also in alignment with the users' longtime activity on Twitter, while concerning the number of friends Gamergate users seem to have the most distinct behavior.
Finally, considering both the number of followers and friends we observe that BBCpay users are more similar to the NBA.

\subsection{Emotional Characteristics}
Studying in more depth the content of the posted tweets, here we study a set of emotional characteristics to have a better sense of the emotional intensity in users' posts.
More specifically, this set of characteristics involves the following: (i) sentiment, (ii) emotions, (iii) offensive, (iv) emoticons, and (v) uppercase.

\descr{Sentiment.}
The overflow of sentiments drives humans everyday actions and behaviors, a fact that is imprinted not only in the offline, but also in the activities that take place in the online world, i.e., the world of Internet.
Under sentiment analysis, texts are analyzed to detect people's opinions, typically falling into the dual polarity of positive or negative, with occasional consideration of a neutral standing.
A variety of methods and tools has been proposed for detecting sentiments out of text sources~\cite{socher2013recursive,khan2015Combining,chatzakou2015dawak,timmaraju2016sentiment}.
Here, we proceed with the SentiStrength tool~\cite{sentistrength} which estimates the positive and negative sentiment on a [-4, 4] scale in short texts, even for informal language often used on Twitter.
First, however, we evaluate its performance by applying it on an already annotated dataset with $7,086$ tweets~\cite{sentimentAnnotatedDataset}.
The overall accuracy is $92\%$, attesting to its efficacy for our purposes.

Figure~\ref{fig:data_sentiment} shows the CDF of sentiment of tweets for the four datasets.
Comparing the baseline with the Gamergate users we observe that around $25\%$ of tweets are positive for both types of users.
However, Gamergate users post tweets with a generally more negative sentiment (a two-sample Kolmogorov-Smirnov test rejects the null hypothesis with $D=0.101$).
In particular, around $25\%$ of Gamergate users' tweets are negative compared to only around $15\%$ for baseline users.
This observation is in line with the Gamergate dataset which contains a large number of offensive posts.

Quite expected is the fact that the BBCpay users tend to post more negative tweets than the baseline and NBA participants - more similar, but a little bit less than the Gamergate users - since it is a trending topic which involves hate-related behaviors but in a sense is less organized than the Gamergate controversy (less duration in time).
Finally, the users of the NBA community seem to post the least negative and more positive tweets than the other user categories which indicates that the users tend to post in a non so hateful and aggressive fashion.

\descr{Emotions.}
Emotional analysis focuses more on humans' sentiments, by tracking and revealing their emotions, e.g., anger, happiness, and disgust.
In the psychological science there is an ongoing debate on selecting a set of emotions (known as basic) which cover the overall spectrum of humans' emotional states.
For instance, Plutchik~\cite{plutchik1980emotionTheories} recognizes eight emotions as basic, i.e., `acceptance, anger, anticipation, disgust, joy, fear, sadness, and surprise,' while Watson~\cite{watson1930Behaviorism} proposes three emotions, i.e., `fear, love, and rage.'
A theory that is often followed in the emotion detection task is that of Ekman's and his colleagues~\cite{Ekman1982}, which identifies six emotions as basic, i.e., `anger, disgust, fear, joy, sadness, and surprise.'
Ekman's theory is considered in the analysis provided next.

In emotional analysis, there are several approaches that can be used, such as lexicon-based or machine learning-based.
Lexicon-based approaches tend to result in high precision and low recall, while machine learning methods do not consider the syntactic and semantic attributes.
So, both methods embed emotions misinterpretation risks~\cite{Chatzakou2015IC}.
To overcome such deficiencies, here, a hybrid approach is followed similar to the one presented in~\cite{dchatzakouESWA2017}.
Based on this work, initially a lexicon-based approach is used to extract two types of features, the sentimental (i.e., the expressed opinions: positive/negative) and the emotional (i.e., the emotional intensity) ones.
In addition to emoticons, both such features, as well as document feature vectors and emotional words are fed to a machine learning process in order to detect the most prevailing emotion of a new text (see Section 3 of~\cite{dchatzakouESWA2017} for more details).

Figure~\ref{fig:data_joy} shows the CDF of joy, where we reject the null hypothesis that the baseline and Gamergate distributions are the same ($D=0.04$).
We are \emph{unable} to reject the null hypothesis for the other five primary emotions.
This is particularly interesting because it contradicts the narrative that Gamergate users are posting virulent content out of anger.
Instead, Gamergate users appear to be less joyful.
This is a subtle but important difference: Gamergate users are not necessarily angry, but they are apparently less happy.
The BBCpay dataset seems to contain the less joyful users which can be justified by the fact that such a controversy has created a lot of frustration and disappointment to the BBC female, and not only, community.
The difference with the other three user categories is statistical significant ($D=0.04$ with baseline, $D=0.02$ with Gamergate users, $D=0.05$ with NBA).

\descr{Offensive.}
We also compare the offensiveness score that tweets have been marked with according to the hatebase (HB)~\cite{hatebase} crowdsourced dictionary.
Each word included in HB is scored on a [0, 100] scale which indicates how hateful it is.
Though the difference is small (the related plot is omitted), Gamergate users use more hateful words than the baseline ($D=0.006$) and NBA ($D=0.005$) participants (the difference in their distributions is statistically significant).
Even though the Gamergate users appear to post more offensive tweets than the BBCpay users, such a difference is not statistical significant ($D=0.005$), which indicates that there is an aggressive connotation of the posts that came out after the revealing of the disparity in pay between the male and female top earners in the BBC, and this aggression is similar to the Gamergate users' posts.

\descr{Emoticons and Uppercases.}
Two common ways to express emotion in social media are emoticons and ``shouting'' by using all capital letters.
Based on the nature of Gamergate, we initially suspected that there would be a relatively small amount of emoticon usage, but many tweets that would be shouting in all uppercase letters.
Based on the corresponding distributions this is not the case.
Gamergate and baseline users tend to use emoticons similarly (we are unable to reject the null hypothesis with $D=0.028$ and $p = 0.96$).
Comparing the baseline users with the BBCpay and NBA related users, even though the differences are subtle (with the baseline users using more emoticons in their tweets) they are statistical significant ($D=0.006$ and $D=0.01$, respectively).

Finally, based on Figure~\ref{fig:data_uppercases} which shows the usage of uppercase letters on the posted tweets, we observe that Gamergate users tend to use all uppercase \emph{less} than baseline users ($D=0.212$).
As seen previously, Gamergate users are quite savvy Twitter users, and generally speaking, shouting tends to be ignored.
Thus, one explanation is that Gamergate users avoid such a simple ``tell'' as posting in all uppercase to ensure their message is not so easily dismissed.
From the same plot we observe that BBCpay users show a more similar behavior with the Gamergate users ($D=0.06$), while the users of the NBA community are more similar to the baseline ones ($D=0.04$) - nevertheless their differences are statistical significant.

\begin{figure*}[!t]
	\centering
	\includegraphics[width=0.65\textwidth]{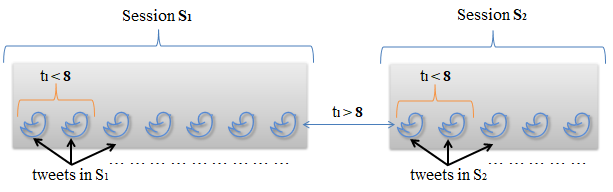}
	\vspace{-0.2cm}
	\caption{Overview of our sessionization process for constructing two consecutive sessions. In each session, the interarrival time between tweets does not exceed a predefined time threshold $t_{l}$.}
	\label{fig:sessioning}
\end{figure*}

\descr{Takeaways.} Overall, the behavior we observe is indicative of Gamergate users' ``mastery'' of Twitter as a mechanism for broadcasting their ideals.
In fact, their familiarity with Twitter could be the reason that Gamergate controversy exploded in the first place.
For instance, they tend to favorite more tweets and share more URLs and hashtags than the baseline users.
We also discovered that while the subject of their tweets is seemingly aggressive and hateful, Gamergate users do not exhibit common expressions of online anger, and, in fact, primarily differ from baseline users in that their tweets are less joyful.
This aligns with the viewpoint of the Gamergate supporters who claim that they never agreed to the aggressive methods used in this campaign~\cite{Mortensen2016}, which can result in a confusing expression of anger manifestation.
Gamergate users tend to be organized in groups, and in fact they participate also in face-to-face meetings to create stronger bonds, which also reflects on the higher number of followers and friends they have in relation to baseline users, despite their seemingly anti-social behavior.
Also, we discover that Gamergate users are seemingly more engaged than baseline Twitter users, which is an indication as to how and why this controversy is still ongoing.

Considering the BBCpay controversy, even though in some cases it shows similar patterns with the Gamergate phenomenon, it also differentiates in some other aspects, e.g., in the number of used hashtags and favorited tweets.
Users involved in the BBCpay controversy tend to be less aggressive in their tweets and they have quite old accounts on Twitter, while their posting activity is also especially intense.
However, in order to extract more concrete conclusions about the BBCpay controversy, and compare them with the Gamergate controversy, a dedicated and more extensive study should be done on this trending topic, to examine the dynamics of the topic and how its community has matured and got organized around key users, their characteristics and behaviors (e.g., abusive, offensive, and sarcastic).

Finally, the users of the NBA community seem to be very popular and with long activity on Twitter, something which is reasonable considering the popularity of the specific sport around the world.
Also, even though the NBA participants are more organized than the baseline users, in the NBA dataset the hate-related behaviors are almost nonexistent and similar to the baseline users.

\section{Ground Truth Building for Cyberbullying Detection}\label{sec:ground-truth}

In the previous section we studied the properties of Twitter users who are active in different groups and post tweets within particular topics.
Having in mind the different semantics of each group or topic, we extracted characterizations for the users in these groups.
Next, we want to perform a more in-depth analysis of user behavior, and how it could be classified as aggressive or normal.
In particular, we are interested in building a machine learning classifier that automatically detects offensive behaviors (bullies and aggressors) and correctly labels users for further (manual) investigation.

In this section, we present the data and the methodology used to build a ground truth dataset that allows us to perform such investigation with machine learning techniques.
To build this ground truth dataset we focus on the Gamergate and baseline data only, for the following two reasons.
First, and based on the analysis presented in Section~\ref{sec:data-analysis}, the users of the Gamergate community appear to be more aggressive and more active (e.g., they use higher number of hashtags and favorites) than the BBCpay and NBA participants.
Second, these users proceed with a higher number of mentions of other users in their tweets, in a possible effort to attack them.
Thus, for the effort to build the machine learning classifier, we focus on the Gamergate dataset, since we consider this community as more suitable than the NBA or BBCpay topics for an in-depth study of cyberbullying and cyberaggressive behavior of Twitter users.
We caution the reader that not all activity in the Gamergate dataset is aggressive or bullying.
It is, however, more probable to detect such incidents of aggression, and that is why we focus on the tweets of this topic.

\subsection{Preparing Data for Crowdsourcing}

In order to automatically characterize users into categories such as bullies, aggressors, normal, or spammers, using a machine learning classification method (Section~\ref{sec:machine-learning}), a ground truth dataset is necessary.
Since no such annotated dataset was already available, we used a crowdsourced approach by recruiting workers who are provided with a set of tweets from a user, and are asked to classify them according to predefined labels.
Here, we initially present the process followed for preparing the data to be used in the annotation process, while the next subsection describes the crowdsourcing labeling approach.

\descr{Sessions.}
Cyberbullying usually involves {\em repetitive} actions.
Thus, we aim to study users' tweets {\em over time}.
Inspired by Hosseinmardi et al.~\cite{Hosseinmardi2015} -- who consider a lower threshold of comments for media sessions extracted from Instagram to be presented in the annotation process -- 
we create, for each user, sets of time-sorted tweets (sessions) by grouping tweets posted close to each other in time.

\begin{figure*}[!t]
	\centering
	\includegraphics[width=0.8\textwidth]{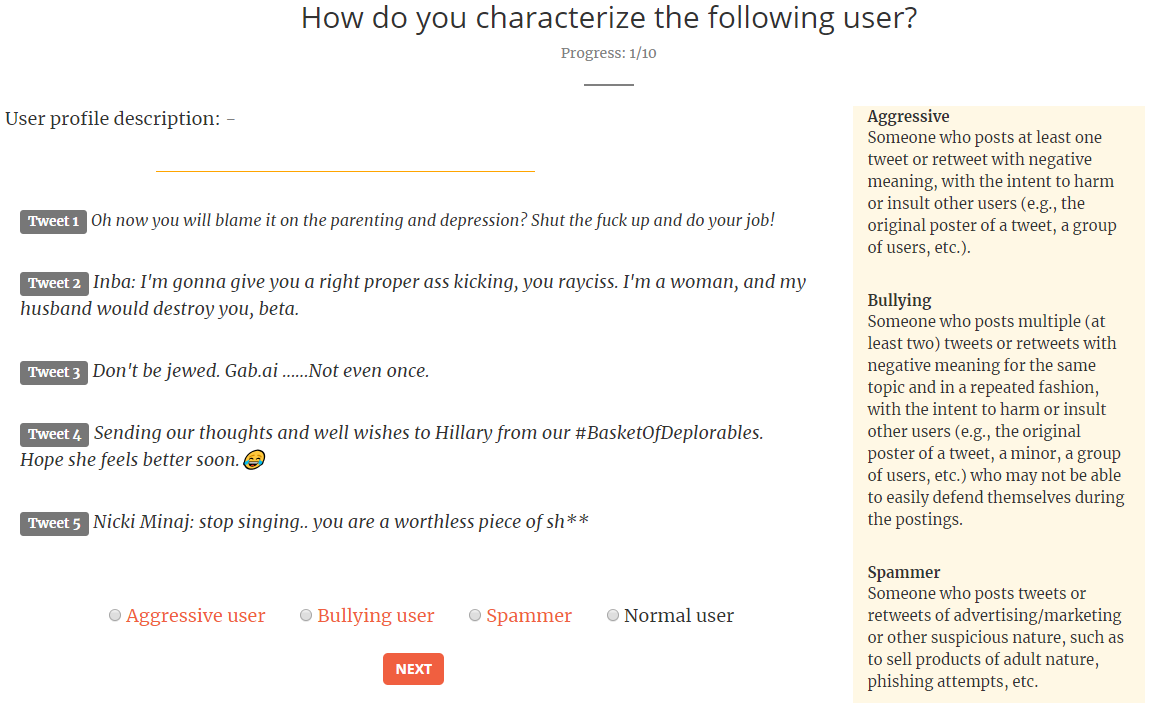}
	\caption{Example of the crowdsourcing user interface.}
	\label{fig:crowdsourcing_tool}
	\vspace{-0.2cm}	
\end{figure*}

First, we remove users who are not significantly active, i.e., tweeting less than five times in the 3-month period. 
Then, we use a session-based model where, for each session $S_{i}$, the interarrival time between tweets does not exceed a predefined time threshold $t_{l}$.
We experiment with various values of $t_{l}$ to find an optimal session duration and arrive at a threshold of 8 hours.
Figure~\ref{fig:sessioning} provides an overview of our sessionization process.
The minimum and maximum length of the resulting sessions (in terms of the number of their included tweets) for the hate-related (i.e., Gamergate) dataset are, respectively, $12$ and $2.6$k tweets. 
For the baseline set of tweets, they are $5$ and $1.6$k tweets.
Based on the sessionization process as it is defined above, the minimum length of a session would be $2$.
But there is always the probability that all sessions from our users in the considered datasets to have a low limit which is higher than 2 - which is the case here - since it totally depends on the users' activity.
From the analysis that is provided in Section~\ref{sec:data-analysis} we observe that the Gamergate users are more active in terms of their posting activity in relation to the baseline users, and thus their sessions are more packed with tweets.

Next, we divide sessions in batches, as otherwise they would contain too much information to be carefully examined by a crowdworker within a reasonable period of time.
To find the optimal size of a batch, i.e., the number of tweets per batch, we performed preliminary labeling runs on FigureEight (formerly known as CrowdFlower)~\cite{crowdflower}, involving $100$ workers in each run, using batches of exactly $5$, $5$-$10$, and $5$-$20$ tweets. 
Our intuition is that increasing the batch size provides more context to the workers to assess if a poster is acting in an aggressive or bullying behavior, however, too many tweets might confuse them.
The best results with respect to labeling agreement -- i.e., the number of workers that provide the same label for a batch -- occur with $5$-$10$ tweets per batch.
Therefore, we eliminate sessions with fewer than $5$ tweets, and further split those with more than $10$ tweets (preserving the chronological order of their posted time).
In the end, we arrive at 1,500 batches.
We also note that we maintain the same number of batches for both the hate-related and baseline tweets.

\subsection{Crowdsourced Labeling}

We now present the design of our crowdsourcing labeling process, performed on the crowdworking platform. 

\descr{Labeling.} Our goal is to label each Twitter user -- {\em not} single tweets -- as \textit{normal}, \textit{aggressor}, \textit{bully}, or \textit{spammer} by analyzing their batch(es) of tweets.
Note that we also allow for the possibility that a user is spamming and has passed our basic spam filtering.
Based on previous research~\cite{Smith2008CyberbullyingNature,Tokunaga2010CriticalReviewCyberbullyingVictimization,grigg2010cyber}, workers are provided with the following definitions of aggressive, bullying, and spam behaviors:

\begin{squishlist}
\item  aggressor: someone who posts at least one tweet or retweet with negative meaning, with the intent to harm or insult other users (e.g., the original poster of a tweet, a group of users, etc.);
\item  bully: someone who posts multiple tweets or retweets ($\ge$$2$) with negative meaning for the same topic and in a repeated fashion, with the intent to harm or insult other users (e.g., the original poster of a tweet, a minor, a group of users, etc.) who may not be able to easily defend themselves during the postings;
\item  spammer: someone who posts texts of advertising / marketing or other suspicious nature, such as to sell products of adult nature, and phishing attempts. 
\end{squishlist}

Similar definitions for distinguishing between cyberbullying and cyberaggression on social media platforms have been used across literature, e.g.,~\cite{Hosseinmardi2015,Dinakar2011ModelingDetectionTextualCyberbullying}.

\descr{Crowdworking Task.}
We redirect employed crowd workers to an online survey tool we developed.
First, they are asked to provide basic demographic information: gender, age, nationality, education level, and annual income.
We then ask workers to label $10$ batches, one of which is a control case (details below).
We also provide them with the user profile description (if any) of the Twitter user they are labeling and the definition of aggressive, bullying, and spam behaviors.
Figure~\ref{fig:crowdsourcing_tool} presents an example of the interface.
The workers rated the instructions given to them, as well as the overall task, as very good with an overall score of 4 out 5.

\descr{Results.}
Overall, we recruited $834$ workers.
They were allowed to participate only once to eliminate behavioral bias across tasks and discourage rushed tasks.
Each batch is labeled by $5$ different workers, and, similar to~\cite{Hosseinmardi2015} and~\cite{Nobata2016AbusiveLanguageDetection}, a majority vote is used to decide the final label.
Out of the $1,500$ batches, $1,303$ batches had majority (3 out of 5 annotators gave the same label), comprising 9,484 tweets in total.
Overall, we had absolute majority (5/5) for $15\%$ of the batches, strong majority (4/5) for $37\%$, and basic majority (3/5) for $48\%$.
About $4.5\%$ of users are labeled as bullies, $3.4\%$ as aggressors, $31.8\%$ as spammers, and $60.3\%$ as normal.
Overall, abusive users (i.e., bullies and aggressors) make up about 8\% of our dataset, which is in line with observations from previous studies (e.g., in~\cite{kayes2015ya-abuse} $9\%$ of the users in the examined dataset exhibit bad behavior, while in~\cite{Blackburn2012Cheaters} $7\%$ of users cheated).
Thus, we believe our ground truth dataset contains a representative sample of aggressive/abusive content from the Twitter-sphere.

\descr{Annotator Reliability.}
To assess the reliability of our workers, we use (i)~the inter-rater reliability measure, and (ii)~control cases.
To estimate the inter-rater reliability, we use the Fleiss' Kappa (Definition~\ref{def:FleissKappa}) which measures the agreement between three or more raters.
We find the inter-rater agreement to be $0.54$, which indicates a moderate strength of agreement between the raters.
Such moderate agreement highlights the difficulty in detecting bullying and aggressive behaviors, even when a manual inspection of the data at hand is considered.

We also use control cases to further assess worker ``quality'' by manually annotating three batches of tweets.
During the annotation process, each worker is given a set of batches to annotate, one of which is a randomly selected control case: the annotation of these control cases is used to assess their ability to adequately annotate for the given task.
We find $67\%$ accuracy overall (i.e., the percent of correctly annotated control cases).
More specifically, $84\%$ accuracy for spam, $54\%$ for bully, and $61\%$ for aggressive control cases.

\theoremstyle{Fleiss Kappa measure}
\begin{FleissKappa}{\textbf{Fleiss' Kappa measure.}}
\label{def:FleissKappa}
Fleiss' Kappa is defined as:
\begin{center}
$\kappa$ = $\frac{p_a - p_{\epsilon}}{1 - p_{\epsilon}}$,
\end{center}
\noindent
where $1 - p_{\epsilon}$ is the degree of agreement that is attainable above chance, while $p_a - p_{\epsilon}$ is the degree of agreement actually achieved above chance. If the raters are in complete agreement, then $\kappa = 1$.
\end{FleissKappa}

\begin{table}[t]
\centering
\resizebox{.5\textwidth}{!}{
\centering
\begin{tabular}{ll}
\hline
{\bf Type}			&	{\bf Feature}							\\
\hline
User				&	avg. \# posts, \# days since account creation, verified account\\
(total: 12)
			    	&	\# subscribed lists, posts' interarrival time, default profile image\\
			    	&	statistics on sessions: total number, avg., median, and STD. of their size\\
				&	location, profile description\\
\hline
Textual			&	avg. \# hashtags, avg. \# emoticons, avg. \# upper cases, \# URLs\\
(total: 15)				&	avg. sentiment score, avg. emotional scores, hate score\\
				&   	avg. word embedding score, avg. curse score, POS\\
				&	\# mentions, unique number of mentioned users, \#retweets\\
				&	avg. words per sentence, avg. word length\\
\hline
Network			&	\# friends, \# followers, hubs,  (d=\#followers/\#friends), authority\\
(total: 11)				&	avg. power diff. with mentioned users,  clustering coefficient, reciprocity \\
				&	eigenvector centrality, closeness centrality, louvain modularity\\
\hline
\end{tabular}
}
\vspace{-0.2cm}
\caption{Features considered in the study.}
\label{tab:features}
\vspace{-0.2cm}
\end{table}

\section{Feature Extraction and Selection}\label{sec:features}

The performance of a machine learning algorithm depends on the input data.
Therefore, it is important to understand the data at hand, given the new label provided by the annotators.
In this section, we perform this in-depth analysis of the dataset (i.e., the annotated dataset that was presented in Section~\ref{sec:ground-truth}) with respect to these four labels, and the different dimensions considered earlier, as well as new ones: user characteristics as extracted from the user profiles, textual characteristics as extracted from the users' tweets, and network properties of users as extracted from the Twitter social network.
We focus on user-, text-, and network-based features so that we can subsequently use them in the machine learning modeling of user behaviors identified in the dataset.

Next, we detail the features from each category, summarized in Table~\ref{tab:features}.
To examine the significance of differences among the distributions presented next, similar to Section~\ref{sec:data-analysis}, we use the two-sample Kolmogorov-Smirnov test and we consider as statistically significant all cases with $p < 0.01$.
Table~\ref{tbl:statsUserTextFeatures} summarizes the median and max values for both user- and text-based features, where necessary.
The work in~\cite{chatzakou2017mean} provides an extended analysis of the Empirical cumulative distribution function for most of the following presented features.

\begin{table}[!t]
\centering
\centering
\resizebox{.5\textwidth}{!}{
\centering
\begin{tabular}{@{}lllll@{}}
\toprule
                           		 & \textbf{bully}    			& \textbf{spam}       			& \textbf{aggressive} 	          & \textbf{normal}     \\ \midrule
Subscribed lists    		& 24 / 428 			& 57 / 3,723 			& 40 / 1,075 		          & 74 / 4,327 \\
Session statistics 		& 3 / 8       			& 3 / 51          		& 3 / 8 		          & 3 / 163           \\
Interarrival time (min)	& 0.0 / 1,068      		& 13.0 / 135k        		& 0.0 / 5,031		& 136 / 135k         \\
Urls                      		& 1 / 1.17 	  		& 1 / 2.38   			& 0.9 / 2    			& 0.6 / 1.38 \\
Hashtags             		& 2.9 / 4.5          		& 2.0 / 5.0             		& 2.9 / 4.4              		& 1.25 / 7.4           \\
Retweets            		&  0.0 / 0.086          	& 0.0 / 0.281             	& 0.0 / 0.139              	& 1.0 / 0.570           \\
Mentions (users)     		&  0.0 / 0.137          	& 0.0 / 0.378             	& 0.0 / 0.186              	& 1.0 / 0.742          \\
Unique mentions (users)	& 0.0 / 0.137          		& 0.0 / 0.346             	& 0.0 / 0.186              	& 1.0 / 0.736           \\
Avg. words per sentence & 12.5 / 12.63         		& 11.8 / 11.72             	& 13.6 / 13.05              	& 12.7 / 12.91           \\
Avg. word length              & 10.01 / 9.54         		& 7.30 / 7.83             	& 8.31 / 8.35              	& 7.38 / 7.76           \\
Adjectives             		&  1.5 / 2.46         		& 2.0 / 3.21             	& 4.0 / 3.46              	& 4.0 / 4.28        \\
Adverbs            		&  1.0 / 1.53        		& 0.0 / 1.52             	& 1.0 / 2.02	            	& 2.0 / 2.61           \\
Nouns             		&  12.0 / 12.43       		& 9.0 / 11.06             	& 15.0 / 14.32              	& 12.0 / 13.35          \\
Verbs		 		&  11.0 / 12.25      		& 5.0 / 6.62	            	& 12.0 / 12.18              	& 10.0 / 11.15           \\\bottomrule
\end{tabular}}
\vspace{-0.2cm}
\caption{Statistical overview of user- and text-based features (median/max).}
\label{tbl:statsUserTextFeatures}
\vspace{-0.2cm}
\end{table}

\subsection{What do Bullies, Aggressors, Spammers, and Normal Users Post About?}
To have a better sense of the four considered user categories, i.e., bullies, aggressors, spammers, and normal users, initially we proceed with an analysis of the content produced from each one of them (e.g., popular topics and top hashtags).
Table~\ref{tbl:user_categories_topics} depicts three popular topics discussed in each one of the four user categories based on the LDA analysis.
We observe that normal users tend to discuss about a variety of topics, such as political and social issues (pro-choice, uniteblue, brexit, etc.), as well as popular music performers (Britney Spears, Selena Gomez, etc.).

Spammers often use Twitter as a tool to send inappropriate content, to post malicious links, or to attract followers in order to extend their network~\cite{McCord2011SpamDetectionTraditionalClassifiers}.
This is also verified when we review the popular topics that are presented in Table~\ref{tbl:user_categories_topics} (and is also depicted in the topics presented in Table~\ref{tbl:spam_topics}), where either inappropriate content is presented or an effort to gain more followers is apparent by discussing attractive topics.
Bully users seem to organize their attack against important and sensitive issues, like the feminism, religion, and pedophiles.
This note is in alignment with earlier observations for the popular topics presented in both Tables~\ref{tbl:gamergate_topics} and~\ref{tbl:gamergate_topics_top_contributors} which are related to the Gamergate controversy.
Often the language used is quite aggressive and in extreme cases insulting.
Finally, aggressive users express their negativity on popular topics, like the `brexit' case, or the situation of Venezuela doctors who fume at official silence on Zika~\cite{venezuela}.
The topics that are presented here tend to be related more with those extracted from the BBC dataset (Table~\ref{tbl:bbc_topics}), where the aggressiveness tends to be lower in relation to the bullying cases.

\begin{table*}[!t]
\centering
\centering
\resizebox{.85\textwidth}{!}{
\centering
\begin{tabular}{@{}ll@{}}
\toprule
\textbf{User category} & \textbf{Topic}                                                                                             \\ \midrule
Normal   & uniteblue, feminism, women, tcot, abortion, gender, imwithher, prochoice, womenrights, otrasheffield      \\
   & mtvstars, britney, spears, lana, great, gomez, selena, demi, lovato, antinwonowtina           \\
   & brexit, voteleave, euref, gamersunite, leaveeu, people, world, voteremain, vote, pushawardsjadines                     \\\bottomrule\bottomrule
Spam   & porn, tweet, boobs, sexy, pics, vids, tits, antinwonowtina, exposes, erol \\
   & love, pushawardskathniels, boobs, retweet, busty, followers, years, girls, again, leaked  \\ 
   & lgbt, dino, love, follow, nowplaying, itunes, giveaway, summer, enter, seconds  \\ \bottomrule\bottomrule
Bully   & feminismisawful, antifeminist, whitegenocide, direction, mtvstars, antifeminism, famous, diversity, hypocrisy, feminista \\
   & action, offend, crowd, comentario, andreiwi, grollizo, hatebritain, jewfnitedstate, feminista, watchmylifegrow \\ 
   & stayandendure, masochist, pigs, feminist, voteremain, paedophiles, genocide, misery, feelthebern, patriarchy \\ \bottomrule\bottomrule
Aggressor   & zionist, groomed, erol, exposes, jews, promisedlanding, misery, heart, world, necessidade \\
   & brexit, leaveeu, more, like, attack, cowards, bluehand, feminismisaw, maga, medical  \\ 
   & feminism, venezuela, hatebritain, ormiga, heard, show, abandon, rioux, brad, safe  \\ \bottomrule
\end{tabular}}
\vspace{-0.2cm}
\caption{Popular topics discussed from bullies, aggressors, spammers, and normal users.}
\label{tbl:user_categories_topics}
\vspace{-0.1cm}
\end{table*}
\begin{table*}[!t]
\centering
\centering
\resizebox{.725\textwidth}{!}{
\centering
\begin{tabular}{@{}lp{11cm}ll@{}}
\toprule
\textbf{User category} & \textbf{Top hashtags}                                                                   & \textbf{Avg. \#hashtags}  \\ \midrule
Normal        & \#mtvstars, \#uniteblue, \#pushawardslizquens, \#pushawardskathniels, \#brexit & 1.81           \\
Spam          & \#boobs, \#ass, \#porn, \#busty, \#milf                                                                    & 2.04           \\
Bully           & \#feminazi, \#hateconsumed, \#fags, \#feminismisawful, \#jewfs                            & 2.86            \\
Aggressor  & \#gay, \#zionist, \#feminismisawful, \#hate, \#brexit                                               & 2.63                        \\ \bottomrule
\end{tabular}}
\vspace{-0.2cm}
\caption{Popular hashtags per user category.}
\label{tbl:user_categories_hashtags}
\vspace{-0.2cm}
\end{table*}

To analyze further the users' tweets, Table~\ref{tbl:user_categories_hashtags} presents five of the most popular hashtags for each one of the aforementioned user categories.
It also shows the average number of hashtags used in the tweets of these user categories.
It seems that both bully and aggressive users tend to use a high number of hashtags in their tweets, in a clear effort to link their tweets with the specific topics covered by the hashtags.
This could also be considered as an attempt to attract more people and create strong communities around the particular topic under discussion.

\subsection{User-based Features}

\descr{Basics.} In this section, various features extracted from a user's profile are examined.
Features in this category include the number of tweets a user has made, the age of his account (i.e., number of days since its creation), the number of lists subscribed to, if the account is verified or not (i.e., acknowledged by Twitter as an account linked to a user of ``public interest''), whether or not the user still uses the default profile image, whether or not the user has provided information about his living place (location), and the length of a user's profile description (if any).
Table~\ref{tbl:statsUserTextFeatures} summarizes the various features analyzed and their corresponding basic statistics.
As a representative example, we note the difference in the participation of groups from each class of users, with normal users signing up to more lists than the other types of users.

\descr{Session Statistics.} Here, we consider the number of sessions produced by a user from June to August and we estimate the average, median, and standard deviation of the size of each user's sessions.
Comparing the distributions among the bullies and aggressors to the normal users, we conclude that the differences are not statistically significant with $D$$=$$0.16052$ and $D$$=$$0.14648$ for bully vs. normal, and aggressors vs. normal, respectively.

\descr{Interarrival Time.} From the analysis presented in~\cite{chatzakou2017mean} and the results of Table~\ref{tbl:statsUserTextFeatures} we can observe that bullies and aggressors tend to have less waiting time in their posting activity compared to the spam and normal users, which is in alignment with the results in~\cite{Hosseinmardi2015} on the Instagram social network.

\subsection{Text-based Features}

For text-based features, we look deeper into a user's tweeting activity by analyzing specific attributes that exist in their tweets.

\descr{Basics.}
We consider some basic metrics across a user's tweets: the number of hashtags used, uppercase text (number of uppercase characters in a word - we exclude the first character of the beginning of a sentence) which can be indicative of intense emotional state or `shouting', number of emoticons, and URLs.
For each of these, we take the average over all tweets in a users' annotated batch.
Furthermore, both the total number of mentions, as well as the unique number of mentioned users, in addition to the number of retweets have been considered as possible factors for better discriminating among the four different user categories.

As we can see from Table~\ref{tbl:statsUserTextFeatures} normal users tend to post fewer URLs than the other $3$ classes.
Also, we observe that aggressive and bully users have a propensity to use more hashtags within their tweets, as they try to disseminate their attacking message to more individuals or groups.

\descr{Word Embedding.}
Word embedding allows finding both semantic and syntactic relation of words, which permits the capturing of more refined attributes and contextual cues that are inherent in human language.
E.g., people often use irony to express their aggressiveness or repulsion.
Therefore, we use Word2Vec~\cite{MikolovWordEmbedding2013}, an unsupervised word embedding-based approach to detect semantic and syntactic word relations.
Word2Vec is a two-layer neural network that operates on a set of texts to: 1)~initially establish a vocabulary based on the words included in such set more times than a user-defined threshold (to eliminate noise), 2)~apply a learning model to input texts to learn the words' vector representations in a $D$-dimensional, user-defined space, and 3)~output a vector representation for each word encountered in the input texts.
Based on~\cite{MikolovWordEmbedding2013}, $50$-$300$ dimensions can model hundreds of millions of words with high accuracy.
Possible methods to build the actual model are: 1)~CBOW (i.e., Continuous bag of words), which uses context to predict a target word, and 2)~Skip-gram, which uses a word to predict a target context.
Skip-gram works well with small amounts of training data and handles rare words or phrases well, while CBOW shows better accuracy for frequent words and is faster to train.

Here, we use Word2Vec to generate features to better capture the context of the data at hand.
We use a pre-trained model with a large scale thematic coverage (with $300$ dimensions) and apply the CBOW model due to its better performance regarding the training execution time.
Finally, having at hand the vector representations of all input texts' words, the overall vector representation of an input text is derived by averaging all the vectors of all its comprising words.
Comparing the bully distribution with the normal one we conclude to $D$$=$$0.0943$ and $p$$=$$0.723$, while in the aggressive vs. normal distribution comparison $D$$=$$0.1105$ and $p$$=$$0.702$, thus in both cases the differences are not statistically significant.
 
\descr{Sentiment and Emotional Characteristics.}
Sentiment has already been considered during the process of detecting abusive behavior in communications among individuals, e.g.,~\cite{Nahar2012SentimentAnalysisDetectionCyberBullying}.
To detect sentiment (similar to Section~\ref{sec:data-analysis}) we use the SentiStrength tool~\cite{sentistrength}.
Comparing the distributions of the aggressive class with the normal we observe that they are statistically different ($D$$=$$0.2743$), while this is not the case for the bully class compared with the normal one ($D$$=$$0.2041$, $p$$=$$0.022$).
We also attempt to detect more concrete emotions, i.e., anger, disgust, fear, joy, sadness, and surprise based on the approach presented in~\cite{Chatzakou2013EmotionallyDrivenClustering} (see also Section~\ref{sec:data-analysis} for Emotional Characteristics).
Comparing the distributions of the abusive classes with the normal, we observe no statistical difference.
For anger, even though the aggressive and normal distributions are statistically different ($D$$=$$0.2152$), the bully and normal users are not ($D$$=$$0.08$ and $p$$=$$0.88$).
These, in fact, are important findings that, at first glance, contradict intuition: users who exhibit bullying behavior, as decided by the majority of annotators, do not demonstrate extreme emotions such as anger, and do not express themselves with intense negative sentiment.
Therefore, a very basic or naive classifier which depends on such emotions would fail to detect subtle expressions of aggressiveness and bullying.

\begin{figure*}[!t]
	\centering
	\begin{subfigure}[b]{0.32\textwidth}
			\captionsetup{font=footnotesize}
			\includegraphics[width=\textwidth]{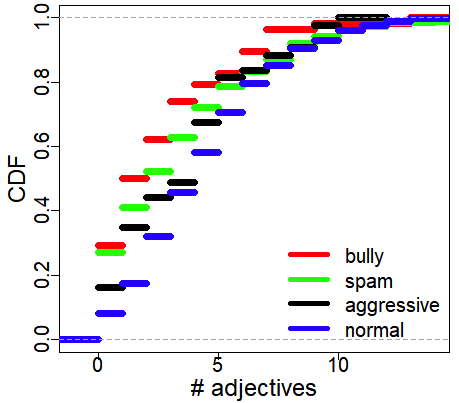}
			\caption{Adjectives.}
			\label{fig:ecdf-adjectives}
	\end{subfigure}
	\begin{subfigure}[b]{0.32\textwidth}
			\captionsetup{font=footnotesize}
			\includegraphics[width=\textwidth]{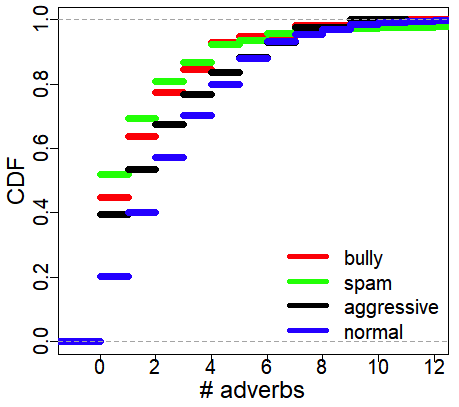}
			\caption{Adverbs.}
			\label{fig:ecdf-adverbs}
	\end{subfigure}
	\begin{subfigure}[b]{0.32\textwidth}
			\captionsetup{font=footnotesize}
			\includegraphics[width=\textwidth]{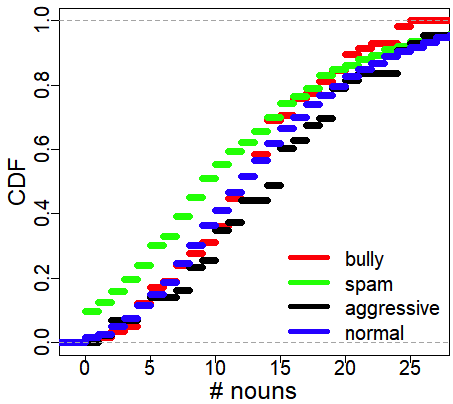}
			\caption{Nouns.}
			\label{fig:ecdf-nouns}
	\end{subfigure}

	\begin{subfigure}[b]{0.32\textwidth}
			\captionsetup{font=footnotesize}
			\includegraphics[width=\textwidth]{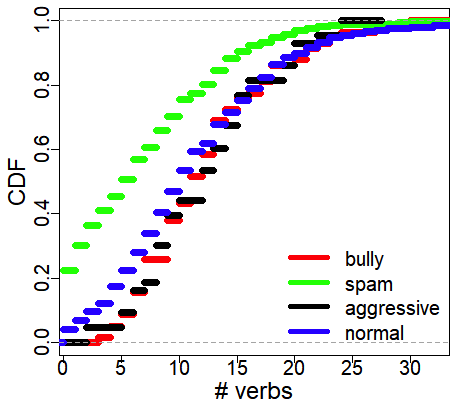}
			\caption{Verbs.}
			\label{fig:ecdf-verbs}
	\end{subfigure}
	\begin{subfigure}[b]{0.32\textwidth}
			\captionsetup{font=footnotesize}
			\includegraphics[width=\textwidth]{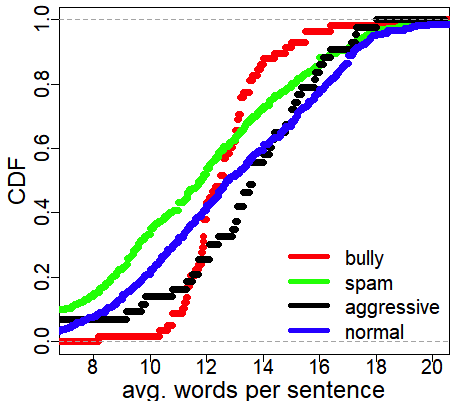}
			\caption{Average words per sentence.}
			\label{fig:ecdf-meanWordsPerSentence}
	\end{subfigure}
	\begin{subfigure}[b]{0.32\textwidth}
			\captionsetup{font=footnotesize}
			\includegraphics[width=\textwidth]{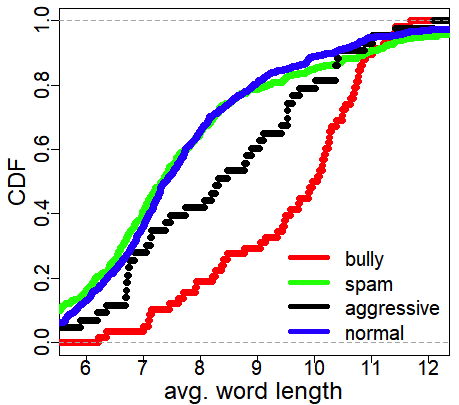}
			\caption{Average word length.}
			\label{fig:ecdf-meanWordLength}
	\end{subfigure}
	\vspace{-0.2cm}	
	\caption{CDF of (a) Adjectives, (b) Adverbs, (c) Nouns, (d) Verbs, (e) Average words per sentece, (f) Average word length.}
	\vspace{-0.2cm}
\end{figure*}

\begin{figure*}[!t]
	\centering
	\begin{subfigure}[b]{0.35\textwidth}
			\captionsetup{font=footnotesize}
			\includegraphics[width=\textwidth]{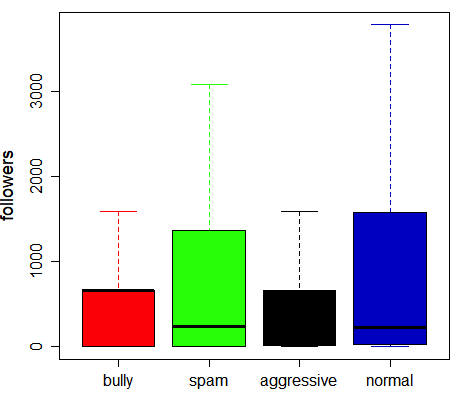}
			\caption{Followers.}
			\label{fig:boxplot-followers}
	\end{subfigure}
	\begin{subfigure}[b]{0.35\textwidth}
			\captionsetup{font=footnotesize}
			\includegraphics[width=\textwidth]{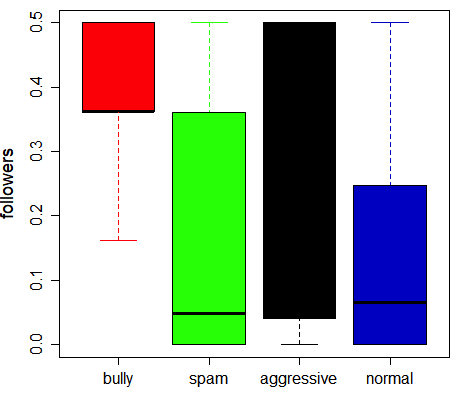}
			\caption{Reciprocity.}
			\label{fig:boxplot-reciprocity}
	\end{subfigure}

	\begin{subfigure}[b]{0.35\textwidth}
			\captionsetup{font=footnotesize}
			\includegraphics[width=\textwidth]{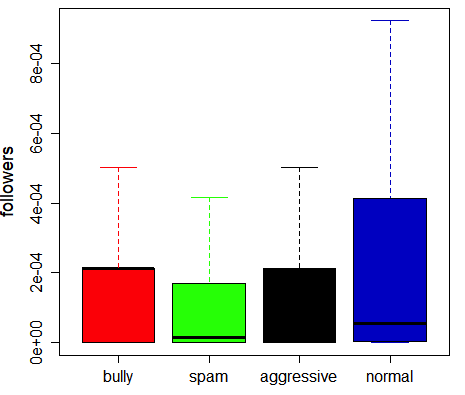}
			\caption{Hubs.}
			\label{fig:boxplot-hubs}
	\end{subfigure}
	\begin{subfigure}[b]{0.35\textwidth}
			\captionsetup{font=footnotesize}
			\includegraphics[width=\textwidth]{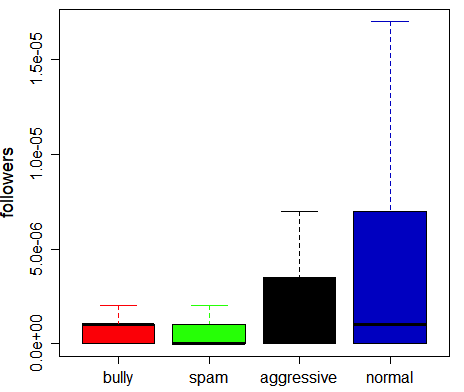}
			\caption{Eigenvector.}
			\label{fig:boxplot-eigenvector}
	\end{subfigure}
	\vspace{-0.2cm}
		\caption{Boxplots of (a) Followers, (b) Reciprocity, (c) Hubs, and (d) Eigenvectors. The data are divided into three quartiles (i.e., first, third, and median quartiles in the data set). The top and the bottom whiskers indicate the maximum and minimum values, respectively. We removed outliers from the plots.}
	\vspace{-0.2cm}
	\end{figure*}

\descr{Hate and Curse Words.}
Additionally, we wanted to specifically examine the existence of hate speech and curse words within tweets.
For this purpose, we use the Hatebase database, as well as a list of swear words~\cite{swearwords} in a binary fashion; i.e., we set a variable to true if a tweet contained any word in the list, and false otherwise.
Even though these lists can be useful in categorizing general text as hateful or aggressive, they are not well suited for classifying tweets as they are short and typically include modified words, URLs, and emoticons.
Overall, we find that bully and aggressive users have a minor bias towards using such words, but they are not significantly different from normal users' behavior.

\descr{Part of Speech (POS).}
POS tagging has been considered during the feature extraction process, to understand better the type of expressions used from the different user categories.
POS tagging is the process of marking up a word in a text corresponding to a particular part of speech, based on both its definition, as well as its context.
More specifically, a POS is a category of words with specific grammatical properties, such as adjectives, adverbs, nouns, and verbs.
To extract the POS tags out of the available textual resources, we built upon the POS tagger provided by the Tweet NLP library~\cite{tweetnlp}.
Figures~\ref{fig:ecdf-adjectives},~\ref{fig:ecdf-adverbs},~\ref{fig:ecdf-nouns}, and~\ref{fig:ecdf-verbs} show the CDF for the number of adjectives, adverbs, nouns, and verbs, respectively, for the four user classes.
Based on the aforementioned figures, and the statistics in Table~\ref{tbl:statsUserTextFeatures}, we observe that bully users tend to use a lower number of adjectives in their posts in relation to other user categories.
A similar pattern is observed in the case of adverbs.
Finally, we observe that spam users tend to use a lower number of nouns and verbs, where especially in the case of the verbs usage the differentiation from the other user categories is quite intense.
Overall, we observe that bully users avoid to use adjective and adverbs in their posts, which are often used to describe a noun/pronoun or to modify nouns, verbs, and adjectives, respectively.
This indicates that they may prefer to perform more straight attacks without adding any additional information which is often provided with the adjectives and adverbs.

\descr{Stylistic Features.}
Finally, we also consider two stylistic features, i.e., the average words per sentence and the average word length, which have been used in stylometric analysis~\cite{bergsma2012stylometric}, i.e., an approach to automatically recognize hidden attributes related to texts' authors.
Figures~\ref{fig:ecdf-meanWordsPerSentence} and~\ref{fig:ecdf-meanWordLength} show the CDF of the average words per sentence and average word length, respectively.
Concerning the average number of words per sentence, it seems that spam users follow a common pattern with the normal users, albeit normal users post tweets with somewhat more words.
This pattern is also confirmed in the average word length, which gives hints to how spammers try to hide their content by imitating normal users and their posts.
Aggressive and bully users have a distinct pattern, with aggressive users posting tweets with the most number of words for the majority of users ($\sim 70\%$), but bully users post tweets with the longest, and perhaps more complex words than aggressive, normal, or spam users.

\subsection{Network-based Features}

Twitter social network plays a crucial role in diffusion of useful information and ideas, but also of negative opinions, rumors, and abusive language (e.g.,~\cite{Jin2013EpidemiologicalModelingNewsAndRumors,Pfeffer2014NegativeWordToMouthDynamics}).
Due to the increased cost for building such a network (Twitter API limits retrieval of information to only $5,000$ user ids per $15$ minutes and per API key~\cite{twitterapinetlimits}), past research has mostly focused on text- and user-based features, as they are easier and faster to access (these data are typically accessible from the JSON files of the downloaded tweets, with no extra crawling required).
Going beyond the state of art, we study the association between cyberbullying or cyberaggressive behavior and the position of users in the social network of friends and followers on Twitter.

For this analysis to take place, we constructed a network of friends and followers on Twitter, by considering the users in our dataset as our seeds for crawling the network.
In fact, a two-hop followers/friends graph was built, where, apart from considering the followers and friends of each user in our dataset, we further extended the network by considering the contacts (followers and friends) of the followers/friends of the initial users.
This way, we were able to expand the network construction beyond the ego-network of each user in our dataset, and thus, better understand the associations that exist between types of users and the formulation of possible communities.

In our case, the constructed network is comprised of about $1.2M$ users and $1.9M$ friend edges (i.e., someone who is followed by a user) or follower edges (i.e., someone who follows a user), with effective diameter of $4.934$, average clustering coefficient of $0.0425$, and $24.95\%$ and $99.99\%$ of nodes in the weakest and largest component, respectively.
Users in such a network can have a varying degree of embeddedness with respect to friends or followers, reciprocity of connections, connectivity with different parts of the network, etc.

\descr{Popularity.}
The popularity of a user can be defined in different ways.
For example, using the number of friends (out-degree centrality), the number of followers (in-degree centrality), or the ratio of the two measures (since Twitter allows users to follow anyone without their approval, the ratio of followers to friends can quantify a user's popularity).
These measures can quantify the opportunity for a user to have a positive or negative impact in his ego-network in a direct way.

Interestingly, in the analysis presented earlier about the Gamergate (Figure~\ref{fig:data_followers}), users involved in that community have been found to have a higher number of followers and friends than the baseline users.
However, when these users are further analyzed into the four annotated categories (Figure~\ref{fig:boxplot-followers}), we find that aggressive and bully users have fewer followers than the other user categories, with normal users having the most followers.
Similar comment applies for the distribution of friends.
This suggests that aggressive behavior does not attract users to follow the aggressors.
To justify that the Gamergate dataset contains an important number of offensive users, we checked how many bully and aggressive users come from the Gamergate dataset.
Out of the $101$ offensive users (i.e., $58$ bully and $43$ aggressive users) the $95\%$ of them derive from the Gamergate dataset, while only a $5\%$ have its origin from the baseline dataset.
This suggests that indeed the Gamergate dataset contains an important number of offensive users and thus correctly has been used from studying abusive phenomena that take place online.

\descr{Reciprocity.}
This metric quantifies the extent to which users reciprocate the follower connections they receive from other users.
Reciprocity as a feature has also been used in~\cite{Hosseinmardi2014TowardsUnderstandingCyberbullying}, but in an interaction-based graph using likes in posts.
Here, being the first to do so in the context of bullying, we investigate the fundamental reciprocity in Twitter friendship.
We have found the average reciprocity in our network to be $0.2$.
In fact, Figure~\ref{fig:boxplot-reciprocity} shows that the user classes considered have different distributions, with the bully and aggressive users being more similar and with a higher number of reciprocities than the normal or spammers.

\descr{Power Difference.}
A recent study~\cite{Pieschl2013241} found that the emotional and behavioral state of victims depend on the power of their bullies, e.g., more negative emotional experiences were observed when more popular cyberbullies conducted the attack, and the high power difference with respect to status in the network has been shown to be a significant characteristic of bullies~\cite{Corcoran2015CyberbullyingOrCyberAggression}.
Thus, we consider the power difference between a tweeter and his mentions an important metric to be analyzed and used as a feature.
In fact, a further analysis of a user's mentioned users could reveal possible victims or bystanders of his aggressive or bullying behavior.
To this end, we compute the difference in power a user has with respect to the mentioned users in their posts, in terms of their respective followers/friends ratio.
Based on the two-sample Kolmogorov-Smirnov test the difference in power between the bully and normal users is statistically significant ($D$$=$$0.31676$), while the same comparison between aggressive and normal users is not ($D$$=$$0.221$, $p$$=$$0.037$).

\descr{Centrality Scores.}
We also investigate users' position in their network through more elaborate metrics such as hub, authority, eigenvector, and closeness centrality, that measure influence in their immediate and extended neighborhood, as well as connectivity.

\descremph{Hubs and Authority.}
A node's hub score is the sum of the authority score of the nodes that point to it, and authority shows how many different hubs a user is connected with~\cite{Kleinberg1999HubsAuthorities}.

\descremph{Influence.}
Eigenvector centrality measures the influence of a user in their network, immediate or extended over multiple hops.
Closeness centrality measures the extent to which a user is close to each other user in the network.

To calculate the last four measures, we consider both the followers and friends relations of the users under examination in an undirected version of the network.
From Figure~\ref{fig:boxplot-hubs} we observe that bullies have lower values in their hub scores (the same applies to the authority scores) which indicates that they are not so popular in their networks.
In terms of influence on their ego and extended network, they have behavior similar to spammers, while aggressors seem to have influence more similar to normal users (Figure~\ref{fig:boxplot-eigenvector}).
We omit the CDF of the closeness centrality measure because we cannot reject the null hypothesis that the distributions are different.

\descr{Communities.}
Past work~\cite{hanish2004bullying} showed that bullies tend to experience social rejection from their environment and face difficulties in developing social relations.
We examine the usefulness of this attribute and calculate the clustering coefficient which shows a user's tendency to cluster with others.
Similar to the above, in our case we also observe that bullies and spammers are less prone to create clusters in contrast to aggressive and normal users.
Finally, we compute communities using the Louvain method~\cite{blondel2011louvain} which is suitable for identifying groups on large networks as it attempts to optimize the modularity measure (how densely connected the nodes within a cluster are) of a network by moving nodes from one cluster to another.
Overall, we observe a few communities with a high number of nodes (especially the network core) resulting in a feature which statistically differentiates bullies vs. normal users ($D$$=$$0.206$), but not aggressive vs. normal users ($D$$=$$0.1166$, $p$$=$$0.6364$).

\descr{Takeaways.}
In this section we analyzed various user, text, and network attributes that could be considered in the classification process in order to detect abusive users, i.e., bullies and aggressors, out of the normal users and spammers.
As it was expected some attributes are more distinguishable among the different user categories and can assist more during the classification process.
Normal users tend to discuss about a variety of topics, such as political and social issues, whereas bully users seem to organize their attacks against important and sensitive issues, such as feminism, religion, and pedophiles, using aggressive and in some cases insulting language.
Aggressive users express their negativity on popular topics, such as the `brexit' case, `maga', and the spread of the zika virus.
Spammers typically post inappropriate content in an effort to gain more followers or attract victims to malicious sites of questionable or malicious content.
We also saw that the number of participated lists, the used URLs as well as hashtags within users' tweets, or the reciprocity, exhibit different patterns for each user category.
In addition, bullies and aggressors tend to have less waiting time in their posting activity compared to the other users.

There are also some cases, where the considered attributes are more distinguishable for specific user categories.
For example, we observed a noticeable distinction in terms of the expressed sentiment from the aggressive users, while we also saw that spam users tend to clearly deviate in terms of the verbs usage in their tweets.
However, spammers try to imitate normal users with respect to the length of the tweet in number of words and length of words used.
Aggressive and bully users have a clearly different behavior than normal or spam users, making them easier to identify.
Also, bully and aggressive users exhibit higher network reciprocity, but bully users are less central in the network than aggressive or normal users.

On the other hand, there are attributes such as whether the account is verified or not, or whether a user has declared his location, where the difference among the considered user categories is not statistically significant, and as a consequence cannot help to distinguish among them during an automatic classification process.
So, overall the analysis that is provided in this section can help in deciding which attributes would be more helpful during the classification process in the effort to detect efficiently abusive users based on their behavior on Twitter.

\section{Machine Learning for Bullying \& Aggression Detection}\label{sec:machine-learning}
In this section we present the effort to model bully and aggressive behaviors on Twitter, using the features extracted and the labels provided by the crowdworkers as presented earlier (Section~\ref{sec:ground-truth}).
Table~\ref{tbl:datasetAnnotated} provides an overview of the dataset used in the following experiments.
The highest number of cases belongs to normal users, followed by spam users, and with the bully and aggressive users to have about the same rate.

\begin{table}[!t]
\centering
\centering
\resizebox{.45\textwidth}{!}{
\centering
\begin{tabular}{@{}lllll@{}}
\toprule
        			& \textbf{bully}	& \textbf{aggressive} 	& \textbf{spam} 	& \textbf{normal}   \\ \midrule
\#users 		& 58   			& 43      			& 415  		& 787     		\\
\#texts (mean) 	& 517 (8.91)   	& 336 (7.81)     		& 3028 (7.29) 	& 5603 (7.11)          \\ \bottomrule
\end{tabular}}
\vspace{-0.2cm}
\caption{Dataset overview (annotated).}
\label{tbl:datasetAnnotated}
\vspace{-0.2cm}
\end{table}

\subsection{Experimental Setup}

We consider various traditional and extensible used machine learning algorithms, either probabilistic, tree-based, ensemble, and neural network classifiers.
Next, a brief overview of the considered machine learning algorithms is presented, while Section~\ref{subsec:classificationresults} presents the corresponding results.

\descr{Probabilistic Classifiers.}
Such classifiers~\cite{Friedman1997BayesianNetworkClassifiers} are used by the machine learning community because of their simplicity and typically good performance.
They are defined as follows:

\theoremstyle{ProbabilisticClassifiers}
\begin{ProbabilisticClassifiers}{\textbf{Probabilistic Classifiers.}}
\label{ProbabilisticClassifiers}
They use Bayes's rule to estimate the conditional probability of a class label $y$, based on the assumption that such probability can be decomposed into a product of conditional probabilities:
\begin{center}$P_r(y|x) = P_r(y|x^1, x^2,..., x^n)$,\end{center}
where $x=(x^1, x^2,..., x^n)$ is the $n-$dimension feature vector.
\end{ProbabilisticClassifiers}

Both the BayesNet (BN) and Naive Bayes (NB) classifiers have been evaluated, with the latter to perform better between the two.

\begin{figure*}[!t]
  \centering
  \includegraphics[width=0.75\textwidth]{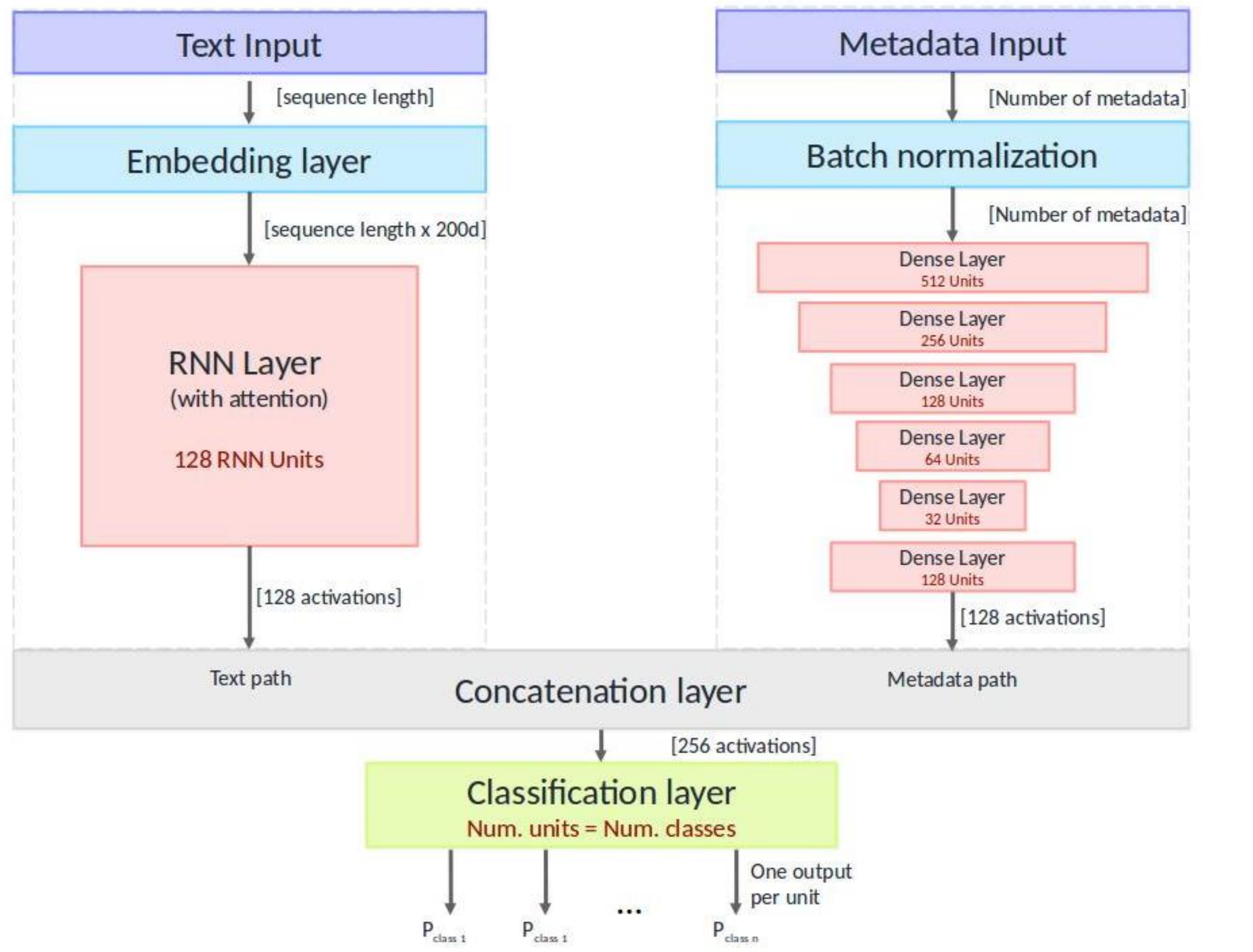}
\vspace{-0.2cm}
  \caption{Overview of the neural network setup for classification of abuse on Twitter.}
  \label{fig:neuralNetworkSetup}
\vspace{-0.2cm}
\end{figure*}

\descr{Tree-based Classifiers.}
Such classifiers~\cite{Quinlan1986DecisionTrees} are considered relatively fast to train and apply, compared to other classification models.

\theoremstyle{TreeClassifiers}
\begin{TreeClassifiers}{\textbf{Tree-based Classifiers.}}
\label{TreeClassifiers}
They construct trees by splitting the set of $x$ attributes into smaller subsets (internal nodes) beginning with $x$ itself (root node) and concluding with the assignment of data to categories (lead node).
Overall, they are comprised of three types of nodes:
\begin{itemize}[noitemsep, topsep=0pt]
\item the \textbf{root} node, with no incoming edges;
\item the \textbf{internal} nodes, with just one incoming and two or more outgoing edges;
\item the \textbf{leaf} node, with one incoming edge and no outgoing edges.
\end{itemize}
The root and each internal node correspond to attribute test conditions (in the simplest form, each test corresponds to a single attribute) for separating data based on their characteristics, while the leaf nodes correspond to the available categories.
\end{TreeClassifiers}

We experimented with various tree classifiers: J48, LADTree, LMT, NBTree, Random Forest (RF), and Functional Tree; we achieved best performance with the Random Forest classifier, which constructs a forest of decision trees with random subsets of features during the classification process.
So, an important advantage of the Random Forest classifier is its ability in reducing overfitting by averaging several trees during the model construction process.
Additionally, Random Forests are quite efficient in terms of the time they need for training a model, while finally they do not require any scaling in the features to be used (e.g., categorical, binary, or numerical).
To build the Random Forest model, we tune the number of trees to be generated as $100$ and the maximum depth unlimited.

\descr{Ensemble Classifiers.}
Such methods~\cite{Dietterich2000EnsembleMethods} are built upon a set of classifiers whose individual decisions are then combined (typically by a (un)weighted voting) to classify new data.
Ensemble classifiers often perform better than single classifiers.
The most important reason behind this performance is that errors from a single classifier can be eliminated by averaging across multiple classifiers' decisions.
Nevertheless, in order the ensemble methods to perform better than any single classifier, the base learners have to be as accurate as possible and as diverse as possible.

We experimented with various combinations of single classifiers from all the previously referred classification methods in order to conclude to the one with the most optimal performance, namely: Random Forest + BayesNet + Naive Bayes + AdaBoost~\cite{hastie2009multi} (with Random Forest).
We proceeded with the voting method which works by creating two or more sub-models.
Each sub-model makes predictions which are then combined based on a \textit{combination rule}, thus allowing each sub-model to vote on what the outcome should be.
In our case the majority vote has been used as a combination rule.

\descremph{Features selection.}
Most of the features presented in Section~\ref{sec:features} are useful in discriminating between the user classes.
Features not useful, i.e., that do not discriminate between classes in a statistically significant level, are excluded from the modeling analysis to avoid adding noise.
Specifically, we exclude the following features from our analysis: \emph{user-based} - verified account, default profile image, statistics on sessions, location, profile description, \emph{text-based} - average emotional scores, hate score, average word embedding, average curse score, number of mentions, unique number of mentioned users, number of retweets, and \emph{network-based} - closeness centrality and Louvain modularity.

\descr{Neural Networks for Detecting Abuse.}
\label{sec:neural-networks}
Even though the traditional machine learning approaches have been extensible used in text classification tasks, they face an important drawback: they cannot successfully combine semantic and cultural nuances of the written language.
For instance, considering the negation of words or sarcastic expressions with traditional machine learning approaches is a quite challenging task, as the structure of the sentence has to be effectively presented in the set of features.
To overcome such difficulties, deep learning algorithms have been proposed that build upon neural networks.
Thus, here we also proceed with a modeling process building upon neural networks.
We build a classification model for abusive behavior that combines together the i) raw text, and ii) the user-, text-, and  network-based features (called metadata). 
In the next paragraphs, we describe these two paths separately at first, but their output is combined together into a single model. 
The whole architecture is shown in Figure~\ref{fig:neuralNetworkSetup}.

\descremph{Text Classification Network.} 
\label{classifier:text}
This part of the classifier only considers the raw text as input.
There are several choices for the class of neural network to base our classifier on.
We use Recurrent Neural Networks (RNN) since they have proven successful at understanding sequences of words and interpreting their meaning.
We experimented with both character-level and word-level RNNs and found the latter to be the most performant in our dataset.

Text preprocessing:
first of all we concatenate all the tweets of a single user into a single text document (referred as \emph{text} from now on).
Before feeding any text to the network, we need to transform each sample to a sequence of words.
As neural networks are trained in mini-batches, every sample in a batch must have the same sequence length (number of words).
Text containing more words than the sequence length is trimmed, whereas text with less words is left-padded with zeros (the model learns they carry no information). 
Ideally, we want to setup a sequence length that is large enough to contain most text from the samples, but avoid outliers as they waste resources (feeding zeros in the network).
Thus, we take the $95$th percentile of length (with respect to the number of words) in the input corpus as the optimal sequence length.
Additionally, we remove any words that appear only once in the corpus, as they are most likely typos and can result in overfitting.
Once preprocessed, the input text is fed to the network for learning.

Word embedding layer:
the first layer of the network performs a word embedding, which maps each word to a high-dimensional vector.
Word embedding has proved to be a highly effective technique for text classification tasks, and additionally reduces the number of training samples required to reach good performance.
We settled on using pre-trained word embeddings from GloVe \cite{pennington2014glove}, which was constructed on more than $2$ billion tweets.
We choose the highest dimension embeddings available, i.e., $200$, as these produce the best results in our dataset (with different dimension embeddings the performance does not vary importantly, i.e., max $-0.03$ in some cases). 
If a word is not found in the GloVe dataset, we initialize a vector of random weights, which the word embedding layer eventually learns from the input data.

Recurrent layer:
the next layer is an RNN with $128$ units (neurons).
As mentioned previously, RNNs learn sequences of words by updating an internal state.
After experimenting with several choices for the RNN architecture (Gated Recurrent Unit or GRUs, Long Short-Term Memory or LSTMs, and Bidirectional RNNs), we find that due to the rather small sequences of length in social media (typically less than $100$ words per post, just $30$ for Twitter), simple GRUs perform as well as more complex units.
Additionally, to avoid overfitting, we use a recurrent dropout with $p=0.5$ (i.e., individual neurons were available for activation with probability $0.5$) as it is proposed in~\cite{srivastava2014dropout}.
Finally, an attention layer~\cite{bahdanau2014neural} is added, as it provides a mechanism for the RNN to ``focus'' on individual parts of the text that contain information that is related to the task.
Attention is particularly useful to tackle texts that contain longer sequences of words (e.g., multiple tweets concatenated together).

\descremph{Metadata Network.}
\label{classifier:metadata}
The metadata network considers non-sequential data.
For example, on Twitter, it might evaluate the number of followers, the number of recent tweets, the location, account age, total number of tweets, etc., of a user.

Metadata Preprocessing: before feeding the data into the neural network, we need to transform any categorical data into numerical, either via enumeration or one-hot encoding, depending on the particulars of the input.
Once this has taken place, each sample is thus represented as a vector of numerical features. 

Batch Normalization Layer: neural network layers work best when the input data have zero mean and unit variance, as it enables faster learning and higher overall accuracy.
Thus, we pass the data through a Batch Normalization layer that takes care of this transformation at each batch.

Dense layers: we use a simple network of several, fully connected (dense) layers to learn the metadata.
We design our network so that a bottleneck is formed.
Such a bottleneck has been shown to result in automatic construction of high-level features~\cite{he2016deep, tishby2015deep}.
In our implementation, we experimented with multiple architectures and decided to use $5$ layers of size $512$, $245$, $128$, $64$, $32$, which provide good results in our dataset.
On top of this layer, we add an additional ($6$th) layer which ensures this network has the same dimensionality as the text-only network; this ends up enhancing performance when we fuse the two networks.
Finally, we use $tanh$ as our activation function since it works better with standardized numerical data.

\begin{table*}[!t]
\centering
\resizebox{.95\textwidth}{!}{
\centering
\tabcolsep=0.11cm
\begin{tabular}{lllllllllllllllllllll}
\hline
                                       & \multicolumn{4}{l}{\textbf{bully}}                                   & \multicolumn{4}{l}{\textbf{aggressive}}                              & \multicolumn{4}{l}{\textbf{spam}}                                    & \multicolumn{4}{l}{\textbf{normal}}                                  & \multicolumn{4}{l}{\textbf{overall (weighted avg.)}} \\ \hline \hline
\multicolumn{1}{l|}{}                  & Prec           & Rec            & F1    & \multicolumn{1}{l|}{AUC}   & Prec           & Rec            & F1    & \multicolumn{1}{l|}{AUC}   & Prec           & Rec            & F1    & \multicolumn{1}{l|}{AUC}   & Prec           & Rec            & F1    & \multicolumn{1}{l|}{AUC}   & Prec      & Rec       & F1       & AUC      \\ \hline
\multicolumn{1}{l|}{\textbf{NB}}       & 0.065          & \textbf{0.902} & 0.120 & \multicolumn{1}{l|}{0.803} & 0.039          & 0.198 & 0.065 & \multicolumn{1}{l|}{0.337} & 0.347          & 0.151          & 0.210 & \multicolumn{1}{l|}{0.577} & \textbf{0.827} & 0.094          & 0.168 & \multicolumn{1}{l|}{0.604} & 0.614     & 0.151     & 0.176    & 0.598    \\
\multicolumn{1}{l|}{(STD)}             & 0.001          & 0.018          & 0.002 & \multicolumn{1}{l|}{0.017} & 0.006          & 0.027          & 0.010 & \multicolumn{1}{l|}{0.157} & 0.018          & 0.008          & 0.009 & \multicolumn{1}{l|}{0.004} & 0.022          & 0.002          & 0.003 & \multicolumn{1}{l|}{0.005} & 0.013     & 0.003     & 0.003    & 0.003    \\ \hline
\multicolumn{1}{l|}{\textbf{RF}}       & \textbf{0.462} & 0.514          & 0.486 & \multicolumn{1}{l|}{0.897} & 0.428          & 0.119          & 0.185 & \multicolumn{1}{l|}{0.780} & 0.697 & 0.548          & 0.613 & \multicolumn{1}{l|}{0.804} & 0.783          & \textbf{0.895} & 0.835 & \multicolumn{1}{l|}{0.828} & 0.729     & 0.742     & 0.728    & 0.822    \\
\multicolumn{1}{l|}{(STD)}             & 0.022          & 0.023          & 0.020 & \multicolumn{1}{l|}{0.009} & 0.096          & 0.021          & 0.033 & \multicolumn{1}{l|}{0.019} & 0.009          & 0.011          & 0.008 & \multicolumn{1}{l|}{0.004} & 0.004          & 0.006          & 0.004 & \multicolumn{1}{l|}{0.003} & 0.006     & 0.005     & 0.005    & 0.004    \\ \hline
\multicolumn{1}{l|}{\textbf{Ensemble}} & 0.405          & 0.731          & 0.521 & \multicolumn{1}{l|}{0.841} & \textbf{0.475} & 0.070          & 0.121 & \multicolumn{1}{l|}{0.534} & 0.697 & 0.557          & 0.620 & \multicolumn{1}{l|}{0.723} & 0.797          & 0.872          & 0.833 & \multicolumn{1}{l|}{0.767} & 0.738     & 0.738     & 0.728    & 0.748    \\
\multicolumn{1}{l|}{(STD)}             & 0.006          & 0.015          & 0.008 & \multicolumn{1}{l|}{0.007} & 0.056          & 0.000          & 0.002 & \multicolumn{1}{l|}{0.001} & 0.006          & 0.008          & 0.006 & \multicolumn{1}{l|}{0.004} & 0.003          & 0.005          & 0.003 & \multicolumn{1}{l|}{0.003} & 0.003     & 0.003     & 0.003    & 0.003    \\ \hline\hline
\multicolumn{1}{l|}{\textbf{NN}}       & 0.270           & 0.640           & 0.380     & \multicolumn{1}{l|}{0.901}  & 0.110           & \textbf{0.300}           & 0.160     & \multicolumn{1}{l|}{0.820}  & \textbf{0.810}           & \textbf{0.680}           & 0.740     & \multicolumn{1}{l|}{0.792}  & 0.610           & 0.570           & 0.59 0    & \multicolumn{1}{l|}{0.745}  & 0.700     & 0.630      & 0.660        & 0.815    \\ 
\multicolumn{1}{l|}{(STD)}             & 0.019          & 0.015          & 0.008 & \multicolumn{1}{l|}{0.006} & 0.010          & 0.010          & 0.003 & \multicolumn{1}{l|}{0.007} & 0.048          & 0.052          & 0.031 & \multicolumn{1}{l|}{0.004} & 0.012          & 0.012          & 0.005 & \multicolumn{1}{l|}{0.005} & 0.002     & 0.006     & 0.004    & 0.003    \\
\hline
\end{tabular}}
\vspace{-0.2cm}
\caption{Results on 4 classes: detecting aggressive vs. bullying behavior.}
\label{tbl:4classes_results}
\vspace{-0.2cm}
\end{table*}

\descremph{Combining the two paths.}
The two classifiers presented above can handle individually either the raw text or the metadata.
To build a multi-input classifier, we combine these two paths using a concatenation layer. 
Finally, we use a fully connected output layer (a.k.a. dense layer) with one neuron per class we want to predict, and a softmax activation function to normalize output values between $0$ and $1$.
The output of each neuron at this stage represents the probability of the sample belonging to each respective class.

\descremph{Training the combined network.}
\label{training}
The simplest approach is to train the entire network at once; i.e., to treat it as a single classification network with two inputs.
However, the performance we achieve from this training technique is suboptimal: the two paths have different convergence rates (i.e., one of the paths might converge faster, and thus dominate any subsequent training epochs).
Furthermore, standard backpropagation across the whole network can also induce unpredictable interactions, as we allow the weights to be modified at both paths simultaneously.

To avoid this issue, we perform interleaved training~\cite{hidasi2016parallel}: at each mini-batch, data flow through the whole network, but only the weights of one of the paths are updated. 
To do so, we train the two paths in an alternating fashion. 
For example, on even-numbered mini-batches the gradient descent only updates the weights of the text path, whereas on odd-numbered  batches the weights of the metadata path.

This results in a more optimal, balanced network as the gradient is only able to change one path at a time, thus avoiding unwanted interactions. 
At the same time, the loss function is calculated over the whole, combined, network.
Notice that the input does pass through the whole network.

\descremph{Neural network setup.}
For our implementation we use Keras~\cite{keras} with Theano~\cite{theano} as back-end for the deep learning models implementation.
We use the functional API to implement our multi-input single-output model.
Finally, we run the experiments on a server that is equipped with three Tesla K40c GPUs.

In terms of training, we use \emph{categorical cross-entropy} as loss function and \emph{Adam} as the  optimization function.
A maximum of $100$ epochs is allowed, but we also employ a separate validation set to perform \emph{early stopping}: training is interrupted if the validation loss does not drop in $10$ consecutive epochs and the weights of the best epoch are restored.

\subsection{Experimental Methodology}\label{sec:experiments}

\descr{Experimentation Phases.}
Four setups are tested to assess the feasibility of detecting abusive user behavior on Twitter:
\begin{itemize}
\item  4-classes classification: bully, aggressive, spam, and normal users.
\item  3-classes after spam removal classification: bully, aggressive, and normal users. This setup examines the case where we filter out spam with a more elaborate technique and attempt to detect the bullies and aggressors from normal users.
\item  3-classes offensive classification: offensive (both bullying and aggressive), spam, and normal users. With this setup, we join aggressive and bullying users into one class, in order to assess how well we can detect abusive behavior overall.
\item  2-classes offensive classification: same as in 3-classes offensive classification but without spam.
\end{itemize}

\descr{Evaluation Metrics.}
For evaluation purposes in all experimental phases, we examine standard machine learning performance metrics:
\emph{precision} (prec), \emph{recall} (rec), \emph{F1-score} (F1), and weighted area under the ROC curve (AUC), at the class level and overall weighted average across classes.
Also, the accuracy value is presented.
For all the experiments presented next, we use the WEKA data mining toolkit and repeated ($10$ times) $10$-fold cross validation~\cite{Kim2009ErrorRate}, providing the relevant standard deviation (STD).
For the Neural Network setup, as already has been stated, we use Keras with Theano.

\subsection{Classification Results}
\label{subsec:classificationresults}

\descr{Detecting Aggressive vs. Bullying Behavior.}
We examine whether it is possible to distinguish between bully, aggressive, spam, and normal users.
Table~\ref{tbl:4classes_results} overviews the results obtained with the considered classifiers, i.e., the best classifier for each family of classifiers (i.e., Probabilistic, Tree-based, Ensemble, Neural Network).
In more detail, we observe that with the Naive Bayes classifier we succeed to detect \textbf{90.2\%} (STD = 0.018) of the bully cases which is especially high considering the small number of bully cases.
Compared with the results presented in~\cite{chatzakou2017mean}, here we observe that by considering additional textual features, such as POS and stylistic features, the overall number of detected bully cases has been increased, i.e., \textbf{+47\%}.
The highest number of the detected aggressive users is achieved again with the Naive Bayes classifier, but yet the rate is quite low (\textbf{19.8\%}, STD=0.027).
On the other hand, the highest precision value for the aggressive case is achieved with the Ensemble classifier, where again we observe an important increase on the achieved precision compared to the one presented in~\cite{chatzakou2017mean}, i.e., \textbf{+18\%}.

As we also observe from Table~\ref{tbl:4classes_results}, the considered classifiers perform differently both across the four classes, as well as the evaluation metrics (i.e., precision vs. recall).
So, depending on the task under consideration the most suitable classifier should be selected.
For instance, if the objective is to achieve high true positive rate (i.e., high recall), then the Naive Bayes classifier should be used, while if the case is to achieve the best results in terms of the true positively predicted bully cases, then the Random Forest classifier should be selected.
The Random Forest classifier strikes a balance among precision and recall across the user classes, while at the same time it surpasses the ensemble classifier in terms of the execution speed.
The average precision with Random Forest is \textbf{72.9\%}, and the recall is \textbf{74.6\%}, while the accuracy is \textbf{74.1\%}.

From the confusion matrix that is presented in Table~\ref{tbl:4classes_confusion_matrix} (showing in percentage the distribution of the cases among the classes based on the repeated - 10 times - 10-fold cross validation - the `total' value indicates the rounded absolute number of cases that belong to each one of the considered classes) we observe that a high percentage of misclassified cases falls in the normal class which aligns with the human annotations gathered during the crowdsourcing phase.
Specifically, a high confusion is observed between the spam and normal classes, which may align with the idea that spammers try to be as close to ``real or normal users''.
Quite interesting is the fact that both bully and aggressive users are often classified as spammers which is in alignment with~\cite{Chen2015SpamTweets} work which indicates that spammers often exhibit behavior that could be considered as aggressive or bullying (repeated posts with similar content, mentions, or hashtags).
We should note here that all the results presented on the confusion matrices in this section have been extracted based on the Random Forest classifier.

\begin{table}[!t]
\centering
\resizebox{.5\textwidth}{!}{
\centering
\begin{tabular}{llllll}
\textbf{a}   & \textbf{b}   & \textbf{c}  & \textbf{d}  & \textbf{total}                     & \textit{$\leftarrow$ classified as} \\ \cline{1-5}
89.52\%      & 9.24\%       & 0.23\%      & 1.02\%      & \multicolumn{1}{l|}{\textit{787}}  & \textbf{a=normal}                   \\
40.48\%      & 54.84\%      & 0.39\%      & 4.29\%      & \multicolumn{1}{l|}{\textit{415}}  & \textbf{b=spam}                     \\
41.16\%      & 26.05\%      & 11.86\%     & 20.93\%     & \multicolumn{1}{l|}{\textit{43}}   & \textbf{c=aggressive}               \\
17.07\%      & 25.34\%      & 6.21\%      & 51.38\%     & \multicolumn{1}{l|}{\textit{58}}   & \textbf{d=bully}                    \\
\textit{900} & \textit{326} & \textit{12} & \textit{65} & \multicolumn{1}{l|}{\textit{1,303}} & \textbf{total}                     
\end{tabular}
}
\vspace{-0.2cm}
\caption{Confusion matrix: detecting aggressive vs. bullying behavior.}
\label{tbl:4classes_confusion_matrix}
\vspace{-0.2cm}
\end{table}

Table~\ref{tbl:4classes_results} also presents the results obtained based on the neural network setup.
As far as the bully case we observe that with the neural network setup we succeed to detect a quite satisfactory number of bully users, i.e., \textbf{64\%}, while at the same time we succeed to detect the highest number of aggressive users, i.e., \textbf{30\%}, which is \textbf{+10.2\%} compared to the Naive Bayes classifier.
So, the neural network setup could be used in cases where the focus is on detecting aggressive users when acting on Twitter.
The overall precision and recall values equal to \textbf{70\%} and \textbf{63\%}, respectively, with \textbf{63.1\%} accuracy.

\begin{table*}[!t]
\centering
\resizebox{.85\textwidth}{!}{
\centering
\tabcolsep=0.11cm
\begin{tabular}{lllllllllllllllll}
\hline
                                       & \multicolumn{4}{l}{\textbf{bully}}                                   & \multicolumn{4}{l}{\textbf{aggressive}}                              & \multicolumn{4}{l}{\textbf{normal}}                                  & \multicolumn{4}{l}{\textbf{overall (weighted avg.)}} \\ \hline\hline
\multicolumn{1}{l|}{}                  & Prec           & Rec            & F1    & \multicolumn{1}{l|}{AUC}   & Prec           & Rec            & F1    & \multicolumn{1}{l|}{AUC}   & Prec           & Rec            & F1    & \multicolumn{1}{l|}{AUC}   & Prec      & Rec       & F1       & AUC      \\ \hline
\multicolumn{1}{l|}{\textbf{NB}}       & 0.096          & \textbf{0.886} & 0.173 & \multicolumn{1}{l|}{0.805} & 0.044          & 0.235 & 0.074 & \multicolumn{1}{l|}{0.088} & 0.957          & 0.149          & 0.257 & \multicolumn{1}{l|}{0.776} & 0.857     & 0.201     & 0.243    & 0.761    \\
\multicolumn{1}{l|}{(STD)}             & 0.002          & 0.017          & 0.004 & \multicolumn{1}{l|}{0.011} & 0.003          & 0.021          & 0.006 & \multicolumn{1}{l|}{0.122} & 0.005          & 0.004          & 0.006 & \multicolumn{1}{l|}{0.011} & 0.004     & 0.004     & 0.006    & 0.010    \\ \hline
\multicolumn{1}{l|}{\textbf{RF}}       & \textbf{0.597} & 0.636          & 0.616 & \multicolumn{1}{l|}{0.907} & 0.359          & 0.126          & 0.186 & \multicolumn{1}{l|}{0.786} & 0.948          & \textbf{0.977} & 0.962 & \multicolumn{1}{l|}{0.873} & 0.896     & 0.913     & 0.902    & 0.871    \\
\multicolumn{1}{l|}{(STD)}             & 0.016          & 0.029          & 0.019 & \multicolumn{1}{l|}{0.005} & 0.038          & 0.012          & 0.017 & \multicolumn{1}{l|}{0.019} & 0.003          & 0.001          & 0.002 & \multicolumn{1}{l|}{0.003} & 0.003     & 0.002     & 0.002    & 0.003    \\ \hline
\multicolumn{1}{l|}{\textbf{Ensemble}} & 0.525          & 0.748          & 0.617 & \multicolumn{1}{l|}{0.892} & \textbf{0.388} & 0.072          & 0.121 & \multicolumn{1}{l|}{0.668} & 0.958 & 0.971          & 0.965 & \multicolumn{1}{l|}{0.892} & 0.902     & 0.913     & 0.901    & 0.881    \\
\multicolumn{1}{l|}{(STD)}             & 0.008          & 0.019          & 0.010 & \multicolumn{1}{l|}{0.009} & 0.052          & 0.007          & 0.012 & \multicolumn{1}{l|}{0.013} & 0.002          & 0.001          & 0.001 & \multicolumn{1}{l|}{0.004} & 0.003     & 0.001     & 0.002    & 0.004    \\ \hline \hline
\multicolumn{1}{l|}{\textbf{NN}}       & 0.270           & 0.710           & 0.390     & \multicolumn{1}{l|}{0.873}  & 0.130           & \textbf{0.300}           & 0.180     & \multicolumn{1}{l|}{0.782}  & \textbf{0.980}           & 0.800           & 0.880     & \multicolumn{1}{l|}{0.899}  & 0.890      & 0.770      & 0.810        & 0.853     \\ 
\multicolumn{1}{l|}{(STD)}             & 0.026          & 0.015          & 0.004 & \multicolumn{1}{l|}{0.007} & 0.013          & 0.014          & 0.001 & \multicolumn{1}{l|}{0.006} & 0.017          & 0.014          & 0.002 & \multicolumn{1}{l|}{0.004} & 0.001     & 0.004     & 0.003    & 0.004    \\ 
\hline
\end{tabular}}
\vspace{-0.2cm}
\caption{Results on 3 classes: classifying after spam removal.}
\label{tbl:3classes_nospam_results_new}
\vspace{-0.2cm}
\end{table*}

\descr{Classifying After Spam Removal.}\label{subsec:3_class}
In this experimental phase, we want to explore whether the distinction between bully/aggressive and normal users will be more evident after applying a more sophisticated spam removal process in the preprocessing step.
To this end, we remove from our dataset all the cases identified by the annotators as spam, and re-run the four already considered models (i.e., Naive Bayes, Random Forest, Ensemble, and Neural Network classifiers).
Table~\ref{tbl:3classes_nospam_results_new} shows that for bully cases there is an increase in both the precision and recall values compared to the results obtained on the four class classification (i.e., detecting aggressive vs. bullying behavior).
More specifically, with the Random Forest classifier there is a \textbf{+13.5\%} increase in the precision and a \textbf{+12.2\%} increase in the recall value.

For aggressors, considering the Random Forest classifier the recall values are almost the same, indicating that further examination of this behavior is warranted in the future.
In the Naive Bayes case where we observe the highest overall AUC in the aggressive case, i.e., \textbf{88\%}, we succeed to detect \textbf{+3.7\%} more aggressive users.
Again, after removing the spam users from our dataset, we succeed to detect the highest number of aggressive users based on the neural network setup, i.e., \textbf{30\%}, with a quite satisfactory AUC value, i.e., \textbf{78.2\%}.
At the same time we succeed to detect an important number of bully cases, i.e., \textbf{71\%}, with \textbf{87.3\%} AUC value.
Overall, the highest precision and recall values are achieved with the ensemble classifier, and equal to \textbf{90.2\%} and \textbf{97.1\%}, respectively, while the accuracy equals to \textbf{90.2\%}.
Considering the AUC of \textbf{88.1\%}, we believe that with a more sophisticated spam detection applied on the stream of tweets, our features 
and classification techniques can perform even better at detecting bullies and aggressors and distinguishing them from the typical Twitter users.

\begin{table}[!t]
\centering
\resizebox{.4\textwidth}{!}{
\centering
\begin{tabular}{lllll}
\textbf{a}   & \textbf{b}  & \textbf{c}  & \textbf{total}                    & \textit{$\leftarrow$ classified as} \\ \cline{1-4}
0.66\%       & 1.63\%      & 97.71\%     & \multicolumn{1}{l|}{\textit{787}} & \textbf{a=normal}                   \\
13.26\%      & 26.74\%     & 60.00\%     & \multicolumn{1}{l|}{\textit{43}}  & \textbf{b=aggressive}               \\
7.68\%       & 61.43\%     & 30.89\%     & \multicolumn{1}{l|}{\textit{58}}  & \textbf{c=bully}                    \\
\textit{812} & \textit{17} & \textit{59} & \multicolumn{1}{l|}{888}          & \textbf{total}                     
\end{tabular}
}
\vspace{-0.2cm}
\caption{Confusion matrix: classifying after spam removal.}
\label{tbl:3classes_nospam_confusion_matrix}
\vspace{-0.2cm}
\end{table}

\begin{table*}[!t]
\centering
\resizebox{.85\textwidth}{!}{
\centering
\tabcolsep=0.11cm
\begin{tabular}{lllllllllllllllll}
\hline
                                       & \multicolumn{4}{l}{\textbf{offensive}}                               & \multicolumn{4}{l}{\textbf{spam}}                                    & \multicolumn{4}{l}{\textbf{normal}}                                  & \multicolumn{4}{l}{\textbf{overall (weighted avg.)}} \\ \hline
\multicolumn{1}{l|}{}                  & Prec           & Rec            & F1    & \multicolumn{1}{l|}{AUC}   & Prec           & Rec            & F1    & \multicolumn{1}{l|}{AUC}   & Prec           & Rec            & F1    & \multicolumn{1}{l|}{AUC}   & Prec      & Rec       & F1       & AUC      \\ \hline \hline
\multicolumn{1}{l|}{\textbf{NB}}       & 0.092          & \textbf{0.914} & 0.168 & \multicolumn{1}{l|}{0.716} & 0.364          & 0.176          & 0.237 & \multicolumn{1}{l|}{0.589} & \textbf{0.832} & 0.105          & 0.186 & \multicolumn{1}{l|}{0.614} & 0.626     & 0.190     & 0.201    & 0.614    \\
\multicolumn{1}{l|}{(STD)}             & 0.002          & 0.015          & 0.004 & \multicolumn{1}{l|}{0.018} & 0.019          & 0.019          & 0.020 & \multicolumn{1}{l|}{0.008} & 0.015          & 0.004          & 0.007 & \multicolumn{1}{l|}{0.008} & 0.012     & 0.005     & 0.006    & 0.007    \\ \hline
\multicolumn{1}{l|}{\textbf{RF}}       & \textbf{0.588} & 0.552          & 0.569 & \multicolumn{1}{l|}{0.877} & \textbf{0.713} & 0.545          & 0.618 & \multicolumn{1}{l|}{0.805} & 0.788          & \textbf{0.891} & 0.836 & \multicolumn{1}{l|}{0.829} & 0.748     & 0.755     & 0.746    & 0.825    \\
\multicolumn{1}{l|}{(STD)}             & 0.021          & 0.023          & 0.020 & \multicolumn{1}{l|}{0.008} & 0.011          & 0.016          & 0.014 & \multicolumn{1}{l|}{0.003} & 0.006          & 0.003          & 0.004 & \multicolumn{1}{l|}{0.003} & 0.008     & 0.007     & 0.008    & 0.003    \\ \hline
\multicolumn{1}{l|}{\textbf{Ensemble}} & 0.556          & 0.648          & 0.599 & \multicolumn{1}{l|}{0.855} & 0.690          & 0.558          & 0.617 & \multicolumn{1}{l|}{0.786} & 0.799          & 0.862          & 0.829 & \multicolumn{1}{l|}{0.825} & 0.745     & 0.749     & 0.744    & 0.815    \\
\multicolumn{1}{l|}{(STD)}             & 0.007          & 0.020          & 0.012 & \multicolumn{1}{l|}{0.010} & 0.013          & 0.010          & 0.010 & \multicolumn{1}{l|}{0.004} & 0.004          & 0.006          & 0.004 & \multicolumn{1}{l|}{0.003} & 0.006     & 0.006     & 0.006    & 0.003    \\ \hline  \hline
\multicolumn{1}{l|}{\textbf{NN}}       & 0.330             & 0.690              & 0.450     & \multicolumn{1}{l|}{0.883}     & 0.610              & \textbf{0.610}             & 0.610     & \multicolumn{1}{l|}{0.767}     & 0.820              & 0.700              & 0.760    & \multicolumn{1}{l|}{0.797}     & 0.710         & 0.670         & 0.690        & 0.816        \\ 
\multicolumn{1}{l|}{(STD)}             & 0.006          & 0.012          & 0.011 & \multicolumn{1}{l|}{0.005} & 0.004          & 0.006          & 0.008 & \multicolumn{1}{l|}{0.006} & 0.004          & 0.005          & 0.008 & \multicolumn{1}{l|}{0.004} & 0.004     & 0.005     & 0.005    & 0.003    \\
\hline
\end{tabular}}
\vspace{-0.2cm}
\caption{Results on 3 classes: detecting offensive behavior.}
\label{tbl:3classes_offensive_results}
\vspace{-0.2cm}
\end{table*}

Finally, from the confusion matrix presented in Table~\ref{tbl:3classes_nospam_confusion_matrix} we observe that an impressive percentage of aggressive cases, i.e., 60\%, has been classified as bullies which indicates an important confusion between them.
In this sense, a more thorough analysis should be conducted in order to select attributes that are clearly separable between the bully and aggressive users.
Concerning the bully cases we see that an important percentage has been correctly classified, while $30.89\%$ of them have been characterized as aggressive users with only $7.68\%$ to be classified as normal.

\descr{Detecting Offensive Behavior.}\label{subsec:3_class_offensive}
Here, the objective is to detect offensive users in general, and thus we combine the bullying and aggressive labels into a new label called \textit{offensive}.
Table~\ref{tbl:3classes_offensive_results} presents the obtained results.
In the offensive case, the highest precision is obtained with the Random Forest classifier (\textbf{58.8\%}), while the best recall is achieved with the Naive Bayes (\textbf{91.4\%}), which is quite impressive for a baseline algorithm.
The ensemble classifier is quite comparable with the Random Forest classifier in terms of precision and recall of the offensive class.
Finally, with the neural network setup we achieve to detect \textbf{69\%} of the offensive users which is the second highest number compared to the other machine learning algorithms.
The highest ROC area for the overall classification is achieved with the Random Forest classifier (\textbf{82.5\%}), which indicates that the Random Forest model can quite successfully discriminate between the offensive, normal, and spam users.

Table~\ref{tbl:3classes_offensive_confusion_matrix} presents the confusion matrix based on the classification results obtained after merging the offensive classes.
Similar to the above analysis (i.e., detecting aggressive vs. bullying behavior), here the misclassified offensive cases fall almost in an equal percent to both the normal (i.e., 23.86\%) and spam (i.e., 20.99\%) classes.
Finally, we observe that almost half of the spam cases have been classified as normal users.

\begin{table}[!t]
\centering
\resizebox{.4\textwidth}{!}{
\centering
\begin{tabular}{lllll}
\textbf{a}   & \textbf{b}   & \textbf{c}  & \textbf{total}                      & \textit{$\leftarrow$ classified as} \\ \cline{1-4}
89.15\%      & 8.87\%       & 1.98\%      & \multicolumn{1}{l|}{\textit{787}}   & \textbf{a=normal}                   \\
39.86\%      & 54.49\%      & 5.65\%      & \multicolumn{1}{l|}{\textit{415}}   & \textbf{b=spam}                     \\
23.86\%      & 20.99\%      & 55.15\%     & \multicolumn{1}{l|}{\textit{101}}   & \textbf{c=offensive}                \\
\textit{891} & \textit{317} & \textit{95} & \multicolumn{1}{l|}{\textit{1,303}} & \textbf{total}                     
\end{tabular}
}
\vspace{-0.2cm}
\caption{Confusion matrix: detecting offensive behavior.}
\label{tbl:3classes_offensive_confusion_matrix}
\vspace{-0.2cm}
\end{table}

\descr{Detecting Offensive Behavior After Spam Removal.}\label{subsec:2_class_offensive}
The simplest form of classification would be to distinguish between offensive and normal users (i.e., without considering spam users).
Thus, similar to `Classifying After Spam Removal' experimental setup, we remove all the cases identified by the annotators as spam, and proceed with a classification setup where only the offensive and normal users are considered.
Overall, the best results are obtained with the Random Forest classifier, with \textbf{93.4\%} precision, \textbf{93.7\%} recall, and \textbf{91.1\%} AUC.
Looking into each class in more detail, the highest number of offensive cases is detected with the Naive Bayes, i.e., \textbf{93.7\%} (with the neural network setup falling behind with \textbf{73\%}), while with the Random Forest we succeed to detect \textbf{97.2\%} of the normal users.
Such basic classification model could be considered as the initial step on the effort to detect offensive behavior, overall.
A multi-class classification modeling can be performed as a second step for a more detailed break-down of this behavior.

\subsection{Machine Learning Classification Takeaways}

Overall, our model performs reasonable in the four considered setups (Tables~\ref{tbl:4classes_results},~\ref{tbl:3classes_nospam_results_new},~\ref{tbl:3classes_offensive_results}, and Detecting Offensive Behavior After Spam Removal setup).
This is justified if we take into consideration the overall AUC of the ROC curves, which are typically used to evaluate the performance of a machine learning algorithm~\cite{davis2006relationship} by testing the system on different points and getting pairs of true positive (i.e., recall) against false positive rates indicating the sensitivity of the model.
The resulting area under the ROC curve can be read as the probability of a classifier correctly ranking a random positive case higher than a random negative case.

Specifically, in the aggressor case, even though the recall value is low, the AUC is quite high because the false positive rate is especially low, with $0.003$ and $0.009$ for the 4-classes  and 3-classes after spam removal classification (based on the Random Forest classifier), respectively.
We note that avoiding false positives is crucial to the successful deployment of any automated system aiming to deal with aggressive behavior.
Overall, given that in the four experimental setups the performance is different in terms of the precision and recall values among the considered classification methods, the selection of the model to be used should be based on the system that one is envisioned to develop.
If the system to be developed will be free of any human intervention (i.e., fully automated), then the platform developer could opt for a model that performs best in terms of the precision.
On the other hand, if the system will contain a second level of checks, which involves human editors (i.e., humans who will monitor the detected/flagged behaviors), then the platform developer could opt for a model with a higher recall value.
In this case, the classification system's results would be fed to a human-monitored system for thorough evaluation of the suspected users and final decision if to suspend them or not, thus reducing false positives.
For instance, in our case, in the four classes classification setup, the neural networks method can be used for detecting aggressive users in a human-monitored system, while on the contrary, the Random Forest classifier could be used to detect bully users when developing a system without any human intervention.

Finally, Table~\ref{tbl:features_evaluation_3setups} shows the top $12$ features for each setup based on information gain.
Overall, in all setups the most contributing features tend to be the user- and network-based, which describe the activity and connectivity of a user in the network.
As far as the new-considered features (i.e., POS and stylistic features) we observe that in the 4- and 3-classes offensive experiments the \textit{number of used verbs} in users' tweets is among the top contributed features.
In the 3- (after removing the spam users from our dataset) and 2-classes experiments the \textit{average words length} feature plays an important role during the classification process, while in the 2-classes experiment also the \textit{average words per sentence} as well as the \textit{number of adjectives} are quite important.
All such features indicate that the study of the linguistic style used to written language can importantly contribute in better distinguishing among the different user behaviors.

\vspace{0.1cm}
\noindent
We should note that for brevity, in this section, we presented the confusion matrices and the top contributed features based on the Random Forest classifier only, since with the Random Forest we succeeded a balance between the precision and recall values across all classes and higher overall AUC value in the three out of four experimental phases.

\begin{table*}[!t]
\begin{center}
\centering
\resizebox{.8\textwidth}{!}{
\centering
\begin{tabular}{ll}
\hline
{\bf Experiment}				&	{\bf Features	(preserving order of contribution)}	\\
\hline\hline
4-classes					& \#friends (11.68\%), reciprocity (11.20\%), \#followers (10.94\%)\\
						& \#followers/\#friends (9.62\%), interarrival (9.38\%), \#lists (9.18\%)\\	
						& hubs (9.13\%), \#URLs (7.74\%), \#hashtags (6.99\%) \\	
						& authority (6.24\%), \#verbs (4.38\%), clustering coef. (3.52\%)\\
\hline
3-classes (spam removal)			& account age (12.54\%), \#followers/\#friends (11.38\%), \#friends (10.90\%)\\
						& hub (10.06\%), interarrival (10.06\%), reciprocity (9.66\%), \#lists (8.71\%)\\
						& \#posts (8.66\%), \#hashtags (5.39\%), avg. words length (4.53\%)\\
						& \#followers (4.34\%), \#URLs (3.77\%)\\
\hline
3-classes (offensive)			& \#friends (11.64\%), reciprocity (10.77\%), \#followers (10.62\%)\\
						& \#followers/\#friends (9.44\%), interarrival (9.16\%), \#lists (8.68\%)\\
						& hubs (8.20\%), \#URLs (7.37\%), acount age (7.84\%), \#hashtags (6.46\%)\\
						& authority (5.87\%), \#verbs (3.95\%)\\
\hline
2-classes (offensive - without spam)	& account age (16.46\%), interarrival (12.21\%), \#friends (12.07\%)\\
						& \#lists (11.81\%), \#followers/\#friends (10.00\%), hubs (8.76\%)\\
						& \#posts (8.13\%), \#hashtags (6.60\%), avg. words length (4.89\%)\\
						& \#URLs (3.66\%), avg. words per sentence (3.46\%), \#adjectives (1.95\%)\\
\hline\hline
\end{tabular}
}
\end{center}
\vspace{-0.4cm}
\caption{Features evaluation for the four different experimental phases.}
\label{tbl:features_evaluation_3setups}
\vspace{-0.2cm}
\end{table*}

\section{Abusive Behavior on Twitter: User vs. Platform Reaction}\label{sec:twitter-reaction}

In the previous sections, we investigated users involved in the Gamergate community of Twitter and compared them with the activity of users involved in the BBCpay and NBA topics.
We studied their posting behavior through various features and showed that it is feasible to train a machine learning algorithm to automatically detect bullies and aggressors on Twitter.
However, it is still unclear how abusers are handled by the Twitter platform.
In particular, in this section we aim to address the following open questions:
\begin{itemize}
\item  What is the Twitter user account status and how can we measure it?
\item  What does this status imply for a user and what is the distribution of different statuses among the user communities examined, i.e., Gamergate, BBCpay, and NBA?
\item  What are the characteristics of users who are suspended from the platform and how do they compare with users who delete their Twitter account?
\item  Can we emulate the Twitter account suspension mechanism using typical machine learning methods?
\item  How effective is the recent Twitter effort for curving abusive behavior?
\end{itemize}

\subsection{Users Communities vs. Twitter Statuses}

A Twitter user can be in one of the following three statuses: (i)~\emph{active}, (ii)~\emph{deleted}, or (iii)~\emph{suspended}.
Typically, Twitter suspends an account (temporarily or even permanently, in some cases) if it has been hijacked/compromised, is considered spam/fake, or if it is \emph{abusive}~\cite{twitterStatuses}.
A user account is deleted if the user himself, for his own personal reasons, deactivates his account.

In order to examine the differences between these three statuses and in relation to the aggressive behavior overall, we initially selected a random sample of $40k$ ($10k$ from each dataset) users from the Gamergate, BBCpay, NBA, and baseline users to check their current (i.e., September 2018) Twitter status.
The status of each user account was checked using a mechanism that queried the Twitter API for each user and examined the error code responses returned: \textit{code $63$} corresponds to a suspended user account and \textit{code $50$} corresponds to a deleted one.

\begin{table}[!t]
\centering
\resizebox{.3\textwidth}{!}{
\centering
\begin{tabular}{@{}llll@{}}
\toprule
          		& \textbf{active} & \textbf{deleted} & \textbf{suspended} \\ \midrule
Baseline  	& 65.71\%           & 25.86\%             & 8.43\%                  \\
Gamergate 	& 71.86\%           & 16.22\%             & 11.29\%                  \\
NBA       	& 78.61\%           & 9.14\%               & 12.25\%                  \\
BBCpay    	& 79.79\%           & 10.17\%             & 10.05\%                  \\ \bottomrule
\end{tabular}}
\vspace{-0.2cm}
\caption{Distribution of Twitter statuses on Sept 2018.}
\label{tbl:statusCheck}
\vspace{-0.2cm}
\end{table}

Table~\ref{tbl:statusCheck} shows the distribution of statuses in the four considered datasets (i.e., baseline, Gamergate, NBA, and BBCpay).
We observe that baseline and Gamergate users tend to delete their accounts by choice rather than get suspended.
However, baseline users are more prone to deletion ($25.86\%$) of their accounts in contrast to the Gamergate users ($16.22\%$).
In contrast, Gamergate users tend to get suspended more often than baseline users ($11.29\%$ vs. $8.43\%$), which aligns with the offensive behavior observed in this community.
For the NBA and BBCpay, the percentage of active users is higher than that of Gamergate and baseline users, indicating a set of users actively involved in the topics of the community, and in particular of the newly formed BBCpay topic.
In fact, these two communities also have a lower rate of deleted users, perhaps indicating more satisfied users.

However, the suspended users of these two communities reach a higher portion than baseline, and closer to Gamergate's level.
Given our previous observations on the posting behavior of Gamergate users, someone would expect a high number of suspensions only in the Gamergate users, who are hyper-focused around a somewhat niche topic.
Though this could also be justified for the BBCpay community, with some users expressing themselves in an overly aggressive or abusive fashion, it would not be expected for the users in the NBA community.
Since the baseline and NBA communities are less prone to post abusive content, the suspension rates in the NBA participants could be due to general spam activity.

\subsection{Active, Suspended, and Deleted User Accounts}
\label{subsec:suspended-deleted}

To understand how active, suspended, and deleted users differ in activity, here we compare each of these user statuses for all the already considered user communities, i.e., Gamergate, BBCpay, and NBA, as well as baseline users based on an indicative subset of user (i.e., number of lists, posts, favorites, account age), text (i.e., number of hashtags, URLs, adjectives, sentiment), and network (i.e., number of followers - the same applies for the number of friends) features.

Since users are suspended because their activity violates Twitter rules, while users delete their account for their own reasons, we would expect suspended users to differ from the deleted users, at least at some of the aforementioned considered attributes.
To examine the significance of differences among the distributions presented next, we use the two-sample Kolmogorov-Smirnov test.
We consider as statistically significant all cases with $p < 0.01$.

\begin{figure*}[!t]
	\centering
	\begin{subfigure}[b]{0.24\textwidth}
		\includegraphics[width=\textwidth]{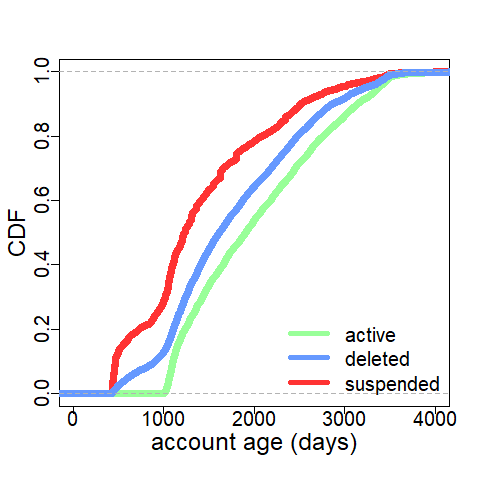}
		\captionsetup{font=footnotesize}
		\caption{Account age}
		\label{fig:sus_del_new_age}
	\end{subfigure}
	\begin{subfigure}[b]{0.24\textwidth}
		\includegraphics[width=\textwidth]{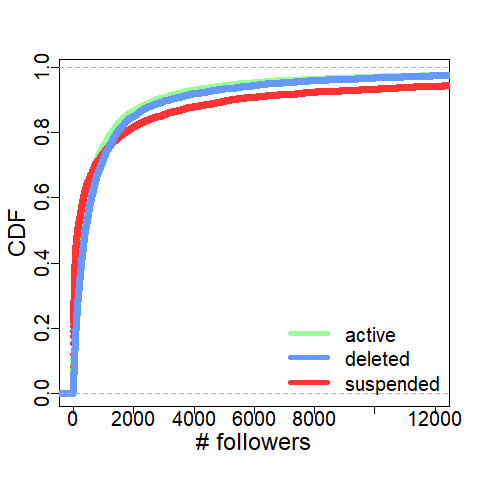}
		\captionsetup{font=footnotesize}
		\caption{Followers}
		\label{fig:sus_del_new_followers}
	\end{subfigure}
	\begin{subfigure}[b]{0.24\textwidth}
		\includegraphics[width=\textwidth]{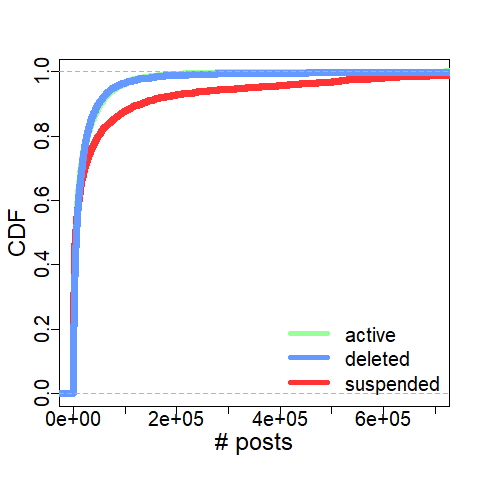}
		\captionsetup{font=footnotesize}
		\caption{Posts}
		\label{fig:sus_del_new_posts}
	\end{subfigure}	
	\begin{subfigure}[b]{0.24\textwidth}
		\includegraphics[width=\textwidth]{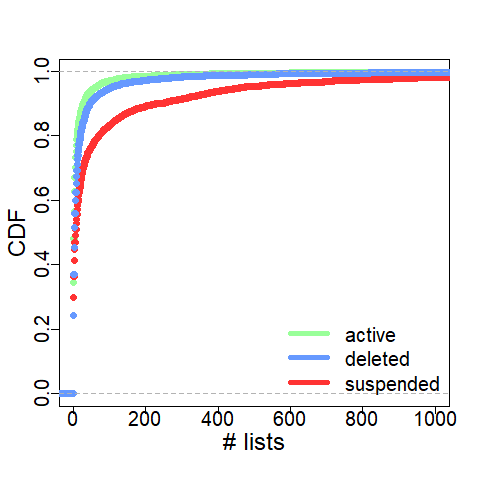}
		\captionsetup{font=footnotesize}
		\caption{Lists}
		\label{fig:sus_del_new_lists}
	\end{subfigure}

	\begin{subfigure}[b]{0.24\textwidth}
		\includegraphics[width=\textwidth]{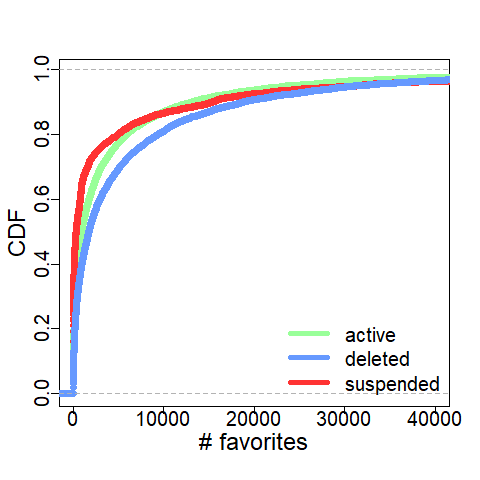}
		\captionsetup{font=footnotesize}
		\caption{Favorites}
		\label{fig:sus_del_new_favorites}
	\end{subfigure}
	\begin{subfigure}[b]{0.24\textwidth}
		\includegraphics[width=\textwidth]{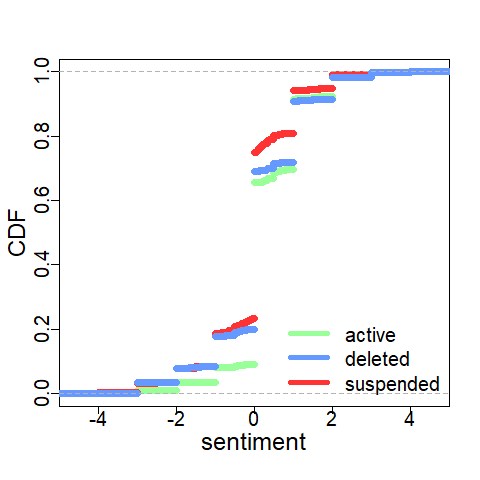}
		\captionsetup{font=footnotesize}
		\caption{Sentiment}
		\label{fig:sus_del_new_sentiment}
	\end{subfigure}
	\begin{subfigure}[b]{0.24\textwidth}
		\includegraphics[width=\textwidth]{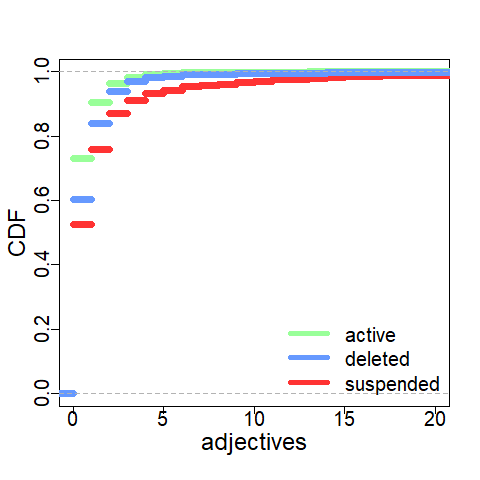}
		\captionsetup{font=footnotesize}
		\caption{Adjectives}
		\label{fig:sus_del_new_adjectives}
	\end{subfigure}
	\begin{subfigure}[b]{0.24\textwidth}
		\includegraphics[width=\textwidth]{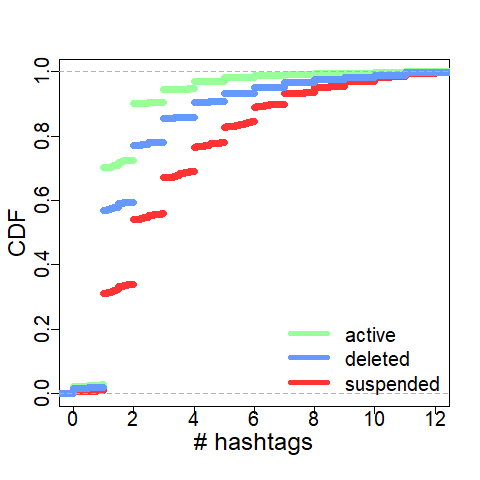}
		\captionsetup{font=footnotesize}
		\caption{Hashtags}
		\label{fig:sus_del_new_hashtags}
	\end{subfigure}	
\vspace{-0.2cm}
	\caption{CDF plots for the active, suspended, and deleted users for the: (a) Account age, (b) Followers, (c) Posts, (d) Lists, (e) Favorites, (f) Sentiment, (g) Adjectives, (h) Hashtags.}
	\label{fig:sus_del_emotional}
	\vspace{-0.2cm}
\end{figure*}

\descr{Account Age, Followers, and Friends.}
Figure~\ref{fig:sus_del_new_age} shows that users who delete their accounts have been on the platform longer than suspended users ($D = 0.21404$).
Surprisingly, for the less amount of time their account was active, suspended users managed to become more popular and thus have more followers (Figure~\ref{fig:sus_del_new_followers}) and friends (the figure has been omitted) than the deleted users.
The difference among their distributions is statistically significant with $D = 0.26208$ and $D = 0.22731$ for the followers and friends count, respectively.
The fact that the deleted users have fewer followers and friends than suspended users, implies they have less support from their social network.
On the contrary, high popularity for suspended users could have helped them attract and create additional activity on Twitter and could be a reason for delaying the suspension of such highly engaging, and even inflammatory, users.

\descr{Posts, Lists, and Favorites.}
Figures~\ref{fig:sus_del_new_posts},~\ref{fig:sus_del_new_lists}, and~\ref{fig:sus_del_new_favorites} show the distribution of the number of posts, participated lists, and favorites, respectively, made by the active, suspended, and deleted users.
We observe the suspended users to be more active than the deleted users, with respect to the number of posted tweets and participated lists ($D = 0.1252$ and $D = 0.15127$, respectively).
Also, the deleted users have an extremely similar behavior with the active users.
However, deleted users exhibit higher activity in terms of the favorited tweets than the suspended ($D = 0.24939$) and active users ($D = 0.1118$).

\descr{Sentiment and Adjectives.}
Figure~\ref{fig:sus_del_new_sentiment} shows similar behavior in terms of the expressed negative sentiments for the suspended and deleted users.
On the other hand, the deleted users tend to express with more positive sentiments than the suspended users, with the active users to express in the most positive way with respect to sentiment.
Overall, the mean value of the expressed sentiment for the active users equals to $0.287$, for the suspended users is $-$$0.046$, while for the deleted users equals to $0.092$, with the difference in their distributions to be statistically significant.

From Figure~\ref{fig:sus_del_new_adjectives} we observe that suspended users tend to use a higher number of adjectives in their posts than the active and deleted accounts.
The mean (STD) values for the active/suspended/deleted users are $0.43$ ($1.63$), $0.96$ ($6.15$), and $15$ ($227$), respectively.
Adjectives are words that modify (describe) nouns and so suspended users maybe use them in higher percentage in their effort to `attack' others.
Again, here the differences in distributions between deleted and suspended users are statistically significant with $D = 0.079577$.

\descr{Hashtags and URLs.}
Figure~\ref{fig:sus_del_new_hashtags} shows that the suspended users use a higher number of hashtags in their tweets than the deleted users ($D = 0.26011$).
Two reasons for the hashtags usage are to: (i) categorize posts and make the content searchable, and (ii) increase the visibility of the posts and encourage the online audience to join the conversation.
Thus, a higher number of hashtags used from the suspended users can be justified by the fact that in most of the cases suspended users are either spammers or exhibit abusive behavior and this is a way to increase the visibility of their messages.
The lowest number of hashtag use is observed from the active user accounts, with the difference in the distributions with the suspended and deleted user statuses to be statistically significant ($D = 0.39277$ and $D = 0.13266$, respectively).

Finally, suspended users tend to post more URLs than the deleted users which is again a way to attract more audience in reading their posts, or during a spamming effort.
The mean (STD) values for the suspended and deleted users are $0.778$ ($0.646$) and $0.563$ ($0.587$), respectively.
The difference in distributions are statistically significant with $D = 0.14646$.
Compared to the other user statuses, similar to the use of hashtags, active users tend to share the lowest number of URLs to their social network (the mean and STD values are $0.395$ and $0.534$, respectively).

\begin{table}[!t]
\centering
\resizebox{.45\textwidth}{!}{
\centering
\begin{tabular}{@{}ll@{}}
\toprule
     \textbf{Status}     & \textbf{Popular hashtags}                                                                                                                                           \\ \midrule \midrule
Active & \begin{tabular}[c]{@{}l@{}} \#trump, \#loveisland, \#nbasummerleague, \#business\\ \#brexit, \#womenshistorymonth, \#racism, \#bullying \end{tabular}              \\ \midrule
Suspended & \begin{tabular}[c]{@{}l@{}}\#gamergate, \#ftm, \#femaletomale, \#whitegenocide\\ \#porn, \#brexit, \#rape, \#stupidity, \#sexcam\end{tabular}              \\ \midrule
Deleted   & \begin{tabular}[c]{@{}l@{}}\#bullying, \#rapefugees, \#racism, \#internationalmensday\\ \#womenshistorymonth, \#voteforteamslayes, \#fakenews\end{tabular} \\ \bottomrule
\end{tabular}}
\vspace{-0.2cm}
\caption{Popular hashtags used from active, suspended, and deleted user accounts.}
\label{tbl:popular_hashtags_statuses}
\vspace{-0.2cm}
\end{table}

\descr{Popular Hashtags per Status.}
Up to now, we discussed what differences exist between active, suspended, and deleted user accounts based on a set of attributes, either user or profile-based, text-based, or network-based.
Next, we investigate in more depth the type of content posted by each type of user (based on account status), with a particular focus on the hashtags included in their tweets.
From Table~\ref{tbl:popular_hashtags_statuses}, we observe that deleted users mainly discuss about important social topics, such as the racism and immigrants.
After a manual inspection of their tweets, it seems that in many cases there is a supportive attitude from the deleted users to the `victims' related to these phenomena (e.g., `The ground you grew on doesn't make you holy - what matters is that you were a decent human being \#racism \#ilmfest').

In the case of the suspended users, based on the most popular hashtags, we observe discussions around topics which often stimulate aggressive behaviors, such as the discussions about the UK leaving the EU (\#brexit).
As it was also expected, the suspended users are often involved in spam behavior, which in some cases is related to the publishing of inappropriate content (e.g., \#porn, \#sexcam).
Therefore, suspended and deleted users tend to follow different directions in the content of the posted tweets.
Finally, active users discuss a variety of topics, such as the NBA summer league, the events that take place under Trump's governance, or about popular TV shows (e.g., \#loveisland).
Also, they seem to touch sensitive issues, such as racism and bullying (e.g., `If someone is \#bullying you, knock them the fuck out. Don't cry about it on the internet.', `Help your child practice what to say to a bully so he or she will be prepared.').
Of course, as it is quite expected, there are active users who still express with aggression (e.g., `Disgusted with @Morrisons after watching \#WarOnWaste! To be a morally corrupt idiots is one thing.') or they post inappropriate content bypassing Twitter's suspension mechanism.

\subsection{Emulating the Twitter Suspension Engine}
\label{subsec:status-classification}

Having gained an overview of the homogeneity or commonalities users have according to their Twitter status, here we investigate if the features we have analyzed so far are meaningful and correlated with account statuses, and more importantly, if they can be used to automatically classify users.
Additionally, our effort to emulate the suspension engine of Twitter constitutes a precursor for the next section which checks for more snapshots how the suspension of Twitter aligns with what our classifier predicts, providing further validation of our annotation and classification effort.
Overall, such efforts lend credence to our feature selection, and the overall classification effort of the previous sections.

To proceed with our analysis, we will use the labels extracted from the account statuses (active, deleted, suspended), in contrast to the previous classification effort which attempted to detect bullies, aggressive, and normal users.
To this end, we perform a supervised classification task using the above features as inputs and using the three statuses as labels, in an attempt to emulate the Twitter suspension engine.
The study is performed on the $1,303$ annotated users (we checked users' statuses on September 2018).
This exercise helps us understand if the features analyzed so far are predictive of the way users select to delete their accounts, or that Twitter decides to suspend users.

For the classification task, we proceed with the Random Forest classifier as it performs very well with respect to accuracy and AUC, as shown in the previous experiments.
It is also very fast to train such a classifier and does not require a large set of examples to do so.
In contrast, Twitter has available millions of examples to be used for training its suspension engine.
Therefore, it could more readily train and test a Neural Network that can outperform typical machine learning algorithms given enough training examples and computational resources.
For the next experiments we use the WEKA data mining toolkit and repeated ($10$ times) $10$-fold cross validation.

\begin{table}[!t]
\centering
\resizebox{.3\textwidth}{!}{
\centering
\begin{tabular}{@{}lllll@{}}
\toprule
               & \textbf{Prec} & \textbf{Rec} & \textbf{F1} & \textbf{AUC} \\ \midrule
active         & 0.979         & 0.996        & 0.987       & 0.995        \\
(STD)          & 0.003         & 0.001        & 0.001       & 0.002        \\ \midrule
deleted        & 0.996         & 0.873        & 0.930       & 0.994        \\
(STD)          & 0.008         & 0.031        & 0.017       & 0.005        \\ \midrule
suspended      & 0.935         & 0.821        & 0.874       & 0.997        \\
(STD)          & 0.008         & 0.023        & 0.013       & 0.001        \\ \midrule \midrule
overall (avg.) & 0.976         & 0.977        & 0.976       & 0.995        \\
(STD)          & 0.002         & 0.002        & 0.002       & 0.002        \\ \bottomrule
\end{tabular}}
\vspace{-0.2cm}
\caption{Status-based classification results.}
\label{tbl:3_class_classification_statuses}  
\vspace{-0.2cm}
\end{table}

\begin{table}[!t]
\centering
\resizebox{.35\textwidth}{!}{
\centering
\begin{tabular}{llll}
\textbf{a} & \textbf{b} & \textbf{c}                   & \textit{$\leftarrow$ classified as} \\ \cline{1-3}
99.56\%    & 0.42\%     & \multicolumn{1}{l|}{0.02\%}  & \textbf{a=active}                   \\
17.92\%    & 82.08\%    & \multicolumn{1}{l|}{0.00\%}  & \textbf{b=suspended}                \\
11.02\%    & 1.69\%     & \multicolumn{1}{l|}{87.29\%} & \textbf{c=deleted}                 
\end{tabular}}
\vspace{-0.2cm}
\caption{Status-based confusion matrix.}
\label{tbl:statuses_confusion_matrix}  
\vspace{-0.2cm}
\end{table}

\begin{table}[!t]
\resizebox{.5\textwidth}{!}{
\begin{tabular}{@{}l@{}}
\toprule
\textbf{Features (preserving order of importance)}                                                                                                                                                                                                                                                                                                              \\ \midrule
\begin{tabular}[c]{@{}l@{}}account age (16.19\%), interarrival (11.95\%), hubs (11.41\%), \#followers (11.28\%)\\ 
\#friends (9.39\%), \#lists (8.31\%), authority (6.73\%), \#favorites (6.48\%), \#posts (6.19\%)\\ 
clustering coef. (4.74\%), eigenvector centrality (4.77\%), \#URLs (2.57\%)\end{tabular} \\ \bottomrule
\end{tabular}}
\vspace{-0.2cm}
\caption{Status-based features evaluation.}
\label{tbl:statuses_features_evaluation}  
\vspace{-0.2cm}
\end{table}

From Table~\ref{tbl:3_class_classification_statuses} we observe that we succeed to detect \textbf{87.3\%} of the deleted accounts and \textbf{82.1\%} of the suspended ones.
The overall precision and recall values equal to \textbf{97.6\%} and \textbf{97.7\%}, respectively, with \textbf{97.6\%} accuracy, $0.888$ kappa, and $99.5\%$ AUC value.
From the confusion matrix (Table~\ref{tbl:statuses_confusion_matrix}), we observe that in both the deleted and suspended statuses the misclassified cases mostly fall in the active class, with the deleted and suspended classes to be almost clearly separable.
Finally, Table~\ref{tbl:statuses_features_evaluation} presents the top $12$ contributed features during the classification process.
Here, we observe again that the user- and network-based features are among the top most contributing.
From these results, we conclude that our features are meaningful in studying abusive behavior and in detecting users that are suspicious for suspension by Twitter, or that are in risk of deleting their account.

\subsection{Suppression of Offensive Behavior by Twitter}

Thus far, we saw what differences exist between Twitter statuses, by analyzing the users involved in the crawled communities for a set of features.
We also tested whether these features are suitable for distinguishing between account statuses in an automatic fashion using a machine learning method.
Next, we investigate what is the association between the labels given to users by our annotators (aggressive, bully, spam, and normal), and the account status that each of these users has (active, deleted, suspended).
In fact, we study this association for three different time periods, of about 10-13 months apart: at the end of November 2016, December 2017, and September 2018.
For this investigation we focus on the data collected from the Gamergate and baseline users.

Tables~\ref{tbl:status_check_nov},~\ref{tbl:status_check_dec}, and~\ref{tbl:status_check_sep} show the breakdown of account statuses for each label for the three time periods.
From the more recent time period (September 2018), we observe that a large number of offensive users has been suspended from Twitter: $55.17\%$ of bullies and $51.16\%$ of aggressors labeled in our dataset have been suspended.
These numbers are in stark contrast to what we have measured in the previous two snapshots in the last two years.
In fact, for the previous snapshot (December 2017) the majority of aggressive or bully users in our dataset had suffered no consequences from Twitter: $53.49\%$ of aggressive and $55.17\%$ of cyberbullying accounts were still active.

However, in the last snapshot, only $5.17\%$ of bully users and only $25.58\%$ of aggressive users are still active.
These results lend credence to our annotation process, the analysis performed, and offer validation to our observations: a large majority of the accounts that had been labeled as bullies or aggressive by our annotators  were indeed deemed inappropriate by Twitter and were suspended.
And another large portion of such accounts was proactive enough to delete their account (perhaps in an effort to evade suspension).

\begin{table}[!t]
\small
\centering
    \begin{subtable}[b]{.45\linewidth}
      \captionsetup{font=footnotesize}
      \centering
      \scalebox{0.85}{
      \begin{tabular}{@{}llll@{}}
      \toprule
      & \textbf{active} & \textbf{deleted} & \textbf{suspended} \\ \midrule
      bully     		& 67.24\%         & 32.76\%          & 0.00\%             \\
      aggressive 	& 65.12\%         & 20.93\%          & 13.95\%            \\
      spam    		& 71.57\%         & 7.71\%            & 20.72\%             \\
      normal    		& 86.53\%         & 5.72\%            & 7.75\%             \\ \bottomrule      
      \end{tabular}
      }    
      \caption{Status check on Nov 2016.}
      \label{tbl:status_check_nov}                
    \end{subtable}
    
        \vspace{0.25cm}
    \begin{subtable}[b]{.45\linewidth}
    \centering
    \captionsetup{font=footnotesize}
    \scalebox{0.85}{
    \begin{tabular}{@{}llll@{}}
    \toprule
    & \textbf{active} & \textbf{deleted} & \textbf{suspended} \\  \midrule
    bully      		& 55.17\%         & 37.93\%          & 6.90\%      \\
    aggressive 	& 53.49\%         & 23.26\%          & 23.25\%     \\
    spam     		& 50.12\%         & 12.53\%          & 37.35\%      \\
    normal     		& 77.38\%         & 8.01\%            & 14.61\%      \\ \bottomrule
    \end{tabular}}
        \caption{Status check on Dec 2017.}
    \label{tbl:status_check_dec}  
    \end{subtable} 

    \vspace{0.25cm}
    \begin{subtable}[b]{.45\linewidth}
    \centering
    \captionsetup{font=footnotesize}
    \scalebox{0.85}{
    \begin{tabular}{@{}llll@{}}
    \toprule
    & \textbf{active} & \textbf{deleted} & \textbf{suspended} \\  \midrule
    bully      		 & 5.17\%           & 39.66\%          & 55.17\%      \\
    aggressive 	 & 25.58\%         & 23.26\%          & 51.16\%     \\
    spam     		 & 49.63\%         & 11.57\%          & 38.80\%      \\ 
    normal     		 & 68.87\%         & 9.02\%            & 22.11\%      \\ \bottomrule
    \end{tabular}}
    \caption{Status check on Sept 2018}.
    \label{tbl:status_check_sep}      
    \end{subtable} 
    \vspace{-0.2cm}
        \caption{Distribution of users' behaviors in Twitter statuses.}
    \label{tbl:twitter_status}    

\vspace{-0.2cm}
\end{table}

The number of the deleted accounts between the three considered time periods, and especially between Dec. 2017 and Sept. 2018, do not exhibit important changes.
Overall, we observe that bullies tend to delete more their accounts proactively (Sept. 2018: $39.66\%$ of bully users over $23.26\%$ of aggressive users) in relation to the aggressive users.
Comparing the statuses of aggressive and bullying users between the first and the last time periods, we see an increase of $37.21\%$ and $55.17\%$, respectively, in the percentage of those suspended.
This is in alignment with Twitter's recent efforts to combat harassment cases~\cite{twitternewrules2017}, for instance, by preventing suspended users from creating new accounts~\cite{CNNtech}, or temporarily limiting users for abusive behavior~\cite{independent}.
Perhaps the increased number of the deleted accounts is an attempt from bully users to prevent their suspension, since they could return later and re-enable their account.
Overall, the high suspension rate of offensive users indicates that indeed Twitter has increased its efforts to better tackle abusive behavior on the platform.

We could also expect a higher suspension rate for the users annotated as spammers.
However, this is not the case here.
Indeed, for the first two snapshots, spammers are suspended in a higher rate than other types of users.
In fact, from November 2016 to December 2017 there is a $16.63\%$ increase of the suspended spam accounts.
However, even though in September 2018 the suspension rate is maintained at the same levels as in December 2017 (39\%), this is lower when compared to the rate observed for bully and aggressive users (55\% and 51\%, respectively).
This result indicates that for the last time period examined, Twitter probably focused more on `cleaning' the platform from abusive behavior and users, rather spam accounts.

\section{Related Work}\label{sec:related-work}

Previous work has presented several methods geared to measure and detect abusive behavior in online social networks and other platforms.
Note that ``abuse'' may indicate a wide range of abnormal behaviors, including cyberbullying, sexual exploitation, aggressive and/or offensive expression.
In this section, we review related work, focusing on two specific forms of abusive behavior, i.e., cyberbullying and cyberaggression, given users' online activity in social media.
 
\descr{Abusive Behavior in Social Media.}
Previous work has focused on detecting abusive behavior by considering online activity patterns.
Djuric et al.~\cite{Djuric2015HateSpeechDetection}  distinguish hate-related comments on Yahoo Finance with word embeddings.
In~\cite{Chen2012DetectingOffensiveLanguage} the authors aim to detect highly probable users in producing offensive content in YouTube comments.
In their method, a set of different features is incorporated, either stylish, structural, or content-specific ones, such as the appearance of words with all uppercases, the ratio of imperative sentences, or users' sexual orientation.
Nobata et al.~\cite{Nobata2016AbusiveLanguageDetection} use a machine learning based method to perform hate speech detection on Yahoo Finance and News data.
Since community-based question answering platforms are rich sources of information around a variety of topics, Kayes et al.~\cite{kayes2015ya-abuse} focus on the reported rule violations in Yahoo Answers and they find that users tend to flag abusive content in an overwhelmingly correct way (as confirmed by human annotators).
Also, some users significantly deviate from community norms, posting a large amount of content that is flagged as abusive. 
Through careful feature extraction, they also show that it is possible to use machine learning methods to predict the users to be suspended.

\descr{Cyberbullying in Social Media.}
Focusing more on cyberbullying behaviors, Dinakar et al.~\cite{Dinakar2011ModelingDetectionTextualCyberbullying} initially decompose such a phenomenon into a set of sensitive topics, i.e., race, culture, sexuality, and intelligence.
Then, they analyze YouTube comments from controversial videos based on a bag-of-words-driven text classification.
Also,~\cite{Dadvar2014ExpertsMachinesAgainstBullies} presents a method which automatically detects bully users on YouTube based on a ``bulliness'' score.
Hee et al.~\cite{Hee2015AutomaticDetectionPreventionCyberbullying} aim to detect fine-grained types of cyberbullying, e.g., threats and insults, with the consideration of linguistic characteristics in cyberbullying related content extracted from Ask.fm.
In the conducted analysis they consider three types of possible participants in a cyberbullying conversation, i.e., victim, harasser, and bystander which is further decomposed to bystander-defenders and bystander-assistants, who support, respectively, the victim or the harasser.
In~\cite{sanchez2011twitter}, Sanchez et al. exploit Twitter messages to detect bullying cases which are specifically related to the gender bullying phenomenon.
Hosseinmardi et al.~\cite{Hosseinmardi2015}, in addition to the comments posted on Instagram, they also consider the corresponding images in an effort to detect and distinguish between cyberbullying and cyberaggression.
Finally, Saravanaraj et al.~\cite{Saravanaraj2016} detect cyberbullying words and rumor texts on Twitter, as well as demographics about bullies such as their name, age, and gender.

\descr{Abusive Incidents in Game Communities.}
The rise of cyberbullying and abusive incidents, in general, is also evident in online game communities.
Since these communities are widely used by people of all ages, such a phenomenon has attracted the interest of the research community.
For instance,~\cite{kwak2015exploringcyberbullying} studies cyberbullying and other toxic behaviors in team competition online games in an effort to detect, prevent, and counter-act toxic behavior.
In~\cite{fox2014Sexism}, Fox et al.~investigate the prevalence of sexism in online game communities finding personality traits, demographic variables, and levels of game-play predicted sexist attitudes towards women who play video games.

\descr{Abusive Detection Methods.}
Various supervised approaches have been used for monitoring different instances of online abusive behaviors.
For instance, authors in~\cite{Nobata2016AbusiveLanguageDetection} use a regression model, whereas~\cite{Dadvar2014ExpertsMachinesAgainstBullies,Dinakar2011ModelingDetectionTextualCyberbullying,Hee2015AutomaticDetectionPreventionCyberbullying} rely on other methods like Naive Bayes, Support Vector Machines (SVM), and Decision Trees (J48).
In contrast, Hosseinmardi et al.~\cite{Hosseinmardi2014TowardsUnderstandingCyberbullying} use a graph-based approach based on likes and comments to build bipartite graphs and identify negative behavior. 
A similar, graph-based approach is also used in~\cite{Hosseinmardi2015}.
In all previous works a variety of attributes has been exploited in an effort to detect accurately harassment instances.
Text-related features, such as punctuation, URLs, part-of-speech, n-grams, Bag of Words (BoW), lexical features that rely on dictionaries of offensive words, and user-related ones, e.g., user's membership duration activity, number of friends/followers, are among the most popular ones.

\descr{Remarks.} 
This article presents in a unified way and, more importantly, extends our previous work on aggressive behavior in Twitter, published in~\cite{chatzakou2017measuring,chatzakou2017hypertext,Chatzakou2017MeanBD}.
Compared to the literature, we advance the state-of-the-art on cyberbullying and cyberaggression analysis and detection along the following dimensions:
\begin{itemize}
\item Propose a robust methodology for extracting user, text, and network features on Twitter, beyond what has been shown in the past.
\item Analyze user tweets, individually and in groups, and extract appropriate features connecting user behavior with a tendency of aggression or bullying.
\item Investigate Gamergate and BBC gender pay controversies and compare their activity and in-depth topics of discussion with users discussing normal topics (e.g., NBA).
\item Compare performance of various types of machine learning algorithms, including random forests and neural networks, for the detection of bullying and aggression on Twitter under different setups.
\item Contrast the performance of the best machine learning method with the suspension and deletion of offensive users from Twitter through time.
\end{itemize}

\section{Discussion \& Conclusion}\label{sec:conclusions}

Although the digital revolution and the rise of social media enabled great advances in communication platforms and social interactions, they also enable wider proliferation of harmful behavior.
Unfortunately, effective tools for detecting harmful actions are scarce, as this type of behavior is often ambiguous in nature and/or exhibited via seemingly superficial comments and criticisms.
Even now that Twitter has increased its efforts to address abusive phenomena in the platform, there are still important cases that stay under the radar, e.g.,~\cite{cesarbomb}.
Aiming to address this gap, in this paper, we analyzed the behavioral patterns exhibited by abusive users and their differences from other Twitter user categories, i.e., random users, community-related, and users posting tweets around a trending topic.
We then presented a novel system geared towards automatically classifying two specific types of offensive online behavior, i.e., cyberbullying and cyberaggression.
Finally, we analyzed Twitter's current mechanism for suspending users, by understanding the differences that exist among the different user categories and then studying Twitter's reaction against aggressive and bullying behaviors.

In the next paragraphs, we highlight the main observations from the present study and discuss how these outcomes help address the open research questions identified in the beginning of the article.

\descr{RQ1: What characteristics differentiate abusive from normal users based on their activity on diverse Twitter communities?}
In order to address this question, we studied the behavioral patterns of the different user categories, analyzing both activity and emotional characteristics.
Specifically, we studied the properties of users tweeting about the Gamergate controversy (an especially hate-prone community), the BBC gender pay controversy (a trending topic with hate and abusive elements), NBA, as well as random cases.
We observed that users who participated in discussions around specific topics, i.e., Gamergate, BBC gender pay controversy, or the NBA, have longer-running accounts on Twitter than baseline or random users, which shows the tendency of users with long activity on Twitter to participate in active and popular topics/issues.

Additionally, these three user groups are more active in terms of their posting activity and used hashtags than baseline users.
Also, users involved in the Gamergate controversy have more hashtags in their tweets, possibly aiming to further disseminate their ideas and views.
We discovered that the subject of tweets involved in the Gamergate is seemingly more offensive than that of the baseline and NBA participants, but more similar to the BBCpay users whose posts have a more aggressive connotation after the revealing of disparity in pay between the male and female top earners in the BBC.
Overall, the Gamergate community seems to be very active and well established through the years, with a more aggressive behavior than the BBCpay and NBA communities, while its activity on Twitter is especially intense.
In general, and using topic detection methods like LDA, users from each community discuss a variety of topics, but especially the one of focus to each community (e.g., BBCpay focuses on inequality in salaries, as well as Brexit and other matters of political leaning, whereas Gamergate focuses on the specific controversy, as well as socio-political issues such as the `Blacklivesmatter' movement, LGBT rights, etc.).

Based on this preliminary analysis of user groups on Twitter, we proceeded with an in-depth investigation of two specific abusive behaviors on Twitter, i.e., cyberbullying and cyberaggression.
We relied on crowdworkers to label $1.5k$ users as normal, spammers, aggressive, or bullies, from a corpus of $\sim$$10k$ tweets they posted over a period of 3 months (distilled from a larger set of $1.6M$ tweets).
For this annotated dataset, we investigated $38$ features from $3$ types of attributes (user-, text-, and network-based) characterizing such behavior.
We found that bully users organize their attacks against important and sensitive societal issues, such as feminism, religion, and politics, using aggressive and in some cases insulting language. Aggressive users express their negativity on popular topics, such as `Brexit', `maga', etc.
Also, such users have distinct behavior with respect to lists they participate, number of URLs and hashtags they use in their tweets.
In addition, aggressive users express with high negative sentiment, and they use different parts of speech (adverbs, nouns and verbs) to emphasize their negativity, in comparison to normal users, or even bullies, who make their arguments using more adjectives and verbs.
Furthermore, aggressive and bully users have a clearly different behavior than normal or spam users with respect to how many words they use in their tweets and how long these words are.

We found that bullies are less popular than normal users (fewer followers/friends, lower hub, authority, eigenvector scores) and participate in few communities.
Interestingly, bullies and aggressive users exhibit higher network reciprocity (i.e., they tend to follow-back someone who follows them), but bully users are the least central in the network, with aggressive users being somewhat similar to normal users with respect to centrality.
Aggressive users show similar behavior with spammers in terms of the number of followers, friends, and hub scores.
Similar to bullies, they also do not post a lot of tweets, but exhibit a small response time between postings, and often use hashtags and URLs in their tweets.
They have also been on Twitter for a long time, however, their posts seem to be more negative in sentiment than bullies or normal users.
On the other hand, normal users are quite popular with respect to number of followers, friends, hubs, authorities.
They participate in many topical lists and use few hashtags and URLs.

\descr{RQ2: Can we design a machine learning methodology to automatically and effectively detect abusive behavior and individuals?}
Following our analysis of user behavior, we proceeded to answer the second research question.
Working towards this goal, we extracted various user, text and network features from the tweeting activity and social network of users.
Even though user- and text-based features have been considered extensively in past works, only a limited number of network features has been used (e.g., number of followers and friends) mainly due to the difficulties in building users' social network based on the available APIs.
In the present work, an extensive list of network-based features was used, such as reciprocity, centrality scores, and community related measures.
Also, new text-based features were considered such as part-of-speech and stylistic ones.

These features were then used to train classifiers for automatic detection of these behaviors.
We compared state-of-the-art methods such as probabilistic, tree-based, and ensemble classifiers like Naive Bayes and Random Forest, as well as deep neural networks, aiming to identify the most optimal one in terms of performance and training time.
We found that traditional methods actually outperform newer ones in most cases, like neural networks, in terms of time and accuracy due to the limited size of the annotated datasets used for such complex tasks.
Moreover, to assess how well we can detect abusive behavior overall, we proceeded with a unified classification process, where the bullying and aggressive cases were merged to the `offensive' class.

While prior work almost exclusively focused on user- and text-based features (e.g., linguistics, sentiment, membership duration), we performed a thorough analysis of network-based features and found them to be very useful, as they actually are the most effective for classifying aggressive user behavior (half of the top-$12$ features in discriminatory power are network-based).
Text-based features, somewhat surprisingly, do not contribute as much to the detection of aggression, with an exception of tweet characteristics, such as number of URLs, hashtags, and average words length. 
However, such features appeared to be useful when the spam class was removed, and the focus on classification was placed on distinguishing between the aggressive, bully, and normal users.
In general, we found that aggressive users are more difficult to be characterized and identified using a machine learning classifier than bullies, since sometimes they behave like bullies, but other times as normal or spam users.
This could be the main reason for their delayed suspension.

\descr{RQ3: How has Twitter addressed the problem of abusive users in its platform?}
To complete our study, we moved on to answer the third research question.
In particular, we were interested in the characteristics of users who were suspended and if we could approximate the suspension mechanism of Twitter with our basic machine learning method.
To better understand how abusive users are currently handled by Twitter, we performed an in-depth analysis of the status of users' accounts who posted about the three topics under investigation (i.e., Gamergate, BBCpay, and NBA) and compared them with random Twitter users.
We did this investigation for three time snapshots.
Surprisingly, we found that baseline and Gamergate users tend to delete more often their accounts by choice rather than get suspended, and this is done more often than NBA or BBCpay users.
However, users in Gamergate tend to get suspended more often than baseline users.
The users from the other two communities tend to be more active, but they also present a high rate of suspension of accounts on par with Gamergate's level.

Furthermore, we investigated users' properties with respect to their account status to understand what may have led to suspension of some of them, but not all of them.
Even though suspended users in Gamergate are expressing more aggressive and repulsive emotions, and offensive language than baseline users, they tend to become more popular and more active in terms of their posted tweets.
In general, suspended users' topics of interest included controversial topics such as gamergate and brexit, as well as spam-related content.
In fact, high popularity for suspended users could have delayed their suspension, as they attract and create additional activity on Twitter.
On the contrary, deleted users have fewer friends and followers than suspended users, which implies they have less support from their social network.
Some of these users seem to express interest in topics such as racism, bullying, and women, but can also be aggressive via content related to handling of refugees.

We also attempted to emulate the suspension mechanism of Twitter by training a machine learning algorithm based on the features used for discriminating among the different user categories.
We discussed the effectiveness of our detection method by comparing prediction results of the examined users with the suspension and deletion of their accounts as observed in the wild.
We found that our features are meaningful enough in studying abusive behaviors on Twitter, as well as detecting users who are likely to be suspended from Twitter, or delete their accounts by choice.

Finally, we examined users' Twitter statuses on three different time periods, i.e., November 2016, December 2017, and September 2018.
We found that in the earlier snapshots of the users' accounts, bullies were not suspended often, but instead, took seemingly proactive measures and deleted their accounts, whereas, aggressors were suspended more often than bullies or normal users.
However, when comparing the suspension rates from November 2016 to December 2017, and then to September 2018, we observed a higher increase in the suspension rates on Twitter in the last snapshot, for both aggressive and bully users, but especially for the latter class.
These results lend credence to our annotation effort, since these accounts were deemed aggressive or bully from our analysis, and approximately two years later Twitter suspended the majority of them (only $\sim5\%$ of the bully users are still active).
This outcome is in line with Twitter's recent efforts to combat harassment~\cite{twitterRecode}, for instance, by preventing suspended users from creating new accounts~\cite{CNNtech} or temporarily limiting users for abusive behavior~\cite{independent}.
Such efforts constitute initial steps to deal with the ongoing war among the abusers, their victims, online bystanders, and the hosting online platforms.

\descr{Future Directions.} In future work, we plan to repeat our analysis on other online social media platforms such as Facebook, Foursquare, and YouTube, in order to understand if our  methods can detect similar behavioral patterns and can help bootstrap their effort to combat them.
Additionally, the proposed detection method could be extended to consider not only user-, text-, and network-related features, but also linguistic attributes of the posted content such as extensive use of idiomatic phrases, active or passive voice, and sarcasm or irony.
Such attributes could make the various behavioral patterns more distinguishable.
Finally, we plan to enhance our techniques by providing real-time detection of abusive behaviors with the aid of properly tuned distributed stream and parallel processing engines.
To this end, different modeling algorithms and processing platforms could be used, e.g., batch platforms like Hadoop vs. stream processing engines like Storm or Flink, if data are processed in batches or in a streaming fashion, respectively.

\descr{Acknowledgements.} The research leading to these results has received funding from the European Union's Marie Sklodowska-Curie grant agreement No 691025 (project ENCASE).
The paper reflects only the authors' views and the Agency and the Commission are not responsible for any use that may be made of the information it contains.
\small
\bibliographystyle{abbrv}
\bibliography{main} 

\end{document}